\documentclass[useAMS,usenatbib]{mn2e}
\usepackage{amsmath,amssymb}
\usepackage{wrapfig}
\usepackage{chngcntr}
\counterwithout{figure}{section} 
\usepackage{graphicx}
\usepackage{float}
\usepackage{hyperref}
\usepackage{multicol}

\usepackage{ragged2e}
 
\usepackage{bbm}
\usepackage{dirtytalk}
\usepackage{xcolor}

\newcommand{\AR}[1]{\textcolor{black}{#1}}

\def\orcid#1{\kern .08em\href{https://orcid.org/#1}{\includegraphics[keepaspectratio,width=0.7em]{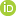}}}

\title[Comparison of AGN with GRMHD Simulations. II: M87
]{On the Comparison of AGN 
with GRMHD Simulations: II. M87} 
\author[R. Anantua et al.]{\parbox{\textwidth}{Richard 
Anantua$^{1,2,3,4,5\orcid{0000-0003-3457-7660}}$\thanks{E-mail: richard.anantua@utsa.edu}, 
\textcolor{black}{
Angelo
~Ricarte$^{2,3\orcid{0000-0001-5287-0452}}$,
George
~Wong$^{6,7\orcid{0000-0001-6952-2147}}$, 
Razieh Emami,$^{3\orcid{0000-0002-2791-5011}}$} 
Roger 
Blandford$^{4\orcid{0000-0002-1854-5506}}$, 
 Lani Oramas$^{1\orcid{0009-0006-4575-5126}}$, 
\textcolor{black}{  Hayley West$^{1\orcid{0009-0003-1798-8406}}$}, 
Joaquin~Duran$^{1\orcid{0009-0003-6244-0271}}$   
\textcolor{black}{and  Brandon~Curd$^{1,2,3\orcid{0000-0002-8650-0879}}$  \hspace{8.2cm} }
}
\\
\\
\\$^{1}$
Department of Physics $\&$ Astronomy, The University of Texas at San Antonio, One UTSA Circle, San Antonio, TX 78249, USA
\\ 
$^{2}$ Black Hole Initiative at Harvard University, 20 Garden Street, Cambridge, MA 02138, USA
\\$^{3}$ Center for Astrophysics $\vert$ Harvard \& Smithsonian, 60 Garden Street, Cambridge, MA 02138, USA
\\
$^{4}$Kavli Institute for Particle Astrophysics and Cosmology, Stanford University\, P.O. Box 20450, MS 29, Stanford, CA 94309, USA
\\
$^{5}$Astronomy Department, University of California, Berkeley, 601 Campbell Hall, Berkeley, CA 94720, USA 
\\ $^{6}$ School of Natural Sciences, Institute for Advanced Study, 1 Einstein Drive, Princeton, NJ 08540
\\ $^{7}$ Princeton Gravity Initiative, Princeton University, Princeton, New Jersey 08544, USA
}

\begin{document}

\date{Accepted ????. Received ; in original form }

\pagerange{\pageref{firstpage}--\pageref{lastpage}} \pubyear{2023}

\maketitle
\label{firstpage}

\begin{abstract}
Horizon-scale observations of the jetted active galactic nucleus M87 are compared with simulations spanning a broad range of dissipation mechanisms and plasma content in three-dimensional general relativistic flows around spinning black holes. Observations of synchrotron radiation from radio to X-ray frequencies can be compared with simulations by adding prescriptions specifying the relativistic electron-plus-positron distribution function and associated radiative transfer coefficients. A suite of time-varying simulations with various spins, \textcolor{black}{plasma magnetizations and turbulent heating and equipartition-based emission prescriptions (and piecewise combinations thereof)} \textcolor{black}{is chosen to represent distinct possibilities for the M87 jet/accretion flow/black hole (JAB) system.} \textcolor{black}{Simulation jet morphology, polarization and variation are then “observed” and compared with real observations to infer the rules that govern the polarized emissivity. Our models support several possible spin/emission model/plasma composition combinations} supplying the jet in M87, whose black hole shadow has been observed down to the photon ring at 230 GHz by the Event Horizon Telescope (EHT). \textcolor{black}{Net linear polarization and circular polarization constraints favor magnetically arrested disk (MAD) models whereas resolved linear polarization favors standard and normal evolution (SANE) in our parameter space.} We also show that some MAD cases dominated by intrinsic circular polarization have near-linear V/I dependence on un-paired electron or positron content while SANE polarization exhibits markedly greater positron-dependent Faraday effects – future probes of the SANE/MAD dichotomy and plasma content with the EHT. This is the second work in a series also applying the “observing” simulations methodology to near-horizon regions of supermassive black holes in Sgr A* and 3C 279.
\end{abstract}
\begin{keywords}
\textcolor{black}{Galaxies: active; MHD  (magnetohydrodynamics), black hole physics, radiative transfer; techniques: interferometric}
\end{keywords}

\section{Introduction}

Over a century ago, M87 was described as a ``curious straight ray" by Heber Curtis \citep{1918PLicO..13....9C} due to its relativistic jet, and nearly three quarters of 
a century ago, it was identified as a discrete radio source \citep{1949Natur.164..101B}. M87 is now the best studied jet/accretion flow/black hole (JAB) system, and the first to be imaged down to the horizon scale by the Event Horizon Telescope \citep{EHT2019I,EHT2019II,EHT2019III,EHT2019IV,EHT2019V,EHT2019VI}.
Throughout the years in which the giant elliptical galaxy M87 has been observed from its lobes to its core, we have learned that it is one of the closest examples of a common physical phenomenon, the production of twin, relativistic jets by accreting, spinning, massive black holes. In recent years, our understanding of how jets form, propagate and radiate has advanced considerably. Much of this progress can be credited to advances in observational capability, throughout the electromagnetic spectrum. In particular, the technique of VLBI (including polarimetry) has been extended to higher frequency, where the angular resolution is finer and depolarization is smaller. Gamma-ray observations have also contributed much. Equivalent progress can be seen in the numerical simulation of non-axisymmetric, general relativistic, hydromagnetic flows where specific models can be evolved dynamically with numerical confidence. The challenge today is to reconcile these two approaches.    

This reconciliation needs to take place at several levels. Radio/mm/submm jets can  
be imaged down to merely tens of 
gravitational radii (defined henceforth to be $M\equiv GM_\mathrm{H}/c^2$, where $M_\mathrm{H}$ is the mass of the hole). Structure has been identified down to $\sim10M$ \citep{2016Galax...4...54F}) and beyond by the Event Horizon Telescope (EHT) project, which 
has made linearly polarized images with resolution limit $\sim5M$ \citep{EHT2021VII,EHT2021VIII}. Jet structure has now been connected with the emitting ring in 3.5mm intensity maps \citep{2023arXiv230413252L}. 
 
The general relativistic regime  has often been simulated under a variety of dynamical assumptions and initial conditions. There has also been progress in adding radiative transfer to these codes to take account of absorption, scattering and Faraday rotation 
\citep{2012MNRAS.421.1517D}. However, in order to achieve the ultimate goal of elucidating jet launch and collimation \citep{2014ApJ...788..120L} and to measure the spin of the black hole, it is necessary to have a better understanding of how high energy electrons are accelerated. 

Most jets are observed on the larger scale, where general relativity is unimportant. In addition to radio observations, optical and X-ray observations extend down to $\sim~0.1$'', $\sim~1$'' respectively. Long term monitoring, using VLBI, has taught us much about the apparent motion of emission sites within jets and accounting for this is a goal too.  It is also necessary to uncover the character of the medium through which the jet is propagating-- inflow, outflow or thick, orbiting torus; fluid or hydromagnetic-- and the interaction with it. In addition, it is necessary to understand the remarkably rapid variability, notably at $\gamma$-ray energies, which cannot originate very close to the black hole and where the jets are totally unresolved. Answering these questions presents an even greater challenge to our current understanding of particle acceleration. 

Relativistic jets are associated with a large and heterogeneous sample of sources and the distribution of their properties is largely determined by the statistics of relativistic beaming. (W\textcolor{black}{e} are primarily concerned with AGN here, but the problems we are addressing are also features of gamma-ray bursts and galactic superluminal sources, from which much can also be learned.)  The orientation of a specific source is a parameter which can be adjusted to match a simulation to a particular source. However, we also know that black hole spin axes are isotropically distributed in space. Furthermore, the fluxes and images should scale with the jet powers and hole masses. All of this implies specific contributions to the overall distributions of total fluxes, apparent expansion speeds, polarization and so on in a complete sample of sources selected according to well-defined criteria. The overall nonthermal radiative efficiency can be determined observationally  
\citep{1982MNRAS.200..115S}. This, too, relates strongly to the particle acceleration.   

There have been many proposals as to how particles are accelerated under these conditions. Strong shocks are commonly invoked, but these are not efficient accelerators in magnetically-dominated flows and may be too slow to account for many observations. Supersonic jets are surely very noisy, and the associated hydromagnetic turbulence can promote second order acceleration. The very existence of fast variability, especially at $\gamma$-ray energy, suggests that unstable electromagnetic configurations lead to very rapid particle acceleration where electromagnetic energy is efficiently converted locally to high energy particles by a process we have called ``magnetoluminescence'' 
\citep{2017SSRv..207..291B}.

Clearly the program that has just been sketched is a massive undertaking and it is premature to try to execute it in full. In this paper, we limit ourselves to a smaller exercise designed to link plasma microphysics to discrete observable AGN features.  
Observing JAB simulations has already been carried out for Sagittarius A* in \cite{Anantua2020b}b 
\textcolor{black}{and \cite{Anantua2023}}, where a turbulent heating model exponentially suppressing emission from high-gas-to-magnetic pressure regions outperformed other distinct phenomenological model classes with respect to image size and spectral constraints anticipating EHT results \citep{EHT2022SgrAI}\footnote{ \textcolor{black}{The Critical Beta two-parameter model was not among the $6\%$ of models  in Sgr A* Paper I passing non-EHT constraints on 86 GHz flux-- with the important caveat that it was explored at a single point in model parameter space for limited inclinations relative to the fiducial models. } }. 
Here, we consider a few 
simulations and apply it to another 
well-observed source, M87 -- with the added flexibility of comparing two turbulent heating prescriptions, using a separate presription for jet regions, and including positrons. We will pay most attention to varying the particle acceleration/emissivity prescription and explore how to see which choices come closest to reproducing the observations and calculating the underlying physical properties. There is no reason to expect the match to be especially good right now. However, better simulations and observations, both imminent, should allow this approach to be followed more productively, now armed with an arsenal of distinguishable models ranging from M87-like ordered electromagnetic jet and magnetosphere to dense, Faraday-thick turbulent plasma more suitable for other AGN.

In Section 2, we review observations of the M87 hole, disk and jet, emphasizing those that are most directly connected to the synchrotron emissivity. Section 3 presents GRMHD simulations with varying spin and accretion mode and describes commonalities and differences in their plasma flow structures.
In Section 4, we introduce 
self-consistent prescriptions for emission (and absorption),
particle acceleration and dissipation, including the essential effects of positron physics. 
In Section 5, we describe the global properties of our GRMHD simulations.
In Section 6, 
we apply our emission prescriptions with positrons to the time-dependent simulations to ``observe''  M87. Our general conclusions and plans for further investigations are collected in Section 7. 
Synchrotron radiation theory calculations for 
alternative emission model prescriptions including positrons, are expounded in the Appendix.

\section{Observations of M87}

Located at the heart of the Virgo Cluster $d=16.7\pm0.6{\rm\ Mpc}$ away \citep{2009ApJ...694..556B} (cf. Table 1), 
the bright active galaxy M87 (3C 274) serves as an exemplary laboratory for the investigation of black hole jets. Observations of the jet on all scales suggest that M87 is a FRI misaligned BL-Lac blazar. M87 has a remarkably prominent jet, with an equally remarkably faint disk centered on a large black hole.

\subsection{Black Hole}

We adopt a black hole mass of 
$(6.5\pm0.7)\times10^9{\rm M}_\odot$ \citep{EHT2019I},
corresponding to length, time, angular and energy scales $10^{13} {\rm\ m}, 9 {\rm\ hr}, 4\ \mu{\rm as}\mathrm{\ and\ } 1.2\times10^{57} {\rm\ J}$, respectively.
The associated Eddington luminosity is $\sim8\times10^{40}{\rm\ W}$. A lower bound of $a>0.2M_\mathrm{M87}$ has been derived for the spin of the hole \citep{2009ApJ...699..513L}, assuming M87's SMBH is surrounded by a prograde, radiatively inefficient accretion flow (RIAF). Hybrid jet/advection-dominated accretion flow (ADAF) models \cite{2017MNRAS.470..612F} have provided an estimate as high as $a=0.98M_\mathrm{H}
$. We first adopt an intermediate angular frequency of $\Omega_\mathrm{H}=0.35
/(GM_\mathrm{H}
/c^3)$ of  $10^{-5}{\rm s}^{-1}$ for M87's black hole. Using the relations 
$J_H=G\textcolor{black}{M_\mathrm{H}}a/c$ and $\Omega_\mathrm{H}=\frac{a}{r_+^2+a^2}$, where $r_+=\frac{1}{2}(r_S+\sqrt{r_S^2-4(J_H/(M_\mathrm{H}c))^2})$ is the radius of the outer Kerr horizon, the corresponding dimensionless black hole spin is $a/M_\mathrm{H}=0.94
$. We also consider lower spin prograde $a/M=0.5$ ($\Omega_\mathrm{H}=0.13
/(GM_\mathrm{M87}/c^3)$) and retrograde $a/M=-0.5$ cases
. Observations of the jets on all scales suggest that they, and by hypothesis, the spin of the hole, are inclined to the line of sight at an angle $\theta=20^\circ$ \citep{Wang01052009,2016MNRAS.457.3801P}. The extractable rotational energy and angular momentum are $\sim2\times10^{56}{\rm J}$ and $\sim4\times10^{61}{\rm kg\ m}^2{\rm s}^{-1}$, respectively -- ample to power the jets observed today for a Hubble time without any accretion. Henceforth, we measure all lengths, angles and times in units $M$ set by M87's black hole, i.e., $GM_\mathrm{M87}/c^2$, $(GM_\mathrm{M87}/c^2)/d$, $GM_\mathrm{M87}/c^3$, respectively (confer Table 2). When the spin direction is 
along the general direction of the angular velocity of the orbiting gas \citep{2013ApJ...770...86W}, it is 
aligned with the receding jet. 

\subsection{Radio-mm-submm Observations}
Following the pioneering observations of \cite{1995AJ....109..500J}, there have been many impressive high resolution observations of the inner jet of M87.
\begin{itemize}
\item A 15 GHz map with beam $600\ \mu{\rm as}\times1300\ \mu{\rm as}\equiv\ 150M\times\ 325M$ measured out to a projected length $Y\sim2\times10^4M$ \citep{2007ApJ...668L..27K} and therefore a length along the deprojected jet $z\sim6\times10^4M$.
\item Pilot monitoring at 22~GHz with resolution $\sim250M\times250M$ extending out to projected length $\sim6000M$ and showing superluminal motion with speed up to $\sim1.6c$ \citep{2017arXiv170602066H}
\item A 43 GHz VLBI time sequence-- 11 maps made over 210 d with a beam FWHM 
$\sim 55M\times115M$, which extends out to $z\sim6000M$ in projected radius \citep{2008JPhCS.131a2053W,2016Galax...4...46W,2016A&A...595A..54M} (Fig.~\ref{CWalkerM87Movie1
}). Apparent speeds up to $\sim2c$ are observed.
\item An 86~GHz image with resolution $\sim20M\times60M$ extending out to $z\sim3000M$ and exhibiting $\sim20$~percent linear polarization and strong Faraday rotation \citep{2016ApJ...817..131H}, and a Global mm-VLBI Array (GMVA) 86 GHz M87 observation 
exhibiting a limb-brightened jet base (see \cite{Kim2018} Fig. 4).
\item Event Horizon Telescope (EHT) observational data made with effective beam of size $\sim10M$ \citep{2015ApJ...807..150A}. Increased coverage with next generation facilities may further bridge the gap between this jet structure and the ringlike EHT 2019 observation around the central supermassive black hole with greater precision, building off of \cite{2023arXiv230413252L}. 
\end{itemize}

These observations confirm that the approaching radio jet is modestly relativistic and collimated within $z\sim30M$. They are also broadly consistent with there being a self-absorbed innermost jet called the core with constant brightness temperature $\sim2\times10^{10}{\rm\ K}$ and flux density $S_\nu\sim 1{\rm\ Jy}$ for $3{\rm\ GHz}\lesssim\nu\lesssim 300{\rm\ GHz}$. At higher frequency, the entire jet appears to be optically thin. The resolved jet structure accounts for a minority of the flux at all frequencies but is strongly edge-brightened and the mean intensity decays 
as an inverse square with distance from the axis. 
There are indications that the southern edge is brighter than the northern edge. The jet is quite variable. The shape of the jet is roughly parabolic for $30M\lesssim z\lesssim7000M$ with the separation of the brightened edges roughly given by $\sim6z^{1/2}$. At larger radii, the jet expansion is closer to linear. 

\subsubsection{Observational Model}

We have made a simple analytical model which captures many of the features of the time-averaged observation over the range of frequencies where the inner jet has been resolved. While this does not do full justice to the observations, it is sufficient for our purpose.

We introduce a Cartesian coordinate system on the sky with $Y$ measuring distance along the jet from the hole and $X$ distance across it (in units of $M$). The intensity satisfies
\begin{equation}
I=\frac{I_0e^{\xi(1-\xi)}}{1+\nu_{300}^{8/3}Y^2},
\end{equation}
where $\xi=X^2/Y$.

\begin{figure}
\includegraphics[height=125pt,trim = 6mm 1mm 0mm 1mm]{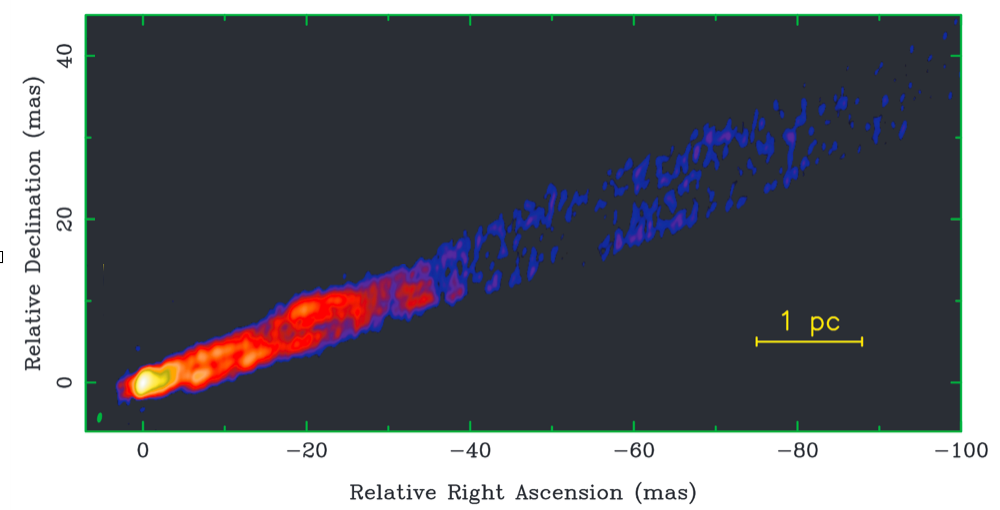}
\caption{VLBA 2 cm image of M87. The swirling jet substructure suggests magnetic Kelvin-Helmholtz instability-- a feature also seen in the simulation movie of the corkscrewing jet \url{http://richardanantua.com/sample-page/jetaccretion-diskblack-hole-movies/}.
Image adapted with permission of Dan Homan, Yuri Kovalev, Matt Lister and Ken Kellermann.
}\label{SwirlingJet}
\end{figure}

\begin{figure}\nonumber
\begin{align}
\hspace{-0.5cm}\includegraphics[height=125pt,trim = 6mm 1mm 0mm 1mm]{M87MNRASVLAFig2a
.png}  & 
\end{align}
\begin{align}
& 
\hspace{-0.5cm}\includegraphics[height=130pt,width=130pt,trim = 6mm 1mm 0mm 1mm]{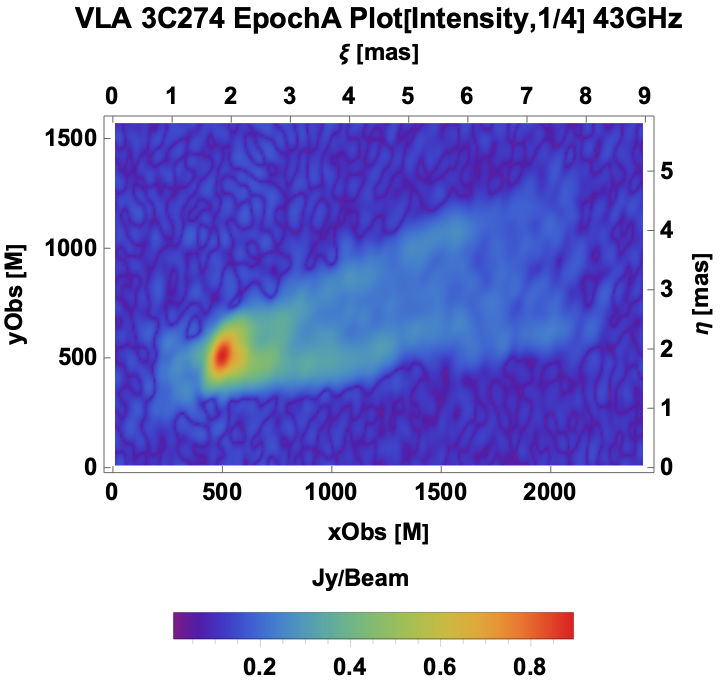
} & 
\includegraphics[height=130pt,width=130pt,trim = 6mm 1mm 0mm 1mm]{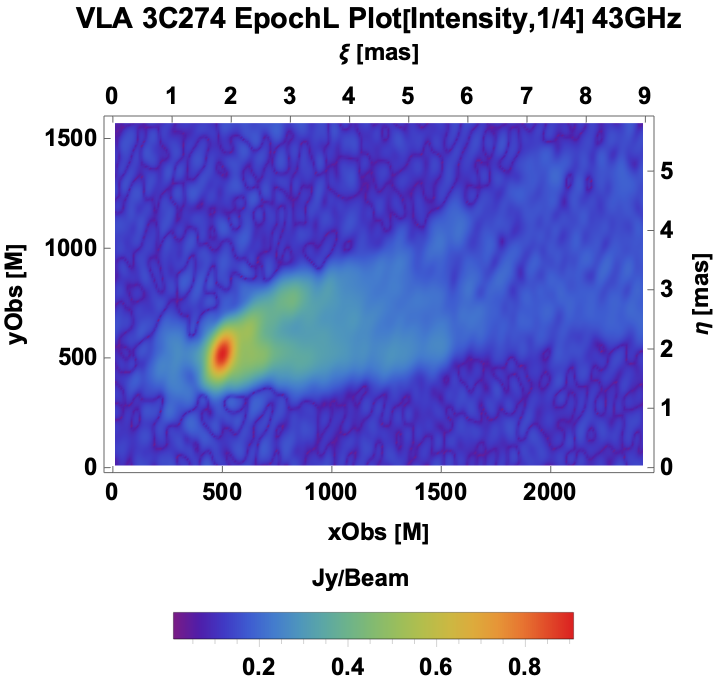
} & 
\end{align}
\caption{M87 VLA observation on linear intensity scale (Top) and observational movie snapshots at $t_\mathrm{Obs0}$ (Bottom Left) side-by-side with a snapshot at $t_\mathrm{Obs0}+10t_\mathrm{Step}$ (Bottom Right) for  
$t_\mathrm{Step} = 21\mathrm{days}\approx 56M_\mathrm{M87}$
monotonically transformed by $(\cdot)^{1/4}$ for visual clarity on the bottom row. The beam dimensions are $4.3\cdot10^{-4}\mathrm{arcs}\cdot 2.1\times 10^{-4}\mathrm{arcs}$. These images can be viewed as a movie sans transformation, courtesy of Craig Walker and his collaborators, here:
\url{http://www.aoc.nrao.edu/~cwalker/M87/}.}\label{CWalkerM87Movie1}
\end{figure}
The observed radio-mm-submm isotropic power is dominated by mm observations and is $\sim10^{34}$~W 
\citep{2016MNRAS.457.3801P}. A beaming corrected guess might be as high as $\sim10^{35}$~W. 

The spectral energy distribution of the central 0.4 arcs (32 pc) of M87 from \cite{2016MNRAS.457.3801P} is $\nu F_\nu\approx 10^{11.6}$ Jy Hz for $10^{10.5}$ Hz $<\nu<$ $10^{14.5}$ Hz. The radio-to-UV bolometric luminosity is $3.6\cdot 10^{-6}L_\mathrm{Edd}=2.7\cdot 
10^{35}$ W \citep{2016MNRAS.457.3801P}.

\subsubsection{Event Horizon Telescope Observations}

In April 2019, the Event Horizon Telescope released the first images resolving the boundary of a black hole \citep{EHT2019I}, ushering in the age of direct observation of horizons. The results have already resolved a wide discrepancy in the black hole mass for M87*--  from stellar dynamical measurements of $6.6\times10^9M_\odot$ \textcolor{black}{from} \cite{2011ApJ...729..119G}, compared to gas dynamical measurements \textcolor{black}{from} \cite{2013ApJ...770...86W} who measured half this mass. Simulations concordant with EHT M87 observations require that the central black hole have nonzero spin in order to explain the presence of the jet powered by the Blandford-Znajek mechanism \citep{BZ1977}, and polarized observations \citep{EHT2021VII,EHT2021VIII} indicating the hole is supplied vertical magnetic flux further support this interpretation.


\begin{figure}\nonumber
\begin{align}
& 
\hspace{-5.1cm}  
\includegraphics[height=150pt,trim = 6mm 1mm 0mm  1mm]{M87EHTObservation11Apr2017Mathematica
.png} & \\ 
\hspace{1.1cm}  
\includegraphics[height=150pt,trim = 6mm 1mm 0mm 1mm]{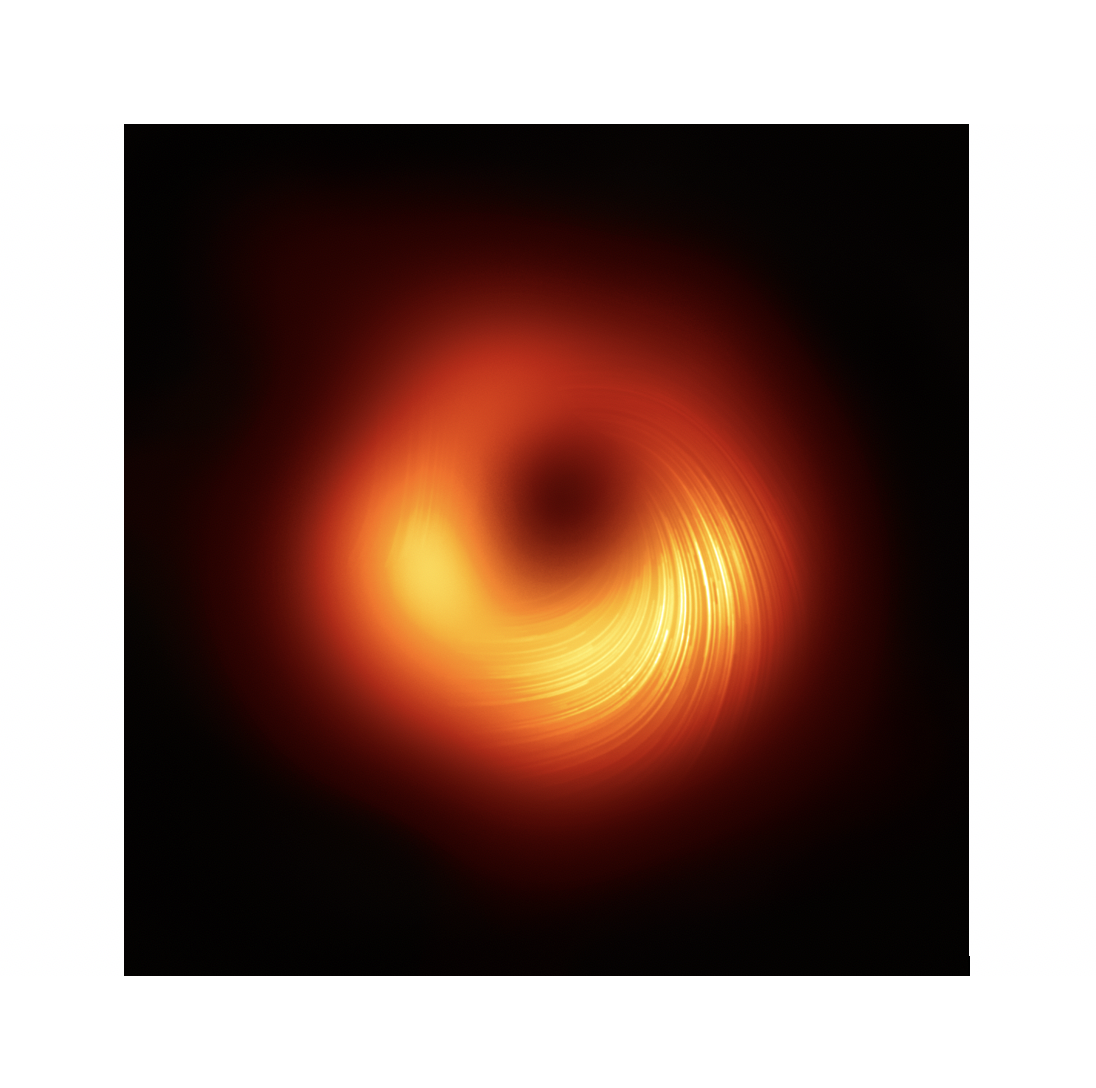} & 
\end{align}  
\caption{EHT M87 2017 observational campaign: intensity map 
on \textcolor{black}{(Top)} 
and linear polarization map 
\textcolor{black}{(Bottom)}
\citep{EHT2021VII}.
}\label{EHTM87Obs}
\end{figure}

\subsection{Optical-Infrared Observations}

The Hubble Space Telescope has provided us with stunning optical band observations of the M87 jets, including knots with superluminal components and flatter spectrum than the rest of the jet \citep{2001ApJ...551..206P}. The most famous feature, HST-1 $\sim 80\ \mathrm{pc}$ from the nucleus, produces blobs that appear to move up to $6c$ on the observer plane \citep{1999ApJ...520..621B}. $\mathit{HST}$-1 has exhibited 40$\%$ variability between 1993 and 1997 \citep{2001ApJ...551..206P}. 
 
 The isotropic radiant power of M87 has a bolumetric luminosity 
 given by $L_\mathrm{bol}\sim3\times10^{35}\mathrm{W}$ \cite{2016MNRAS.457.3801P}. Using the quiescent spectral energy distribution from \cite{2016MNRAS.457.3801P}, it is inferred that the upper limit to disk power is $L_\mathrm{disk}\le3.4\cdot 10^{41}$erg/s. At 10$\%$ efficiency, $L=\eta\dot{m}c^2$ implies an upper bound to the mass accretion rate of $3.8\cdot 10^{21}$\ g/s\ =\ $6\times 10^{-5}M_\odot/\mathrm{yr}$. The twofold change in M87's observed bolometric luminosity from its quiescent state value to ($L=5.4\times 10^{42}$ergs/s) during its 2005 outburst \cite{2016MNRAS.457.3801P} suggests a Doppler boosting factor of 8-16 given the mass accretion rate upper limit.



\subsection{X-ray Observations}

M87 has been observed at X-ray wavelengths by $\mathit{Chandra}$ \citep{2002ApJ...568..133W}. The equipartition magnetic field value for the knot HST-1 was found to be $\sim 3\times 10^{-4}$ G  \citep{1989ApJ...340..698O}.

The steady jet isotropic luminosity in 2-10  keV X-rays is $\sim3\times10^{34}$~W \citep{2016MNRAS.457.3801P}. However, the variable source HST-1 can be roughly ten times brighter including 
observations at optical wavelengths. As 
HST-1 also displays features moving towards us exhibiting apparent  speeds $\sim6c$, 
we suppose that this is a small part of the flow that, unlike the main body of the jet, is directed along our line of sight. It may not contribute significantly to the true integrated jet bolometric power.

M87 resides at the center of the $\sim10^{15}{\rm M}_\odot$ Virgo cluster of galaxies, surrounding M87 in a cooling flow region with X-ray luminosity  $10^{36}$~W \citep{2001ApJ...554..261C}. The cooling time is short compared with the flow time, suggesting that the hot gas is maintained in rough dynamical equilibrium by mechanical heating associated with the jets (though other possibilities have been widely discussed). If this is the case, then each jet must carry a total power that is significantly larger than this and which is mostly carried off by buoyant bubbles. On this rather uncertain basis, we estimate a total jet power of $L_{\rm jet}=5\times10^{36}$~W, noting that M87 could be in a relatively dormant state right now. 

\subsection{Gamma-ray Observations}

The $\mathit{Fermi}$ Large Area Telescope (LAT) has seen M87 as a gamma ray point source with variable TeV emission on yearly timescales \citep{Aharonian1424} and no observed yearly or decadal variability at MeV-GeV ranges \citep{2009ApJ...707...55A}.

The observed $>100$ MeV flux is $2.54\cdot 10^{-8}$ photons cm$^{-2}$ $s^{-1}$ and the corresponding luminosity is $4.9\cdot 10^{41}$ erg $s^{-1}$ \citep{2009ApJ...707...55A}. 

\subsection{Galaxy and Cluster}

The largest scale observations \citep{2000ApJ...543..611O,2012A&A...547A..56D,2017ApJ...844..122F} show that the M87 jets interact strongly with the surrounding medium. The jet orientation changes at a radius $\sim2{\rm\ kpc}$. The jets have inflated large, buoyant bubbles that probably produce enough dissipation to balance radiative cooling loss. M87 has been active but quite variable for several Gyr, although the underlying mass supply rate may have varied significantly over this time. Analyzing the most recent activity leads to a conservative estimate of the current power per jet $\sim10^{37}{\rm\ W}$.

\subsection{Summary}

M87 is fairly extreme in many respects. It has the most massive black hole we can study in detail. Its disk luminosity is $\lesssim3\times10^{-7}L_{\rm Edd}$ while it creates jets with total power $\gtrsim3\times10^{-4}L_{\rm Edd}$ which have been shown to be collimated on scales $\lesssim10M$. Most importantly, we
have recently begun to learn
much more about the region within $\sim100M$ 
from EHT observations. This makes it an excellent source to model.

\section{3D GRMHD Simulations}

\subsection{Historical Overview}

We now give a brief synopsis of the development of GRMHD simulations to contextualize the principal simulation used in this work.

Koide and Meier pioneered GRMHD simulations \citep{2000ApJ...536..668K} of jet formation in the magnetosphere of rapidly rotating Kerr black holes. Their code solves the GRMHD equations for conservation of particle number, momentum and energy in a Kerr metric for $0.75 r_S<x^1<20r_S$ (where $x^1$ is the radius $r$ in Boyer-Lindquist coordinates). 

The next major advance was the high-accuracy relativistic magnetohydrodynamics (HARM) code of Gammie 
McKinney and Toth \citep{2003ApJ...589..444G}. This conservative numerical scheme for integrating the GRMHD equations is guaranteed to obey shock jump conditions at discontinuities in the fluid variables.

HARM led to a number of applications, such as Dexter and Fragile's disk simulations of Sgr A* \cite{2012JPhCS.372a2023D} and \cite{2016MNRAS.463.3437M} and 
\cite{2009MNRAS.394L.126M} simulations of stable relativistic jets.

Later simulations by Farris and Gold account for strong gravitational 
curvature near black hole binaries \citep{2012PhRvL.109v1102F} or describe models with magnetorotational (MRI) disk instability and magnetically arrested disks (MAD) \citep{2016arXiv160105550G}. The simulations in \cite{2016arXiv160105550G} that model the disk -- which is governed by distinct emission mechanisms from the jet -- require an evolution equation for proton temperature $T_p$ and electron temperature $T_e$. 
\textcolor{black}{Later simulations by Aloy et al. \citep{2015ASPC..498..185A} have merged the Multi-Scale Fluid-Kinetic Simulation Suite with the high resolution 3D RMHD code MR-GENESIS.}


\begin{table*}
\caption{M87 Geometry
}
    \begin{tabular}{| c | c | c | c |c|
    }
    \hline \hline
Distance from  Earth$^1$& Schwarzschild  radius$^2$& Apparent angular width$^3$& Jet opening angle$^4$& Jet viewing angle$^5$\\ 
\hline \hline                              
 $(16.7\pm 0.6)$MPc &  $(6.35\times 10^{-4}\pm3.69\times10^{-5}$)pc  & 3.9$\mu$as &  $5^\circ$ (at 100pc) & $10^\circ - 19^\circ$\\
($5.15\times10^{25}\pm2.78\times10^{24}$)cm & ($1.94
\times 10^{15}\pm2.09
\times10^{14}$)cm &&$60^\circ$ (at the core)& \\
\hline
\hline
\end{tabular}
$^1$\cite{2009ApJ...694..556B}$\qquad$$\qquad$$\qquad$$\qquad$$\qquad$$\qquad$$\qquad$$\qquad$$\qquad$$\qquad$$\qquad$$\qquad$$\qquad$$\qquad$$\qquad$$\qquad$$\qquad$$\qquad$$\qquad$$\qquad$$\qquad$$\qquad$\\
$^2$\cite{EHT2019I} $\qquad$$\qquad$$\qquad$$\qquad$$\qquad$$\qquad$$\qquad$$\qquad$$\qquad$$\qquad$$\qquad$$\qquad$$\qquad$$\qquad$$\qquad$$\qquad$\\
$^3$\cite{2011ApJ...729..119G}$\qquad$$\qquad$$\qquad$$\qquad$$\qquad$$\qquad$$\qquad$$\qquad$$\qquad$$\qquad$$\qquad$$\qquad$$\qquad$$\qquad$$\qquad$$\qquad$$\qquad$$\qquad$$\qquad$$\qquad$$\qquad$$\qquad$\\
$^4$\cite{2012Sci...338..355D}$\qquad$$\qquad$$\qquad$$\qquad$$\qquad$$\qquad$$\qquad$$\qquad$$\qquad$$\qquad$$\qquad$$\qquad$$\qquad$$\qquad$$\qquad$$\qquad$$\qquad$$\qquad$$\qquad$$\qquad$$\qquad$$\qquad$\\
$^5$\cite{Wang01052009}$\qquad$$\qquad$$\qquad$$\qquad$$\qquad$$\qquad$$\qquad$$\qquad$$\qquad$$\qquad$$\qquad$$\qquad$$\qquad$$\qquad$$\qquad$$\qquad$$\qquad$$\qquad$$\qquad$$\qquad$$\qquad$$\qquad$
\label{M87GeometryTable}  
   \end{table*}


\begin{table}\label{MUnits}
\caption{Code scale $M$ to physical units for M87}
    \begin{tabular}{| c | c | c |c|}
    \hline \hline
  Unit type      & $M$  \\ \hline \hline                  
Mass& $(6.6\pm 0.4)\times 10^9M_\odot$\\
Length & 
$(3.2\times10^{-4}\pm1.8\times10^{-5})$pc
\\ 
Angular width at $d_\mathrm{M87}$ & 
$(3.7\pm0.3)\mu$as
\\
Time &
$(9.1\pm0.8)$h
\\
\hline \hline
    \end{tabular}
\label{M87codeunits}  
   \end{table}

\subsection{Fiducial Simulations }

\subsubsection{Overview}
\label{section:grmhd_model_overview}

 
In this work, we use a set of three numerical GRMHD simulations of black hole accretion. The fluid simulations were produced with the KHARMA code, a GPU-based descendant of {\tt{}iharm}, a conservative second-order explicit shock-capturing finite-volume
code for arbitrary stationary spacetimes 
\citep{2003ApJ...589..444G,Prather2021}. The governing equations of ideal GRMHD can be written as a set of conservation laws; in a coordinate basis, they are
\begin{align}
    \partial_{t}\big(\sqrt{-g}\rho u^{t}\big) &= -\partial_{i}\big(\sqrt{-g}\rho u^{i}\big),  \\
    \partial_{t}\big(\sqrt{-g}T^{t}_{~\nu}\big) &= -\partial_{i}\big(\sqrt{-g}T^{i}_{~\nu}\big) + \sqrt{-g}T^{k}_{~\lambda}\Gamma^{\lambda}_{~\nu k}, \\
    \partial_t\big(\sqrt{-g}B^{i}\big) &= -\partial_{j}\big(\sqrt{-g}(b^{j}u^{i}-b^{i}u^{j})\big),
\end{align}
along with a no-monopoles constraint $\partial_i \left( \sqrt{-g} B^i \right) = 0$. Here, the rest-mass density of the fluid is $\rho$, $u^\mu$ is the fluid four-velocity, $b^\mu$ is the magnetic induction four-vector, and the magnetohydrodynamic stress--energy tensor is
\begin{align}
    T^{\mu\nu} = \left( \rho + u + P + b^2 \right) u^\mu u^\nu + \left( P + b^2 / 2\right) g^{\mu\nu} - b^\mu b^\nu,
\end{align}
where $u$ and $P$ are the internal energy of the fluid and its pressure, which is related to the internal energy via an ideal gas law equation of state with constant adiabatic index $\hat{\gamma}$ via $P = \left( \hat{\gamma} - 1 \right) u$. The effects of spcaetime are accounted for in the usual way, with $g = \sqrt{-{\rm det} g_{\mu\nu}}$, the determinant of the covariant metric and $\Gamma$ a Christoffel symbol encapsulating derivatives of the metric.
More detail about the end-to-end simulation procedure can be found in 
\cite{Wong2022}

The simulations all used outflow boundary conditions at both the inner and outer radial edges, located within the event horizon and at $1,000 GM_\mathrm{H}/c^2$ respectively. Each simulation was run from $t = 0\,GM_\mathrm{H}/c^3$ until $30,000\,GM_\mathrm{H}/c^3$ in order to provide a converged characterization of the source, although we use snapshots from the latter $25,000\,GM_\mathrm{H}/c^3$ of each simulation. In particular, we consider two MAD simulations with dimensionless black hole spins $a/M_\mathrm{H} = -0.5$ and $+0.94$. We also consider one SANE simulation with spin $a/M_\mathrm{H} = -0.5$. In MAD accretion, magnetic flux carried by the accretion flow builds up near the event horizon until the magnetic pressure near the hole is large enough to counterbalance the inward ram pressure of the accreting material. MAD accretion thus proceeds in chaotic bursts of isolated, thin plasma streams beginning far from the hole, with the overall flow characterized by occasional violent magnetic eruption events. In contrast, SANE accretion proceeds in a turbulent but consistent, disk-like flow. \textcolor{black}{Further information about the KHARMA GRMHD library can be found in \cite{EHT2022SgrAV}}.

\begin{figure}\nonumber
\begin{align} 
\includegraphics[height=150pt,trim = 6mm 1mm 0mm 1mm]{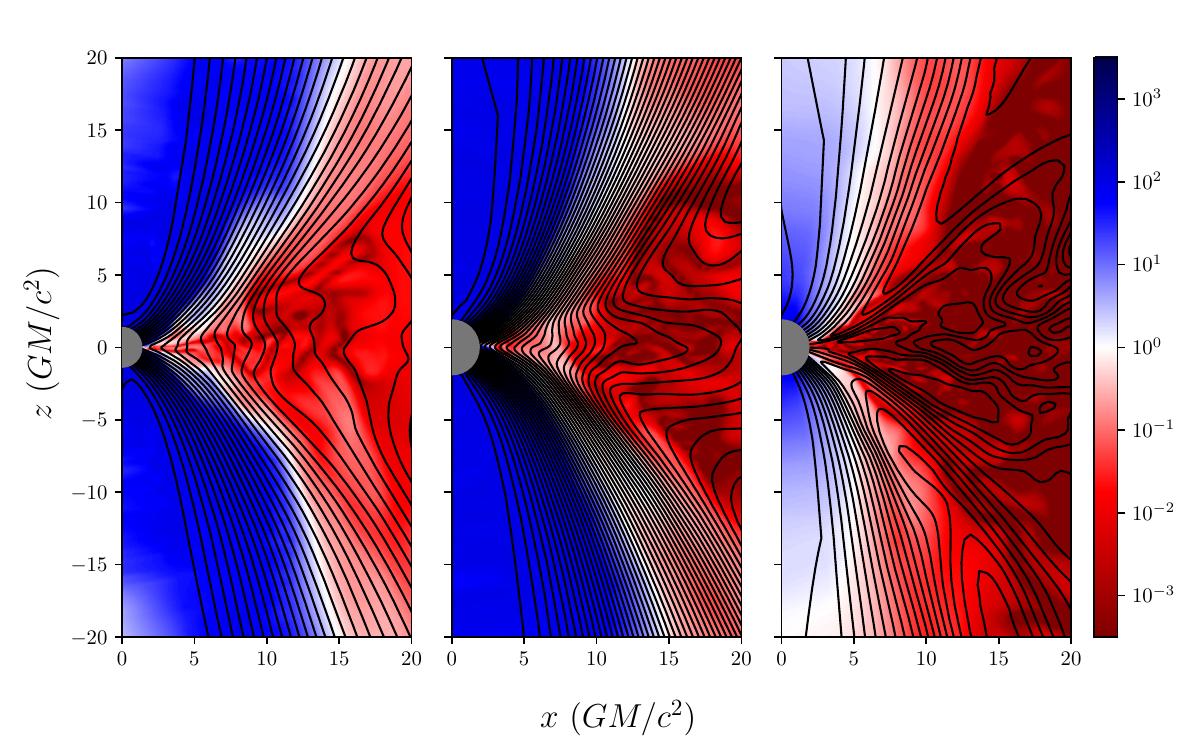} & 
\end{align}
\caption{Magnetization $\sigma \equiv b^2 / \rho$ in azimuthal slices for fiducial simulations MAD $a=0.94$ (Left) MAD $a=-0.5$ (Middle) and SANE $a=-0.5$ (Right).}\label{SigmaSlicesFiducialSims}
\end{figure}
\begin{figure}\nonumber
\begin{align}
\includegraphics[width=250pt,trim = 6mm 1mm 0mm 1mm]{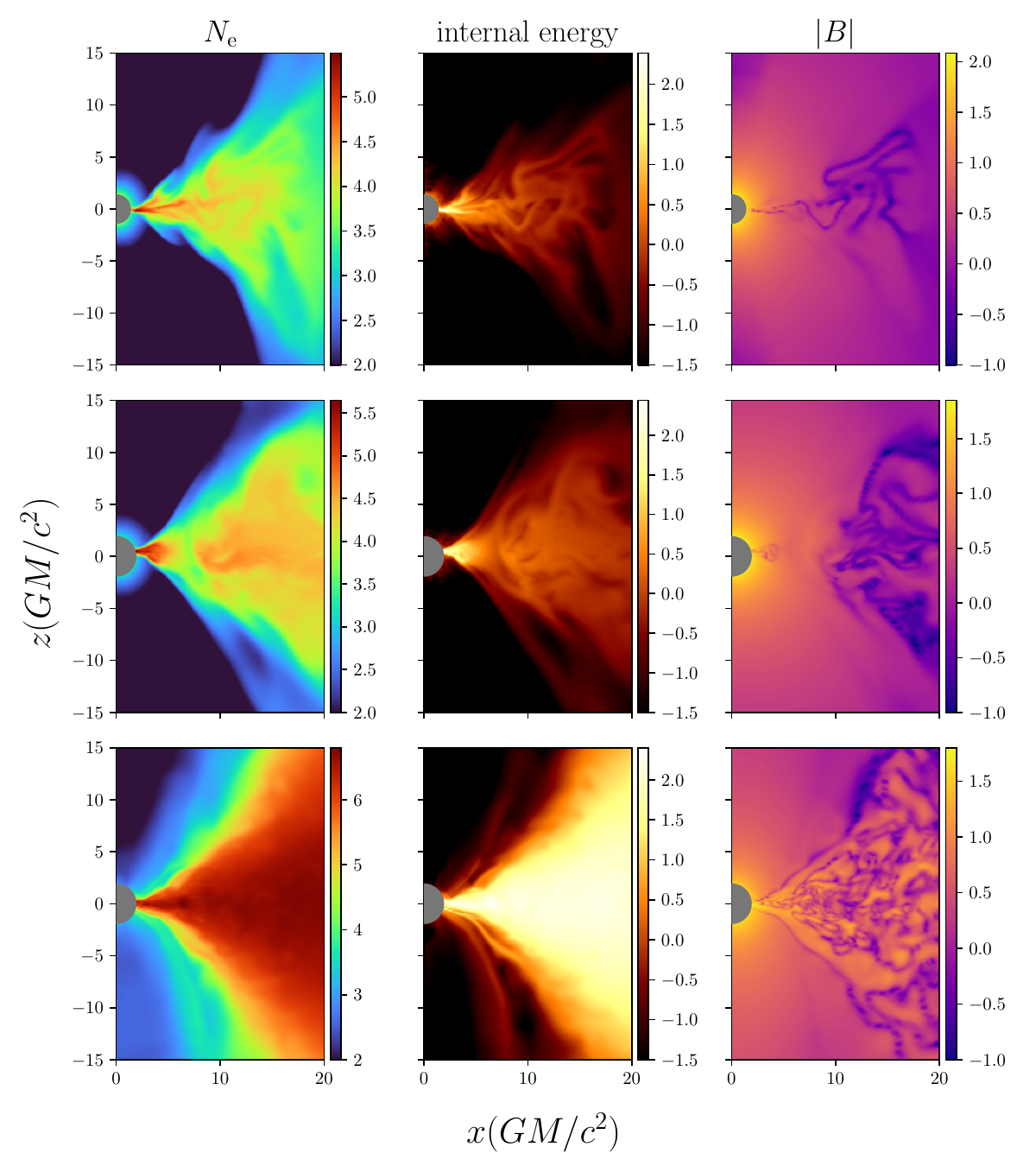} 
\end{align}
\caption{Vertical slices for electron number density $N_e$, internal energy $U$, and magnetic field strength $|B|$ in cgs units for fiducial simulations MAD $a=0.94$ (Top) MAD $a=-0.5$ (Middle) and SANE $a=-0.5$ (Bottom).  
}\label{BNeUSigmaSlicesFiducialSims}
\end{figure}
\begin{figure}
\includegraphics[trim = 0mm 0mm 0mm 0mm, clip, height=150pt
]{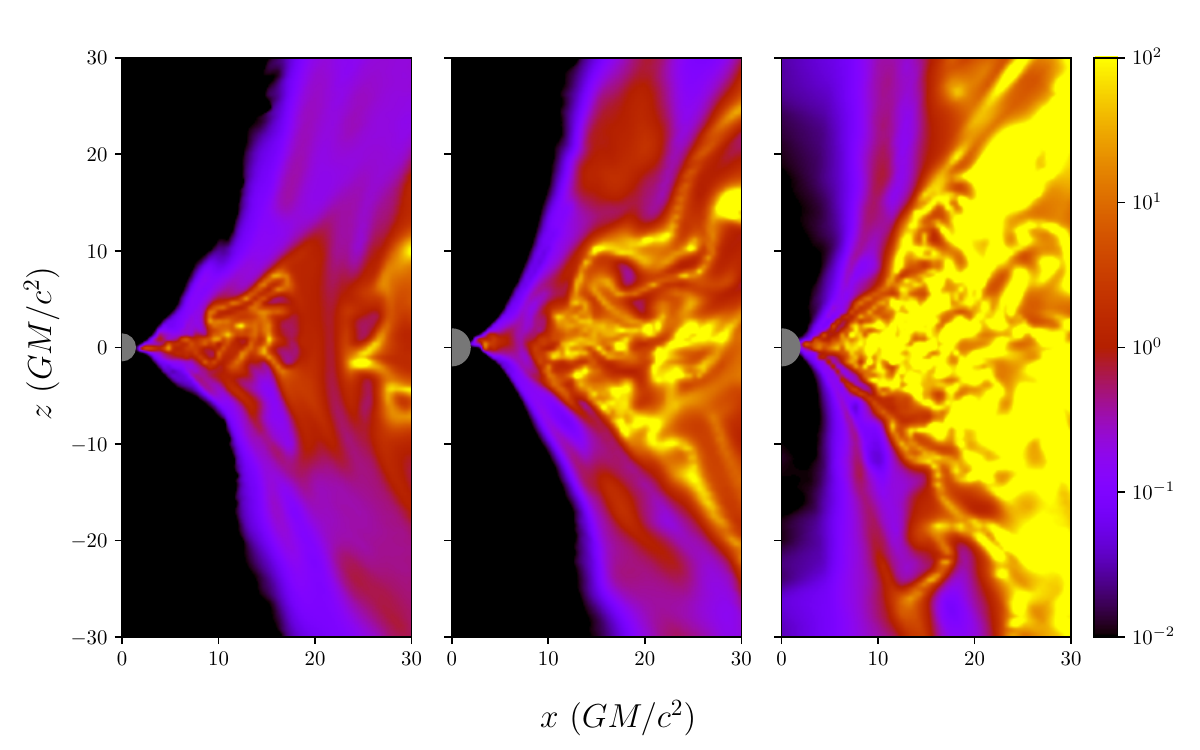} 
\caption[Simulation Gas to Magnetic Pressure]{Simulation plasma $\beta$ parameter $\equiv P_{\rm gas} / P_{\rm mag}$ on azimuthal slices for fiducial models MAD $a=0.94$ (Left) MAD $a = -0.5$ (Middle) and SANE $a=-0.5$ (Right).}\label{BetaModelPOnbSq}
\end{figure}

Figure
\ref{SigmaSlicesFiducialSims} shows xz- (poloidal) slices of the fiducial simulations snapshots at $T=25,000M$ for MAD $a/M_H=0.94$, $a/M_H=-0.5$ and SANE $a/M_H=-0.5$ for the key quantity magnetization $\sigma=\frac{b^2}{\rho}$. Expectedly, the MAD simulations are more highly magnetized-- particularly along the polar regions where there is a relativistic outflow. The columns in Figure
\ref{BNeUSigmaSlicesFiducialSims} representing electron number density, internal energy and magnetic field strength for the same simulations show that turbulence in the equatorial inflow-- such as that driven by the magnetorotational \textcolor{black}{instability}--  is particularly prominent for the SANE case. The MAD/SANE magnetic substructure and field strength dichotomy is also apparent in slices of plasma $\beta$ in Fig. \ref{BetaModelPOnbSq}.

\subsubsection{Mass Accretion Rate}

The mass accretion rate in the code is an adjustable parameter with which the flux scales. In this work, our target flux for synthetic images is 0.5 Jy. Now that we have described the simulations we are using to test the radiative properties of M87's JAB system, we add the key physics governing  energy transfer from the GRMHD plasma to the high energy particles 
responsible for the observed emission.

\section{Model Prescriptions for Emission, Particle Acceleration and Dissipation}

\subsection{General Considerations}

We suppose that the radio and mm emission is synchrotron radiation 
due to particle acceleration arising from a number of different mechanisms discussed here and in the  
Appendix. 
We expand upon the synchrotron prescriptions implemented for jet models in \cite{Anantua2018,Anantua2020a}a, where 
 the relativistic electron distribution function is a power law with slope 2, which implies that the emissivity in the comoving (primed) frame $j_{\nu',\Omega'}'(\nu',{\bf n}')\propto{\tilde P}_e|{\bf B}'\times{\bf n}'|^{3/2}\nu'^{-1/2}$, where ${\tilde P}_e$ is the (presumed isotropic) partial pressure of the electrons emitting at the frequency $\nu'$. 
 The choice to scale near-horizon jet emissivity in terms of magnetic pressure is motivated by the observation  
 that the jet becomes increasingly simple and electromagnetic --- high $\sigma$ --- as the horizon is approached and as exhibited by the RMHD simulations. However, we note the assumptions underlying  some of the previous simulations make no provision for the particle acceleration and transport resulting in the spatial and temporal variation of ${\tilde P}_e$. In this work, we use 
 a more generic electron pressure 
 formalism distinguishing the highly magnetized outflow from the less magnetized inflow to prescribe emission models. It is the primary objective of this investigation to explore how much this matters and if indeed {\sl any} prescription for the particle acceleration is compatible with existing and anticipated VLBI imaging.

Perfect MHD defines a set of reference frames in which the plasma, treated as a fluid, is at rest. The motion perpendicular to the magnetic field has a velocity ${\bf v}_\perp={\bf E}\times{\bf B}/B^2$ in the simulation frame. In general, the component of the fluid velocity resolved along the field ${\bf v}_\parallel$ is problematic when the inertia of the plasma is ignorable as we are implicitly assuming here. It should be emphasized that the minimum charged particle density needed to support the space charge and current 
\citep{1969ApJ...157..869G}
is orders of magnitude smaller than what is needed to account for the radio and $\gamma$-ray emission. It should also be stressed that efficient and progressive pair production and particle acceleration is to be expected in AGN jets as modeled here. The motional potential difference across an electromagnetic jet near the black hole should be $\sim1-300\,{\rm EV}$, many orders of magnitude greater than the $\sim1\,{\rm MV}$ minimum needed to create positrons or the $\sim1\,{\rm GV}$ needed to accelerate electrons to the $\lesssim1\,{\rm GeV}$ energies associated with the mm emission. The numerical simulations express none of this physics and, in any case, introduce a ``floor'' to the electron density for purely numerical reasons. The simulation particle density should not be trusted within the inner jet.

The composition of the plasma is also uncertain. Close enough to the event horizon, the plasma must fall inward and be connected to a source as jets are outflowing at larger radius. The simplest assumption, which we shall adopt, is that pairs are continuously produced in the inner magnetosphere. Plasma can also be entrained from the surrounding medium. This is expected to play a large role in the dynamical evolution and emission of the jet at large radii. The simulations suggest that it is unimportant within, say, $\sim1000M$ and we shall suppose that the associated radio emission is a locally accelerated pair plasma where $\gamma$-ray production is balanced by annihilation and other quantum electrodynamical processes.  

\subsection{Phenomenological Jet Models}

Given all this uncertainty, we adopt a phenomenological approach where we adopt a set of simple prescriptions for electron and positron pressure and temperature
that can lead to quite different simulated observations. We now develop a compendium of formulas linking plasma variables to radiation phenomenology, starting with jet models 
for the electromagnetically regions of a force-free, relativistic plasma. 

\subsubsection{Constant Electron $\beta$ Model}

The simplest prescription is the Constant $\beta_e$ Model, that the pressure of synchrotron radiating electrons is a constant fraction of the magnetic pressure 
\begin{equation}
P_e=\beta_{e0}P_B
\end{equation}
Close to the hole, the total pressure $P$ is dominated by the magnetic pressure 
$P_B=b^2/2\mu_0$ but at large radius, there is gas pressure contributed by the entrained gas. If we adopt $\beta\lesssim0.01$, close to the hole, then this is the sub-``equipartition'' prescription that is often invoked when interpreting jet observations \textcolor{black}{e.g., } \cite{Anantua2020b}b.

\subsubsection{Current Density ($j^2$) Model}

Another type of model 
by which electromagnetic jets can dissipate power employs currents. This type of model can be implemented using  
the field gradient --- the current density $j'$ --- and introducing a resistivity $\eta=\mu_0 L_j$ where $L_j$ is a length scale which we choose to be a fixed fraction of the jet width. The dissipation rate is then $W'=\eta j'^2$. This approach is partly motivated by particle-in-cell (PIC) simulations of relativistic reconnection. This model has been compared to the 43 GHz M87 jet in \cite{Anantua2018} as a mechanism for generating limb brightening. Though we do not make images of the Current Density Model here, we note the physical significance of jet currents as a spatially inhomogeneous source of dissipation.

\subsection{JAB System Models}

We now consider the entire inflow-outflow structure governed by the supermassive black hole. We have previously described the relativistic polar outflow as a Blandford-Znajek jet.  Beyond the outflow or jet funnel, astrophysical plasmas experience discontinuities in pressure and density. When there is a sufficient velocity gradient, Kelvin-Helmholtz instabilities produce gaseous swirling features-- notable in M87 up to 1 kpc from the black hole \citep{Pasetto2021}. The enveloping corona is loosely bound to the JAB system. The inflowing disk is supported against its own inertia by magnetic and thermal pressure and momentum transport-- the latter of which may lead to the magneto-rotational instability.

\textcolor{black}{The property of turbulent heating to preferentially energize electrons in magnetically dominated regions and protons in gas pressure dominated regions has been explored in \citet{Howes2010}. There are several ways of parameterizing this behavior \citep{Moscibrodzka2011, Anantua2020b}b.}

\subsubsection{$R-\beta$ Model}
We start our JAB emission modeling by noting  the tendency of plasma turbulence to preferentially heat electrons at low $\beta$ and ions at high $\beta$, as was originally conceptualized in the context of the solar corona \citep{Quataert1999,Howes2010}. Applied to JAB systems, the $R-\beta$ turbulent heating model takes the form
\begin{equation}
R=\frac{T_i}{T_e}=\frac{\beta^2}{1+\beta^2}R_\mathrm{high}+\frac{1}{1+\beta^2}R_\mathrm{low}
\end{equation}
It is the primary model used by the Event Horizon Telescope \citep{EHT2019V,EHT2021VIII} and developed by \cite{Moscibrodzka2016}.

\subsubsection{Critical $\beta$ Model}
\begin{figure}\nonumber
\begin{align}
& 
\includegraphics[height=200pt,width=200pt,trim = 6mm 1mm 0mm 1mm]{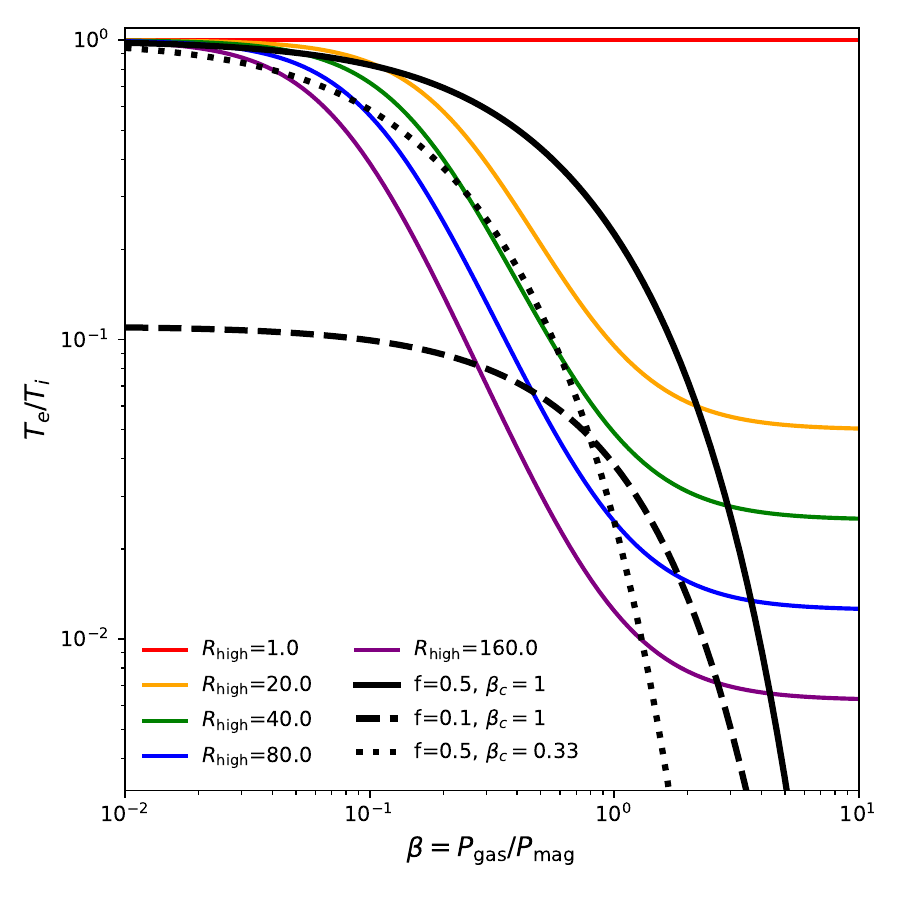} & 
\end{align}
\caption{Comparison of R-$\beta$ (solid lines) and Critical Beta (dashed lines) models for reasonable parameter values.}\label{RBetaCritBetaComparison}
\end{figure}

The Critical $\beta$ Model is an alternative turbulent heating model with an exponential parameter $\beta_c$ controlling the transition between electron- and ion-dominated heating
\begin{equation}
\frac{T_e}{T_e+T_i}=fe^{-\beta/\beta_c}
\end{equation}
This model was developed in \cite{Anantua2020b}b. The models are compared for reasonable parameter values in Fig. \ref{RBetaCritBetaComparison}. We see at low $\beta$, the electron-to-ion temperature ratios in the $R-\beta$ and the Critical Beta Models have similar asymptotic behavior. However, at high $\beta$, the Critical $\beta$ Model $T_e/T_i$ exponentially falls to 0, which by preliminary indications may reduce the bremsstrahlung contribution (though we reserve a more extensive investigation modeling emission processes beyond synchrotron for future work). The Critical $\beta$ Model also has transition behavior at intermediate $\beta$'s controlled by an exponential parameter $\beta_c$, leading to greater variety of intermediate $\beta$ emission regions probed between the same range of electron-to-ion temperature ratios compared to the $R-\beta$ Model.

\subsubsection{Multi-Zone Emission Models}

We have seen how plasma inertial and electromagnetic properties within and across our JAB simulations differ by orders of magnitude through $\beta$ and $\sigma$. Moreover, differences in the plasma velocity towards and away from the black hole leads to plasma mixing regions and whose behavior may not be amenable to the existing smooth emission parametric models.  We thus refine our JAB system emission models by combining the $\beta$-dependent turbulent heating models with jet regions radiating by conversion of magnetic to particle energy. We define the jet region using a transitional value of magnetization $\sigma$ of 1/2. Armed with these models of particle thermodynamics, we turn to the emitted radiation.

\subsection{Radiation} 


An electromagnetically-dominated jet outflow must be continually converting electromagnetic energy into high energy pairs through a $-{\bf E}\cdot{\bf j}$ interaction. These particles will radiate through the synchrotron and Compton processes. We can use the observations to draw some inferences about the variation of the distribution function along the jet.

At sufficiently small radius, the jet will become optically thick to synchrotron self-absorption at a given frequency given by
\begin{multline}
\nu_{\rm SA}=
\\
22\left(\frac {L_{\rm EM}}{10^{44}{\rm erg\ s}^{-1}}\right)^{0.1}\left(\frac S{\rm1Jy}\right)^{0.4}\left(\frac r{50M}\right)^{-0.8}\left(\frac\Gamma3\right)^{-0.4}{\rm GHz}
\end{multline}
Also, at a given radius there is a characteristic frequency where the synchrotron cooling time of the emitting electrons is equal to the expansion time scale 
\begin{equation}
\nu_{\rm cool}=\left(\frac{L_{\rm EM}}{10^{44}{\rm erg\ s}^{-1}}\right)^{-1.5}\left(\frac r{50M}\right)^{-0.2}\left(\frac\Gamma3\right)^6{\rm THz}
\end{equation}
If a power law is accelerated, the local spectrum should break by $\Delta\alpha=0.5$ at this frequency. However, note the extreme sensitivity to the bulk Lorentz factor $\Gamma$. This probably controls what is actually observed.

The entire nuclear spectrum within $r\sim3\times10^5M$ has been carefully determined by \cite{2016MNRAS.457.3801P}. They find a sharp break at $\nu=\nu_b\sim150$~GHz. Presumably the flatter $\alpha\sim0.2$ spectrum at $\nu\lesssim\nu_b$ is attributable to the superposition of a radial sequence of spectra with a frequency to radius mapping as defined above. The lowest frequency considered, $\sim3\ {\rm GHz}$, should originate at $r\sim1000M$.

\subsection{Emission Modeling with Positrons}
\subsubsection{Particle Production and Acceleration}

Though the vast majority of GRMHD simulations consider ionic plasma without an explicit fluid for electron-positron pairs, it is unknown whether jet matter is typically dominated by an ion-electron plasma or a pair one. In the latter case, black hole-powered jets 
generally get their initial mass-loading through $\gamma-\gamma$ pair production.
The means of particle acceleration within AGN jets, however,
is not understood. There have been several mechanisms invoked, including those involving strong shock fronts, both non-relativistic and relativistic, magnetic reconnection, stochastic acceleration though wave-particle interaction and electrostatic acceleration -- either along a magnetic field or perpendicular to the field as a consequence of drift motion. Recent observations, especially when seen in the context of observations of extreme acceleration in pulsar wind nebulae and Gamma Ray Bursts, point to the need for new and more rapid approaches. The need is best exemplified by $\gamma$-ray observations which can show that electric fields as strong as magnetic fields (setting $c=1$) may be needed in order for electrons to attain the required energies in the face of strong radiative loss through synchrotron emission and Compton scattering.

\subsubsection{
Positron Production Modeling and Radiative Transfer}

Much of the plasma in the accreting component of the RIAF system is likely a mixture of ionized hydrogen and helium from stellar winds and the interstellar medium. Since the conductivity is high near the event horizon, particles in the plasma are forced to follow magnetic field lines, so the jet, which is canonically magnetically disconnected from the accretion disk, cannot be directly supplied with plasma from the disk. If electron--positron pairs are produced in these regions, then they may be the dominant matter source.
Electron--positron pairs may be produced by pair cascades in charge-starved magnetospheres (like in evacuated jet funnels) or in the disk coronae{}.

In the systems we study, electron--positron pairs are produced mainly via the Briet--Wheeler process, i.e., as a result of photon--photon collisions. In order to create a pair, the center-of-momentum energy of the photons must exceed the rest-mass energy of a pair $\approx 1\,$MeV $\approx 2 \times \left(1.2 \times 10^{20}\right)\,$Hz. The cross-section peaks near this threshold value, and the participating photon couples lie over a spectrum of energy ratios: some pairs of photons having approximately the same frequency while others are matched with low/high frequencies.
Pair-producing processes are often differentiated based on the photon source and whether the newly created pairs radiate and contribute (non-negligibly) as a new photon source.

\emph{Pair drizzle} occurs when the pairs are produced by photons from the background radiation field (due to synchrotron and bremsstrahlung emission and Compton upscattering) and typically exhibits variation on timescales associated with the plasma fluid. Drizzle pair production has been studied in a variety of scenarios ranging between stellar-mass to supermassive black hole accretion \cite{Moscibrodzka2011,Lau2018,Kimura2020,Wong2021,yao2021}.  
In the alternative scenario, high energy photons with frequencies $\gg 10^{20}\,$Hz can interact with background (low energy) photons from the disk to undergo pair production. Here, the high-energy photons are produced when unscreened electric fields accelerate stray charges 
\citep{Beskin1992}; when the acceleration is large, the leptons radiate the requisite high energy photons. Often, the newly created pairs are born in the same region with unscreened fields and are thus themselves accelerated, restarting the process in a cascade of pair creation. The short timescales associated with pair cascades means they may explain the ultra-rapid high-frequency radio emission flares from AGN jets. Pair cascades have been studied with a variety of analytic, semi-analytic, and numerical methods, e.g.,  
\cite{Williams1995,Fragile2009,Broderick2015,Parfrey2019}. Positrons not only effect images at the level of emission, but also through radiative cooling, e.g., \cite{Fragile2009,Yoon2020}.

Given the uncertainty in jet positron fraction, we focus on the special cases of a sparse ionic jet with electron number density $n_{e0}$ and a jet where all sources of pair production result in a plasma with an equal number density of ionic and pair plasma ($f_\mathrm{pos}\equiv n_\mathrm{pairs}/n_{e0}=1$). 

There are benchmarks for positron content in the Literature, such as the estimate by \cite{Ghisellini2012} of the fraction of a jet (opening angle $\psi$, distance from black hole $R_0$) converted to positrons,  as $f=0.1\min\{1,\frac{\ell_0}{60}\}$ where the compactness is $\ell = \frac{\sigma_TL_0}{\psi R_0\mathcal{D}^4m_ec^3}$.
%
%
More possibilities for painting positrons on jets can be found in Appendix.


\section{Observing a Time-dependent Simulation}

\subsection{Anatomy of a Time-Dependent KHARMA Jet Simulation}

\begin{figure}\nonumber
\hspace{1pt}
\begin{align}\includegraphics[trim = 0mm 0mm 0mm 0mm, clip, width=220pt
]{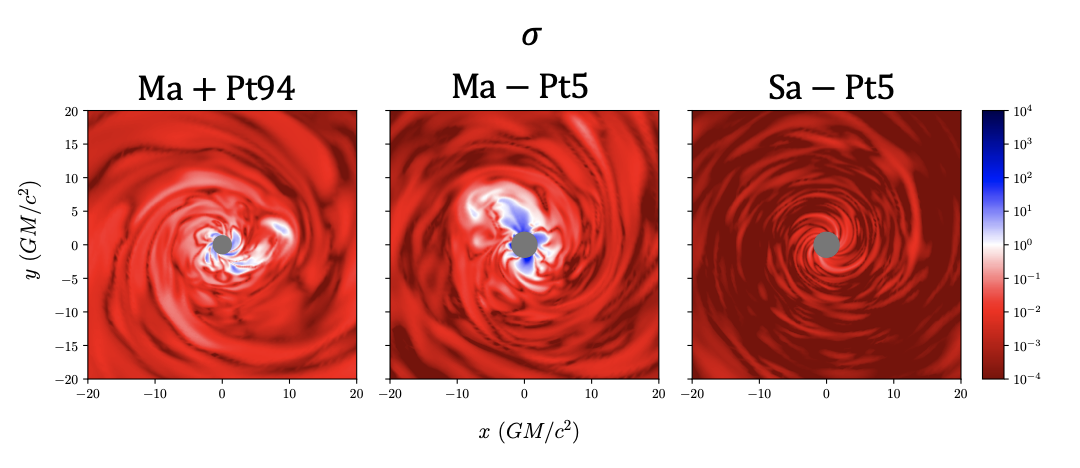
}\end{align}
\begin{align}\includegraphics[trim = 0mm 0mm 0mm 0mm, clip, width=220pt
]{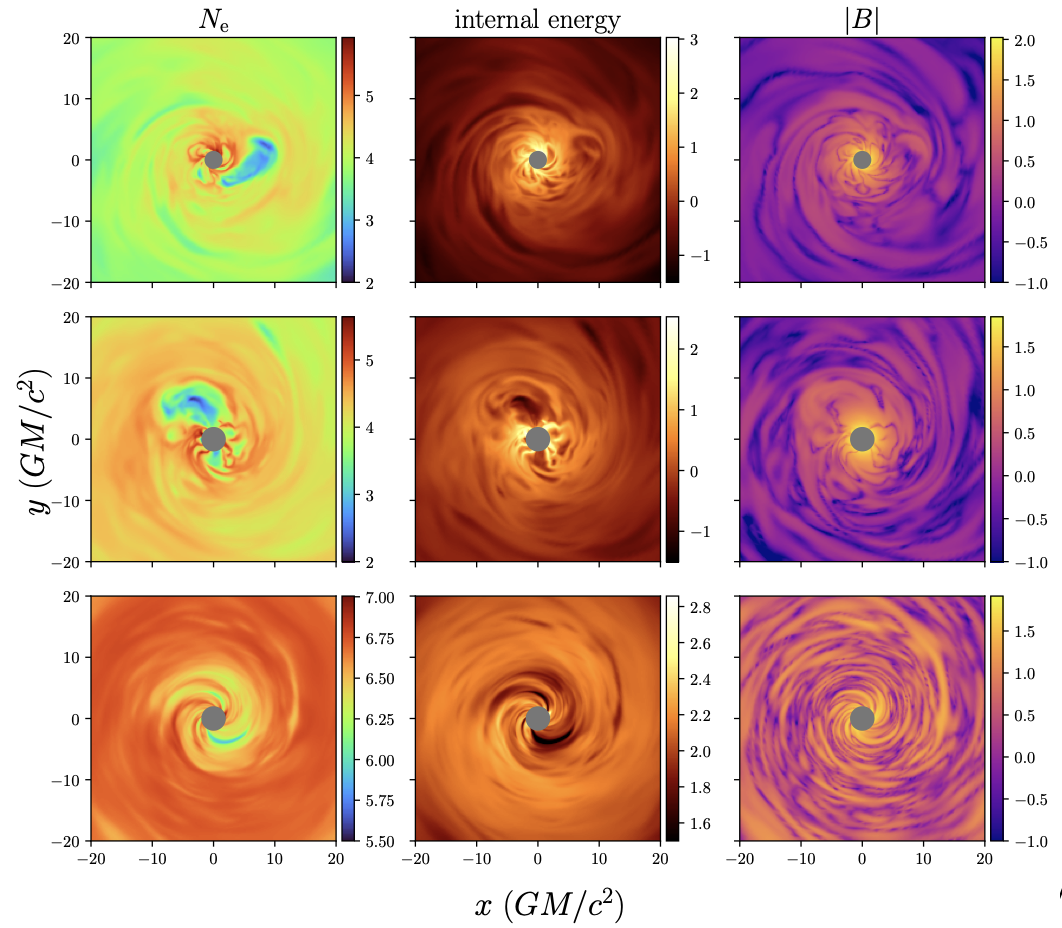
}\end{align}
\begin{align}
\includegraphics[trim = 0mm 0mm 0mm 0mm, clip, width=220pt
]{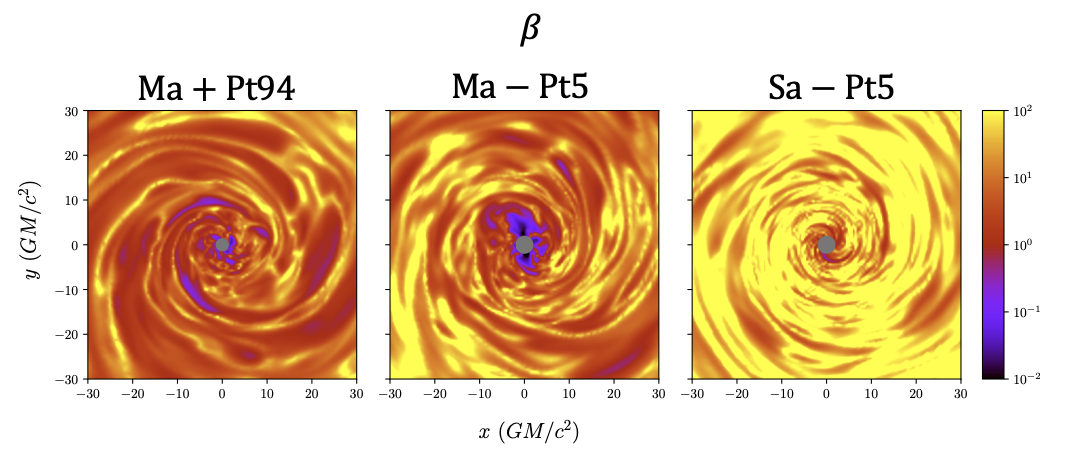
} \end{align}
\caption[Polar Variation 15Deg to 20Deg]{Equatorial slices for fiducal MAD $a/M=0.94,-0.5$ and  SANE $a/M=-0.5$ of $\sigma$, $N_e$, internal energy, $|B|$ and $\beta$. 
}
\label{AzimuthalAndPolarVariation15And20DegViewingAngleKHARMA}
\end{figure}

We have outlined a methodology for combining jet emission prescriptions with detailed, 3D time-dependent simulations.
To emphasize the 3D nature, we take transverse slices in the equatorial plane in Figure \ref{AzimuthalAndPolarVariation15And20DegViewingAngleKHARMA}. There, we see even among two MADs (spins $a/M=0.94,-0.5$) there are large, azimuthal symmetry-breaking patches of high magnetization emanating in different directions from the black hole. Electron number density exhibits a similar pattern of asymmetry, however, internal energy, magnetic field strength and plasma $\beta$ are relatively azimuthally symmetric.

We now describe the process of ray tracing the resulting emission.

\subsection{Radiative transfer with IPOLE: Azimuthal and Polar Variation}


GRMHD simulations can be ray-traced using general relativistic radiative transfer (GRRT) codes to simulate surveys of sources throughout the sky. In this work, we use the GRRT code {\tt{ipole}} \cite[][see also \citealt{Wong2022}]{2018MNRAS.475...43M} to produce polarimetric images of the GRMHD simulations described in Section~\ref{section:grmhd_model_overview}. The {\tt{ipole}}~code solves for the evolution of the polarized intensities at each point along a geodesic with a two-stage operator splitting method. 
{\textcolor{black}{In the first stage, the covariant coherency tensor (which can be written in terms of the invariant Stokes parameters and Pauli matrices) is parallel transported along the direction of the geodesic. In the second stage, the Stokes parameters are updated using the analytic solution to the explicit, general polarized transport with constant emission, absorption, and rotation coefficients, which are computed in the local orthonormal tetrad defined by the fluid and magnetic field.}}

Each image comprises a square grid of $N$x$N$ pixels over a $160\,\mu$as field of view, with each pixel reporting the Stokes intensities for $I, Q, U$, and $V$. Since producing images requires evaluating transfer coefficients in physical units, it is necessary to specify scales for the mass-density of the accreting plasma and the size of the black hole as well as the orientation of the observer (i.e., the software camera) with respect to the black hole. We list the physical M87 black hole parameters (and references) in Table~\ref{M87GeometryTable}, and in Table~\ref{M87codeunits} we report the ``code scale'' values corresponding to the black hole mass identified above.


Note that GRMHD codes are generally unable to accurately evolve the fluid state in regions with high magnetization $\sigma \equiv b^2 / \rho$ and artificially inject mass and energy in these regions. Since the plasma density (and temperature) are therefore typically unphysically high in regions of large $\sigma$, ray-tracing codes like {\tt{ipole}} normally introduce a so-called $\sigma$ cutoff, where the plasma density in regions with $\sigma > \sigma_{\rm cutoff}$ is explicitly set to zero before computing the transfer coefficients. In this work, we set \textcolor{black}{ $\sigma_{\rm cutoff} = 2$}, consistent with the typical values used for such flows (see, e.g., \cite{EHT2019V}. 

\section{COMPARISON OF SIMULATIONS WITH OBSERVATIONS}\label{Sec:SimVsObs}

We now present a suite of 3D-GRMHD simulation images spanning 
various plasma compositions and prescriptions for electron-positron thermodynamics. JAB systems are often modeled as electron-proton plasmas with a single function linking electron temperature to plasma variables such as $\beta$ throughout the inflow/outflow system \cite{EHT2019V,EHT2021VIII}. We start with models using this approach with the $R-\beta$  and Critical $\beta$   turbulent heating prescriptions, 
and then we refine the models by prescribing jet funnel emission in a region between $\sigma_\mathrm{min}=1/2$ and $\sigma_\mathrm{max}=2$ and adding pairs. Unless otherwise stated, the images are raytraced at 230 GHz \textcolor{black}{and the inclination angle is 17$^\circ$.} \textcolor{black}{Note that in the tables referenced in this section, comparisons are made with snapshots in the GRMHD space. Due to this, model performance with respect to observations may not hold when comparisons to windows of simulations are made such as those in \cite{EHT2019I}.} 

\textcolor{black}{The image library presented here greatly expands the $a=-0.5M$ MAD and SANE snapshots from \cite{Anantua2023} to include the highly prograde spin $a=0.94$, temporal evolution, extreme positron fractions $n_\mathrm{pairs}/n_{e0}=50,100$, varying frequency from 230 GHz to 86 GHz and varying inclination up to $40^\circ$ viewing angle. Here, the SANE-MAD dichotomy manifest in the image library is also made quantitative by the tabulation of Faraday conversion and rotation depths and comparison to M87 linear polarization data. }

\subsection{SANE Positron Effects}

\subsubsection{SANE R-$\beta$}
\begin{figure} 
\hspace{-0.25cm}
\includegraphics[height=150pt,width=250pt,trim = 6mm 1mm 0mm 1mm]{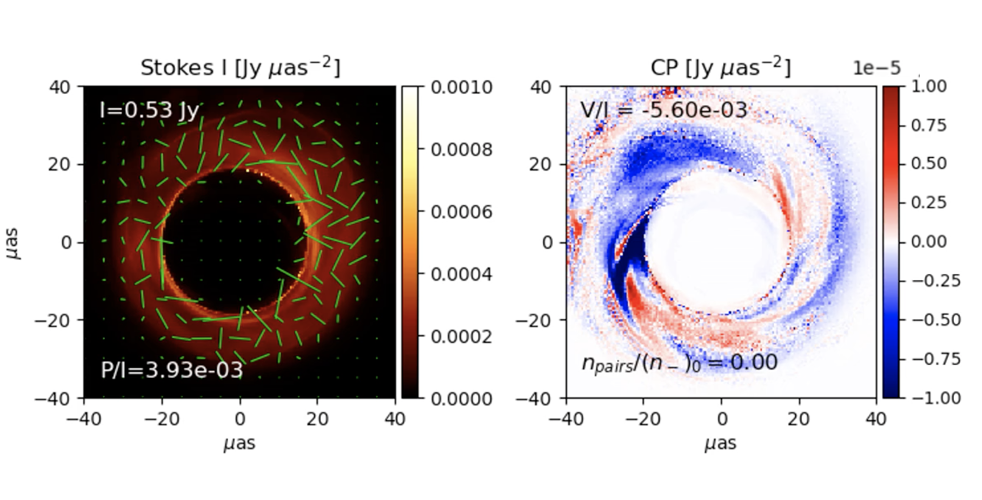}
\begin{align}\nonumber 
& \hspace{-0.25cm} \includegraphics[trim = 6mm 1mm 0mm 0mm,  height=150pt,width=250pt
]{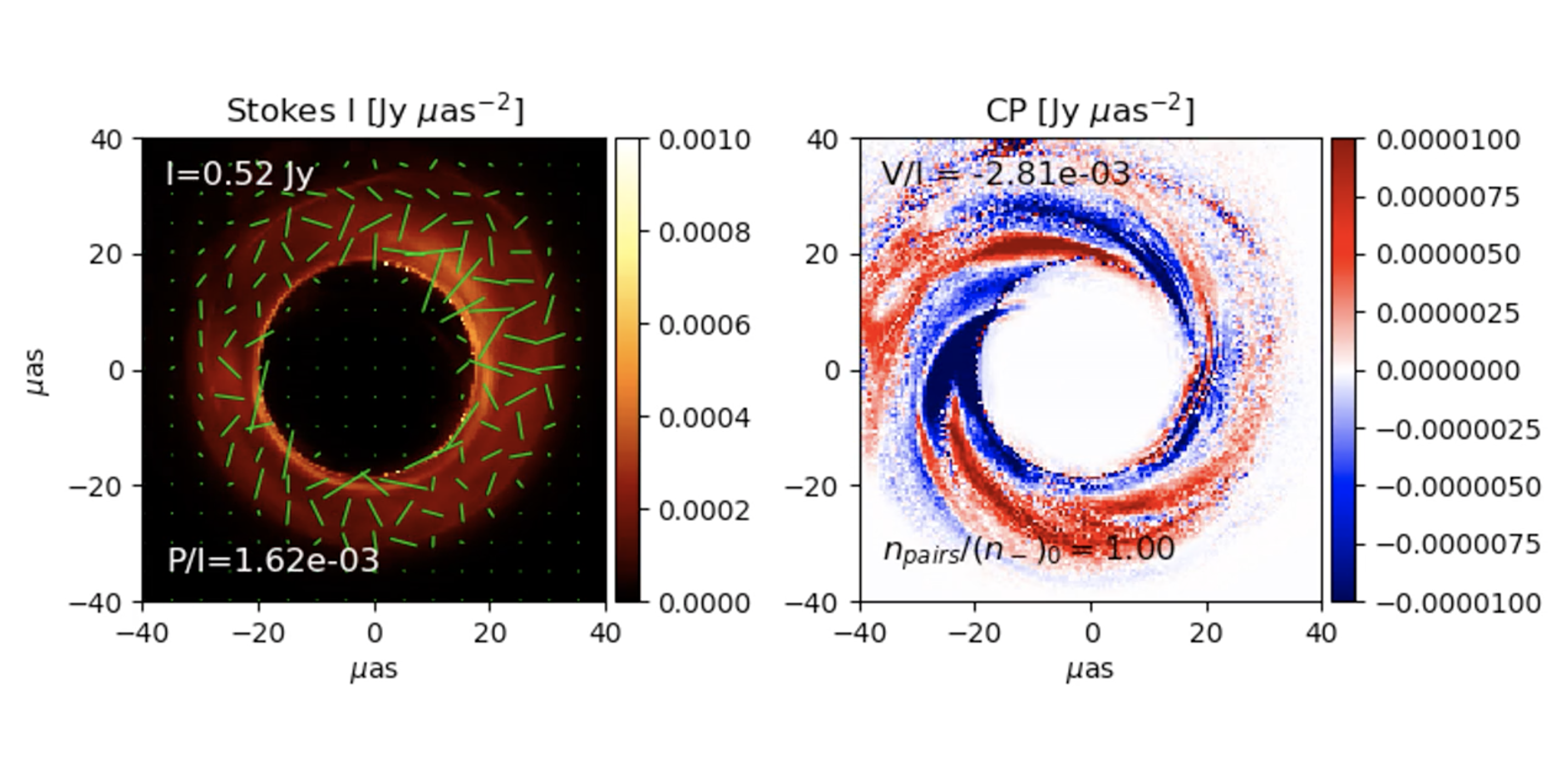}
\end{align}
\caption[Critical Beta Positron Effects]{For the $a=-0.5$ SANE at $T=25,000M$: Top panel: R-Beta at 230 GHz without positrons.
Bottom panel: R-Beta at 230 GHz with for an even mix of pair and ionic plasma with ions and positrons each accounting for 1/4 the plasma number density and electrons accounting for the remaining 1/2.}\label{PositronVariationRBeta}
\end{figure}

Fig. \ref{PositronVariationRBeta} shows intensity with electric vector polarization angle (EVPA) and circular polarization maps for the SANE $a=-0.5$ simulation in the $R-\beta$ model. This model has asymptotic ion-to-electron temperature ratios $R_\mathrm{low}=1$ and $R_\mathrm{high}=20$.  Here, the total flux is greatest near the photon ring immediately surrounding the central depression and slowly decreases radially becoming broadly distributed through the equatorial annulus. Polarization oriented at the EVPA is spread throughout the equatorial annulus in the  $40\mu$as x $40\mu$as field of view. To the ionic plasma in the Top Panels, an equal number density of pairs as the original electron number density are added in the Bottom Panels (while renormalizing $m_\mathrm{unit}$ to maintain a 0.5 Jy image flux). The added positrons significantly rotate EVPAs, as the Faraday rotation measure depends sensitively on the positron fraction for SANEs.  

\subsubsection{SANE Critical $\beta$} 
\begin{figure}
\hspace{-0.25cm} 
\includegraphics[height=150pt,width=250pt,trim = 6mm 1mm 0mm 1mm]{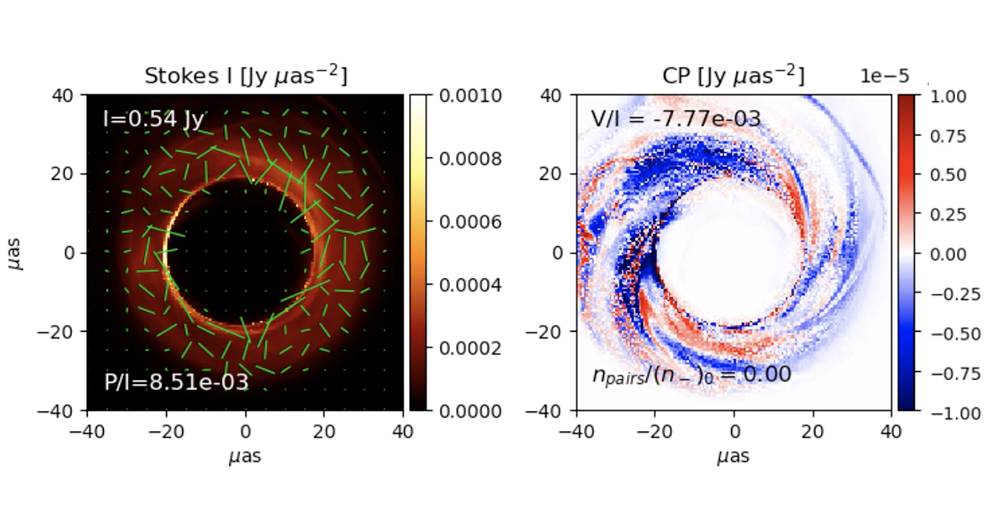}
\begin{align}\nonumber
& \hspace{-0.25cm} \includegraphics[trim = 6mm 1mm 0mm 0mm,  height=150pt,width=250pt
]{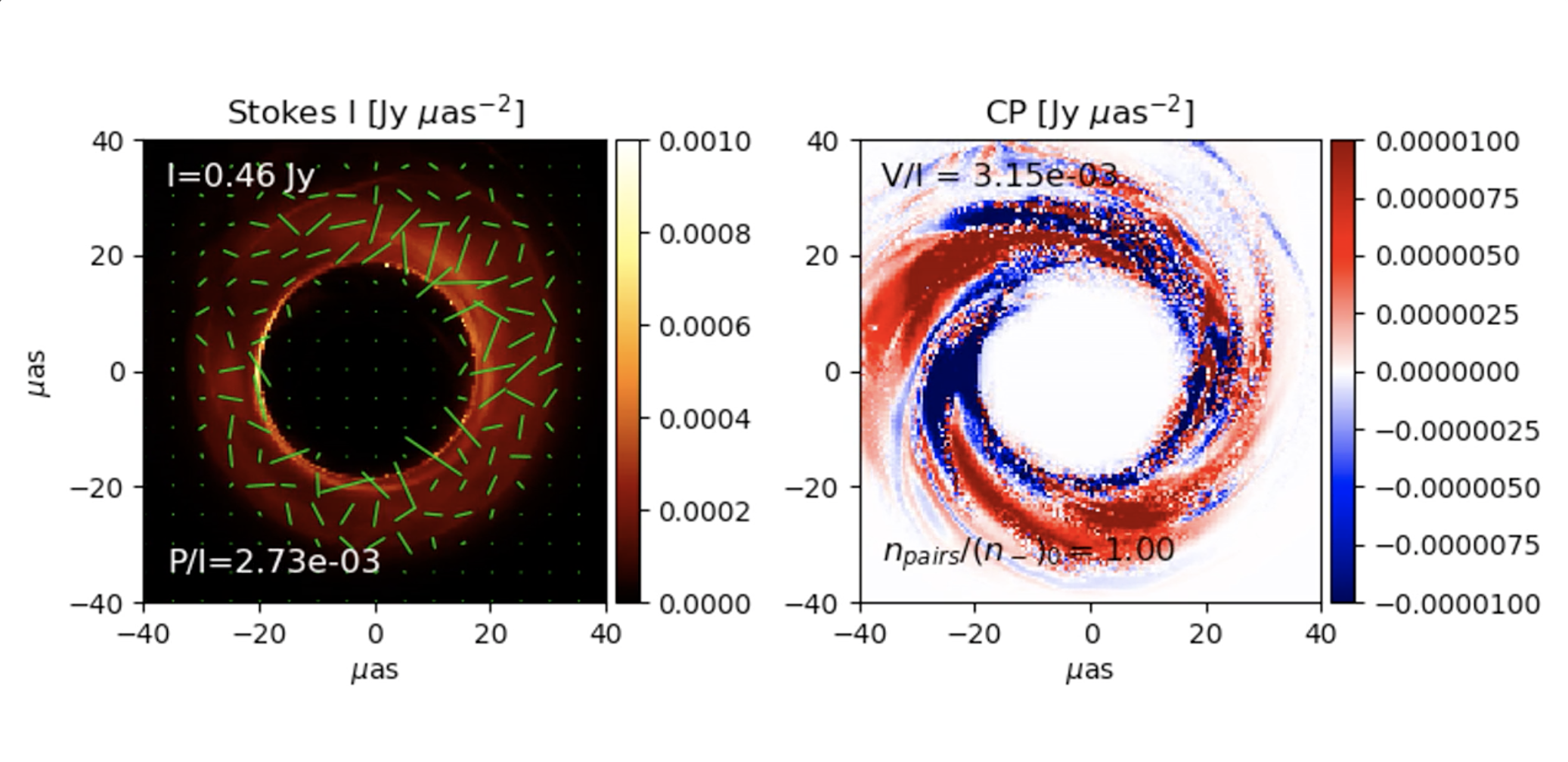}
\end{align}
\caption[Critical Beta Positron Effects]{\textcolor{black}{For the $a=-0.5$ SANE: Top panel: Critical Beta at 230 GHz without positrons
Bottom panel: Critical Beta at 230 GHz for an even mix of pair and ionic plasma. 
}
}\label{PositronVariationCriticalBeta}
\end{figure}

In Fig. \ref{PositronVariationCriticalBeta} we show our other turbulent heating model-- the Critical Beta Model. This model controls the transition from preferential electron heating to preferential ion heating through the exponential parameter $\beta_c$, which for higher values smooths the transition by allowing a larger range of betas to include radiating electrons. For model parameters temperature ratio prefactor $f=0.5$ and $\beta_c=0.5$, total flux is concentrated in a ring at $\sim20\mu$as and regions along lines of sight close to the polar axis, though polarization morphology trends remain similar to the $R-\beta$  case. Note our $R-\beta$ models are more linearly depolarized than Critical $\beta$ Models even with lower contributions to intrinsic emission at high $\beta$ in the latter models. This is one of several examples of Faraday effects we will see in this Section. 

\subsubsection{SANE R-$\beta$ with Constant $\beta_E$ Jet}

\begin{figure}
\hspace{-0.25cm}
\includegraphics[height=150pt,width=250pt,trim = 6mm 1mm 0mm 1mm]{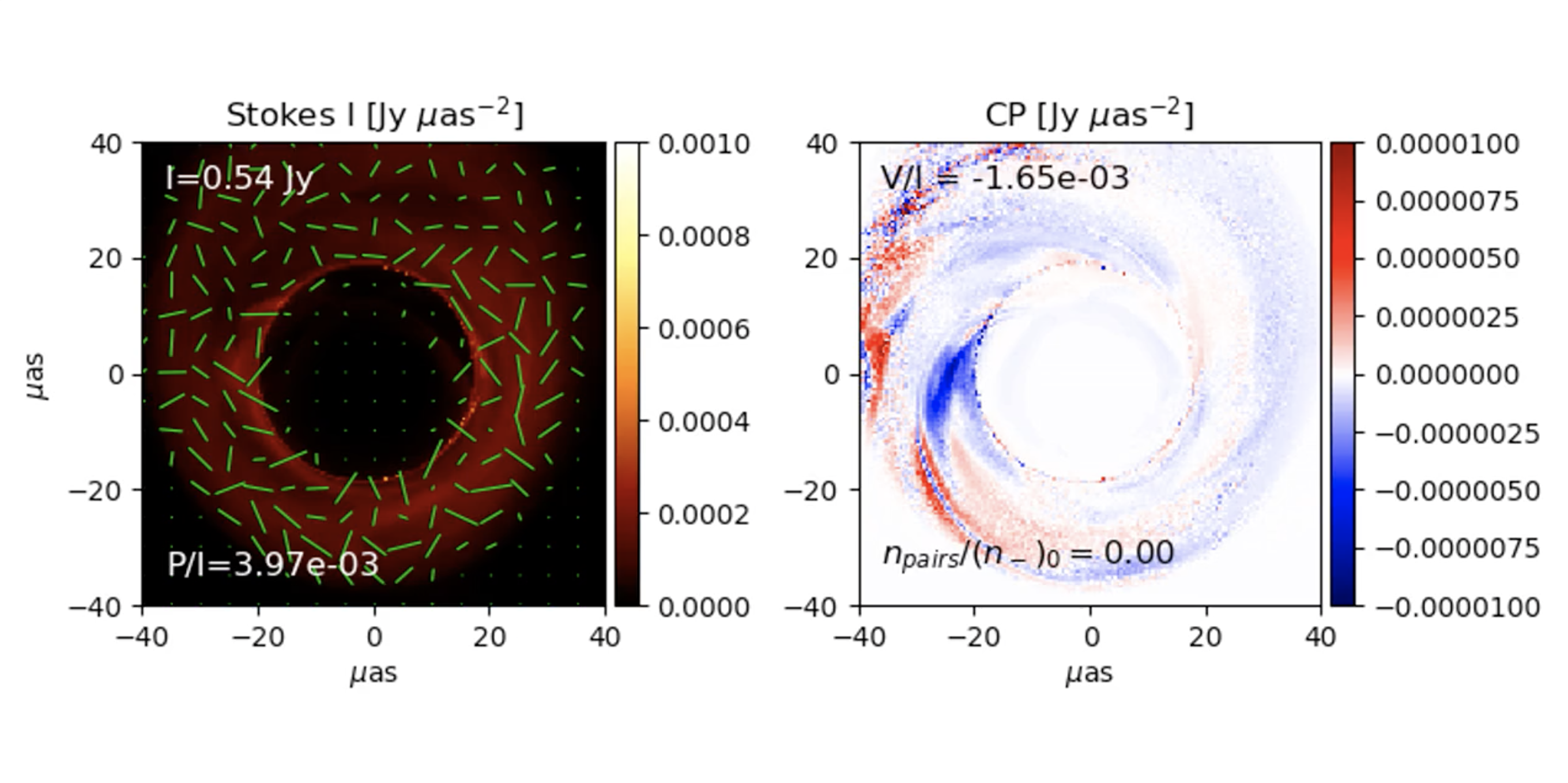}
\begin{align}\nonumber
& \hspace{-.5cm} \includegraphics[trim = 6mm 1mm 0mm 0mm,  height=150pt,width=250pt
]{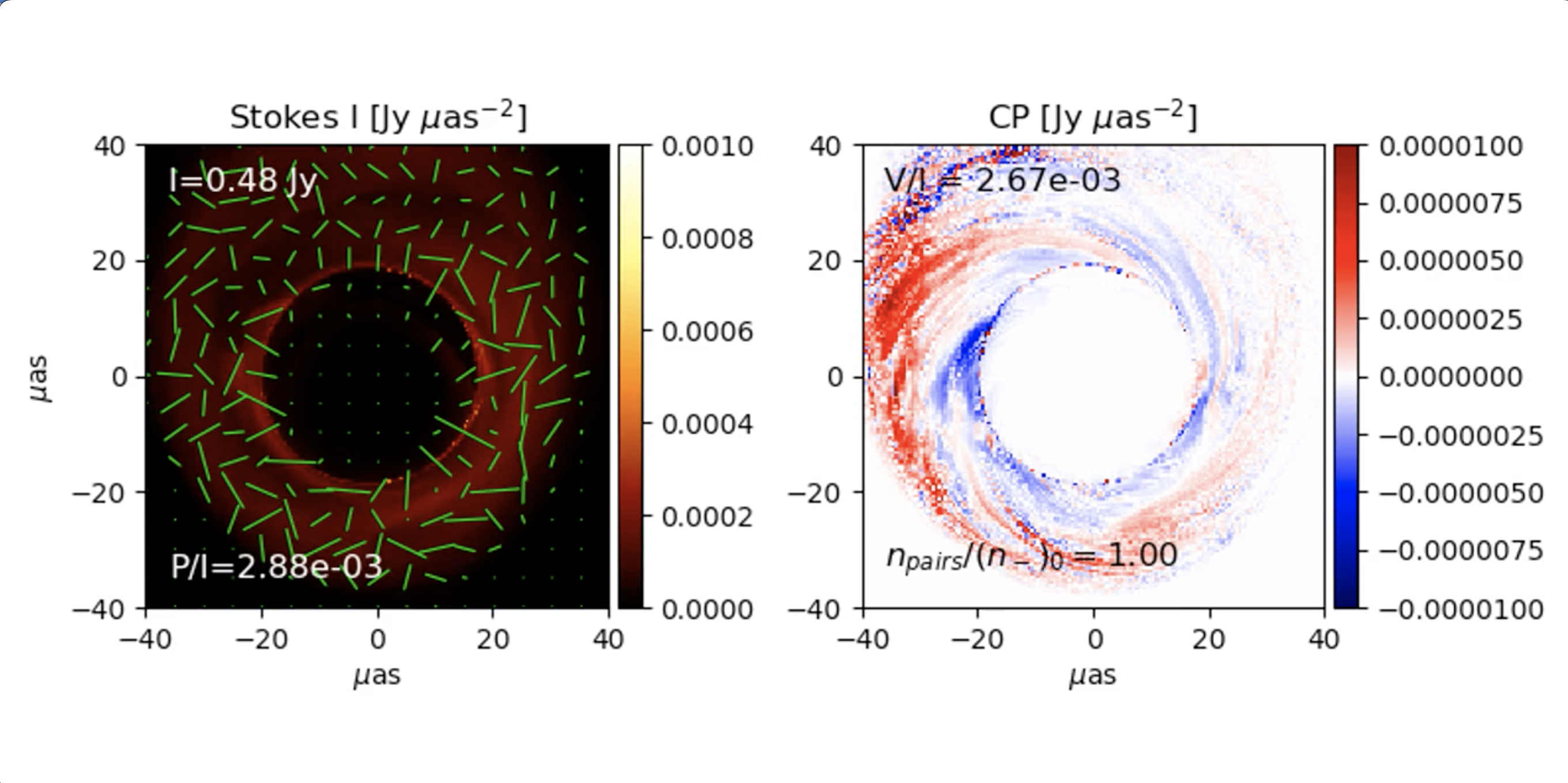}
\end{align}
\caption[Critical Beta Positron Effects]{\textcolor{black}{For the $a=-0.5$ SANE at $T=25,000M$: Top panel: R-Beta with $\beta_{e0}=0.01$ jet at 230 GHz without positrons
Bottom panel: R-Beta at 230 GHz for an even mix of pair and ionic plasma. 
}}\label{PositronVariationRBetaWithJet}
\end{figure}

In Fig. \ref{PositronVariationRBetaWithJet} we add a jet region where the energy of relativistic electrons is directly derived from the magnetic pressure to the $R-\beta$ model. The emission is extended more broadly and evenly throughout the field of view as it is projected from a broader region of the outflow paraboloid governed by the transitional value of $\sigma$ separating the constant $\beta_e$ jet from the turbulently heated plasma.
 
\subsubsection{SANE Critical $\beta$ with Constant $\beta_E$ Jet}

\begin{figure}
\hspace{-.25cm}
\includegraphics[height=150pt,width=250pt,trim = 6mm 1mm 0mm 1mm]{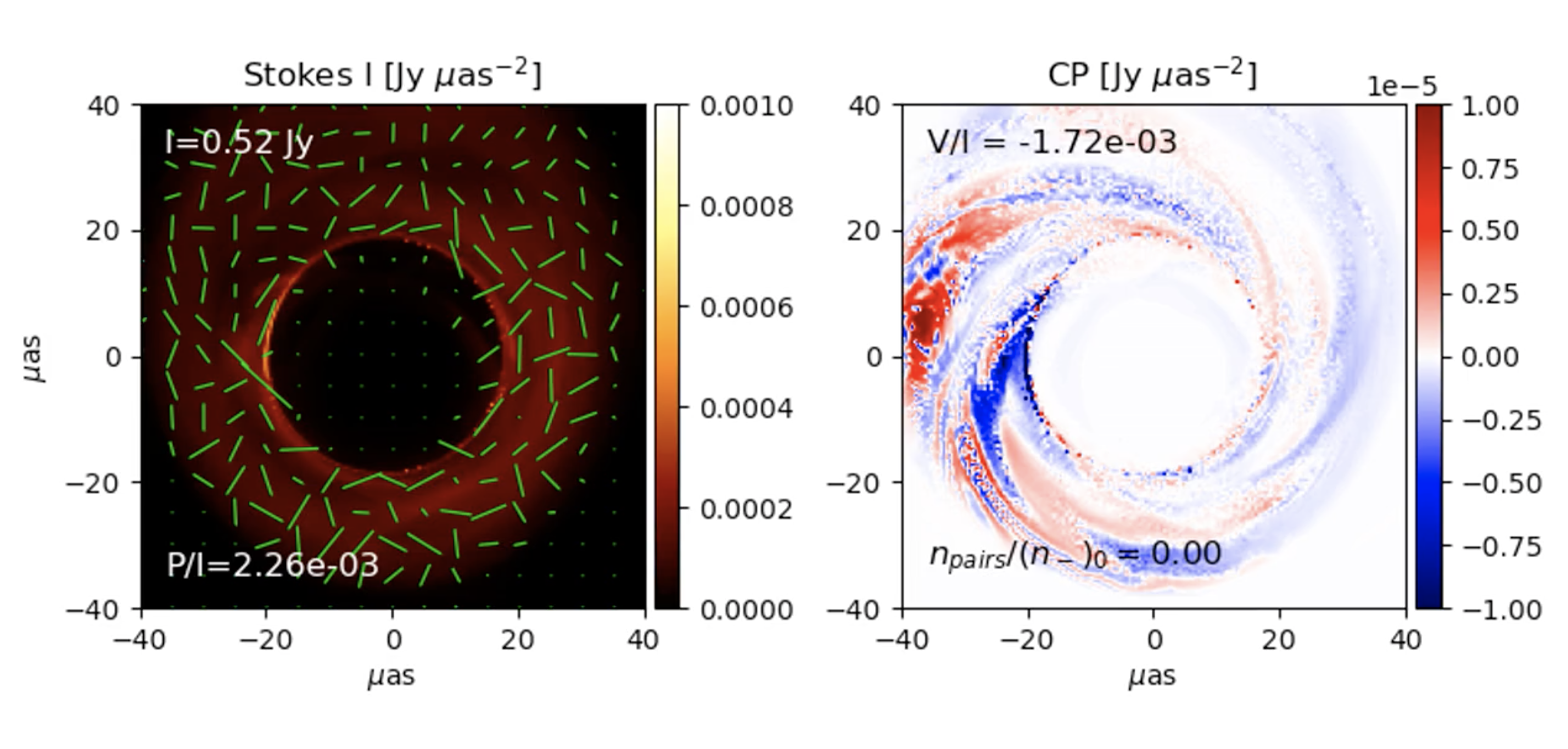}
\begin{align}\nonumber
& \hspace{-0.25cm} \includegraphics[trim = 6mm 1mm 0mm 0mm,  height=150pt,width=250pt
]{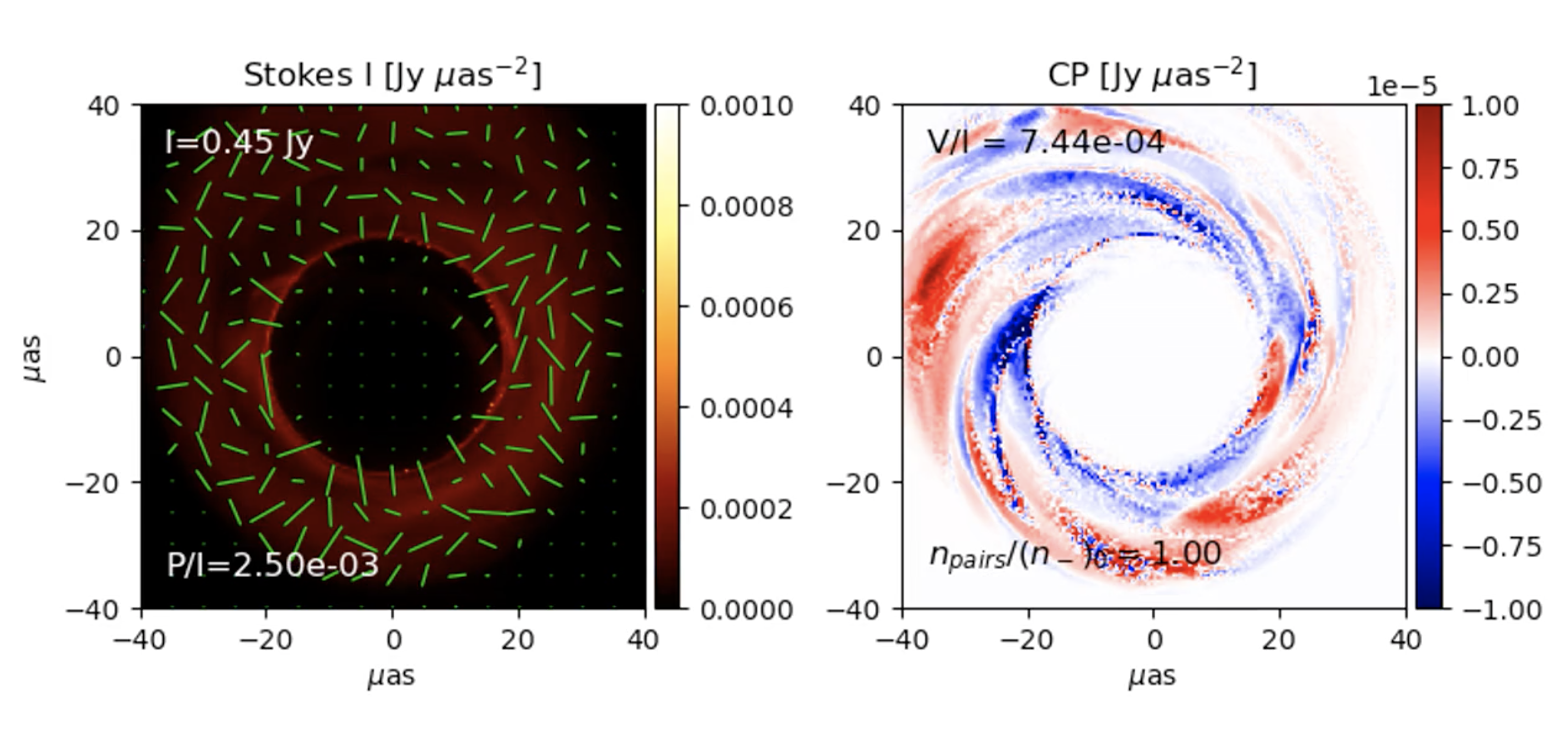}
\end{align}
\caption[Critical Beta Positron Effects]{\textcolor{black}{For the $a=-0.5$ SANE at $T=25,000M$: Top panel: Critical Beta with $\beta_{e0}=0.01$ jet at 230 GHz without positrons. 
Bottom panel: 
Critical Beta at 230 GHz for an even mix of pair and ionic plasma. 
}}\label{SANEPositronVariationCriticalBetaWithJet}
\end{figure}

\textcolor{black}{In Fig. \ref{SANEPositronVariationCriticalBetaWithJet} we add a jet region of magnetic-to-particle energy conversion to the Critical Beta model. In the SANE case, the jet does not appreciably change the image morphology. Moreover, polarization does not vary monotonically with the addition of positrons across different emission models. }
 
\subsection{\textcolor{black}{MAD Positron Effects}}

\textcolor{black}{The MAD images from our fiducial time are nearly indistinguishable when Faraday effects are turned off. In these MAD images, we see a prominent flux tube in a loop extending towards the lower left. In these \AR{particular} MAD images, whose circular polarization is dominated by the intrinsic, we see another dramatic polarization effect: linear 
increase in the magnitude 
of $V/I$ \textcolor{black}{(confer Section 6.5 for polarimetric quantity definitions)} as a function of synchrotron emitters not in pairs (which is maximal for the ionic plasma case).}

\subsubsection{MAD R-$\beta$}
\begin{figure}
\hspace{-0.25cm}
\includegraphics[height=150pt,width=250pt,trim = 6mm 1mm 0mm 1mm]{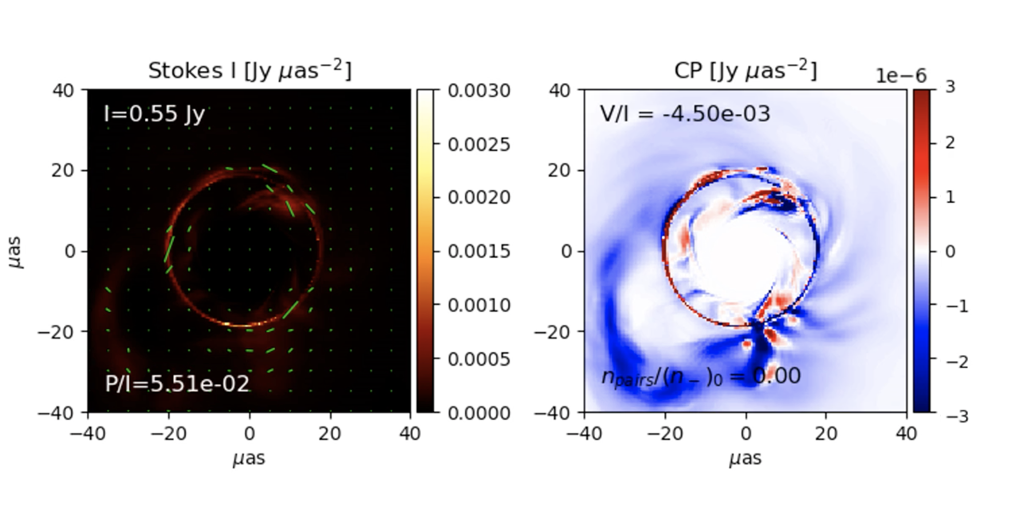}
\begin{align}\nonumber
& \hspace{0.0cm} \includegraphics[trim = 6mm 1mm 0mm 0mm,  height=150pt,width=250pt
]{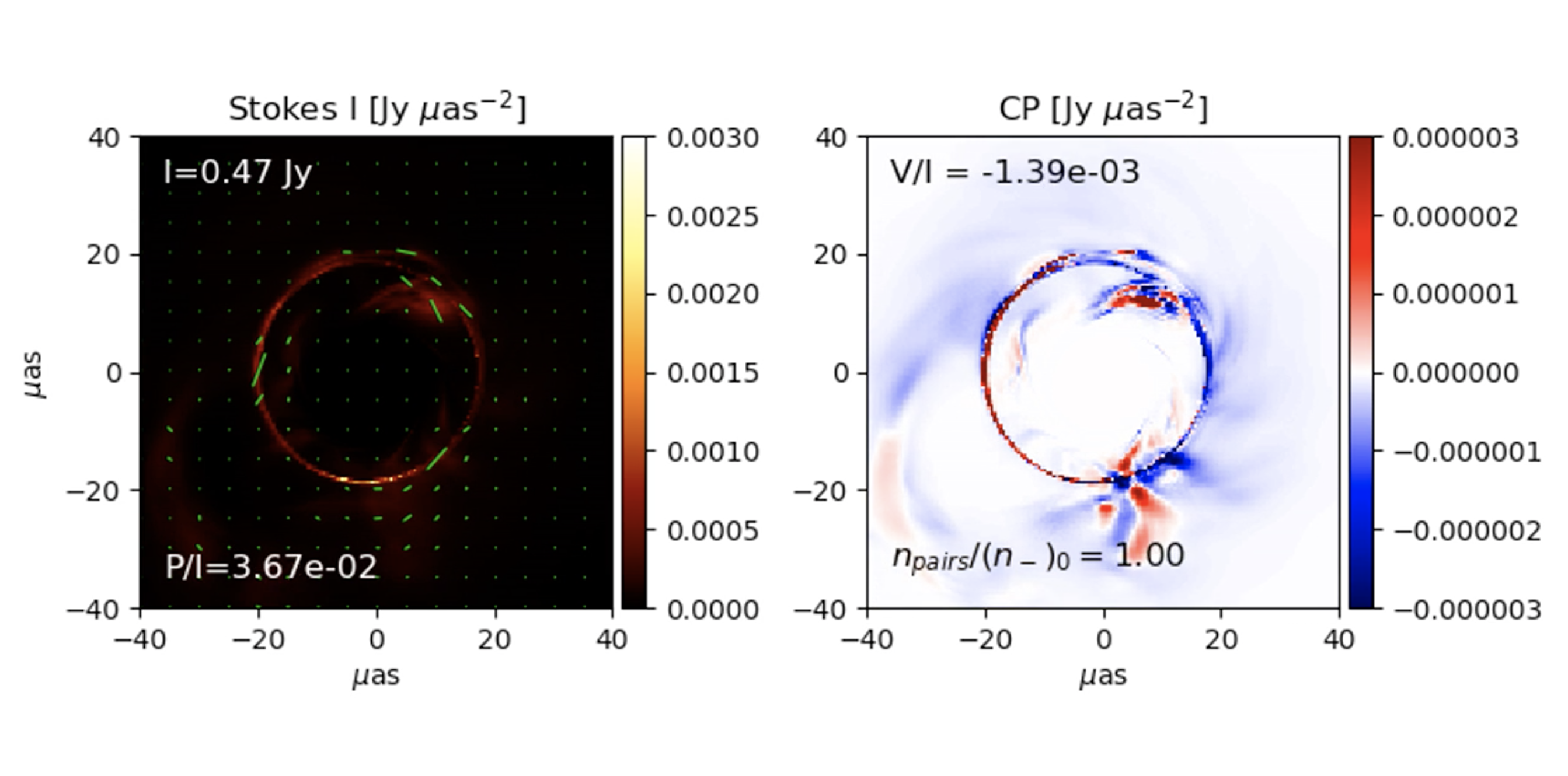}
\end{align}
\caption[R Beta Positron Effects]{\textcolor{black}{For the 
$a=-0.5$ MAD at $T=25,000M$: Top panel: R-Beta at 230 GHz without positrons
Bottom panel: R-Beta at 230 GHz for an even mix of pair and ionic plasma. 
}}\label{PositronVariationMADRBeta}
\end{figure}

\textcolor{black}{Starting in Fig. \ref{PositronVariationMADRBeta} with the R-$\beta$ model, the linear polarization ticks oriented at the EVPA for the 0-positron case remain in their orientations with minimal angular displacement when positrons are added to form the mixed plasma in the ray tracing step. However, in radiative transfer using coefficients for the mixed plasma in which 1/2 the particles are electrons, 1/4 of the particles are positive ions and 1/4 of the particles are positrons (i.e, 2/3 of the synchrotron emitting leptons are paired), we have the degree of circular polarization $V/I$ diminishing to 1/3 of the positron-free value. The addition of positrons also reverses the polarity of the bottom left portion of the flux eruption loop.}
 
\subsubsection{MAD Critical $\beta$} 

\begin{figure}
\hspace{-0.25cm}
\includegraphics[height=150pt,width=250pt,trim = 6mm 1mm 0mm 1mm]{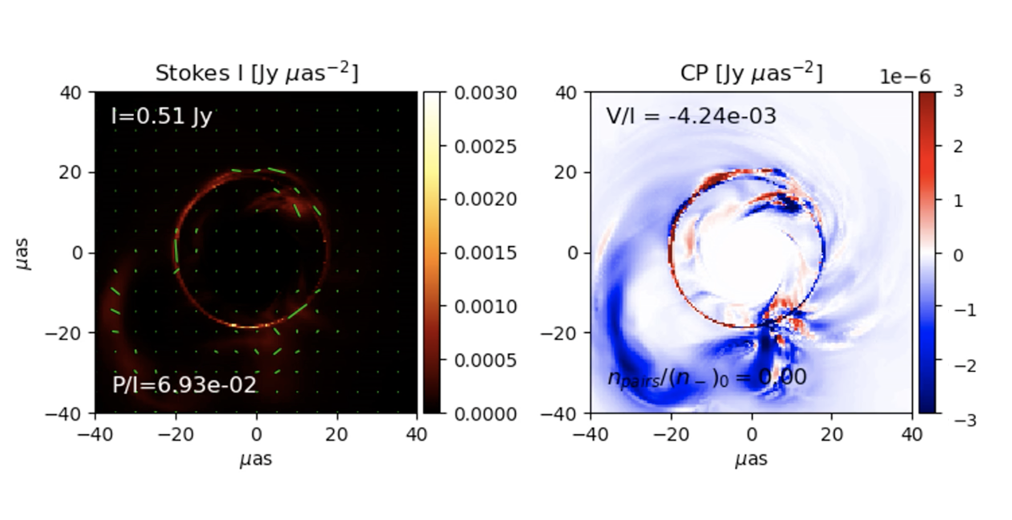}
\begin{align}\nonumber
& \hspace{-0.25cm} \includegraphics[trim = 6mm 1mm 0mm 0mm,  height=150pt,width=250pt
]{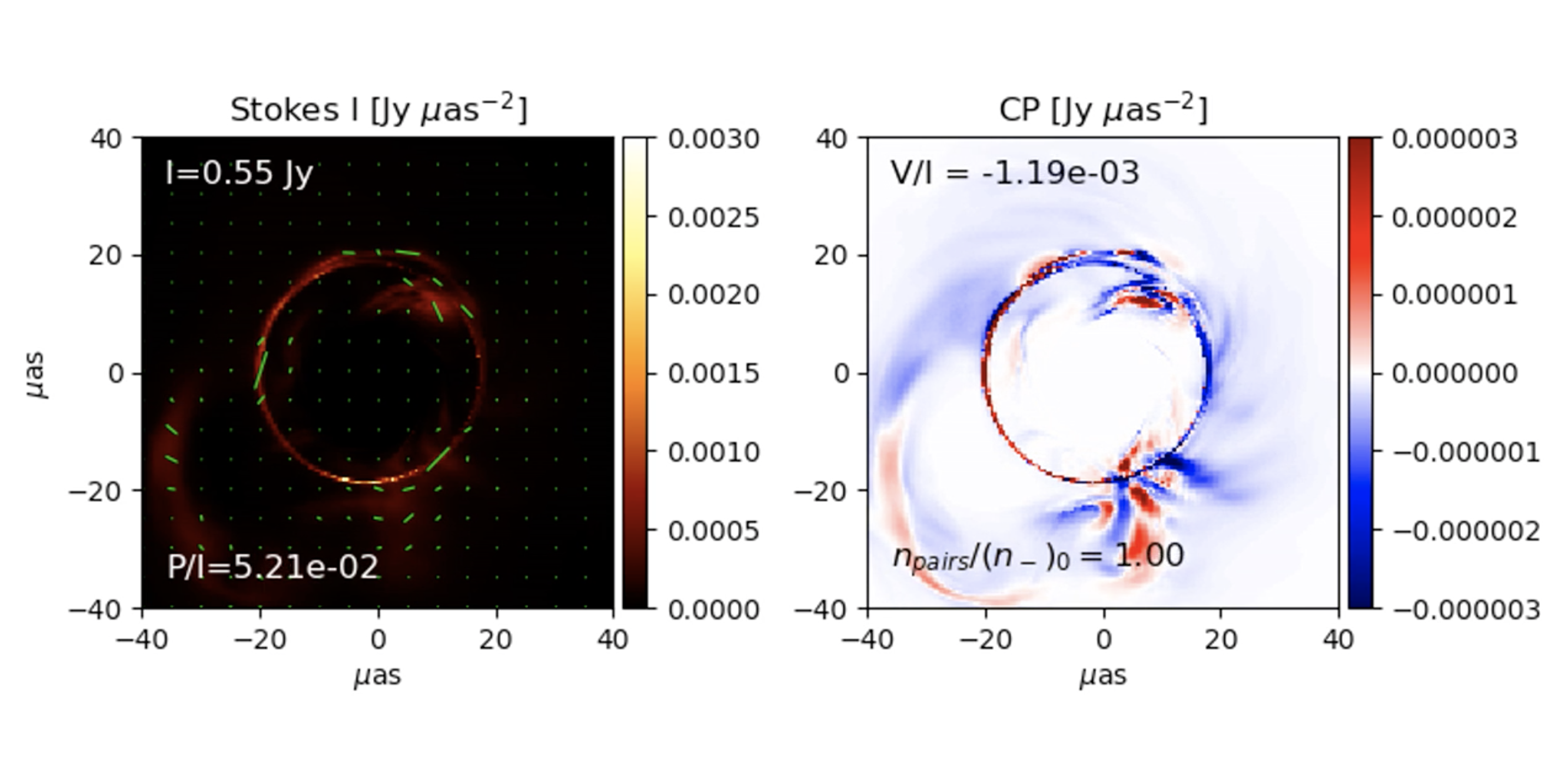}
\end{align}
\caption[Critical Beta Positron Effects]{\textcolor{black}{For the $a=-0.5$ MAD at $T=25,000M$: Top panel: Critical Beta at 230 GHz without positrons
Bottom panel: Critical Beta at 230 GHz for an even mix of pair and ionic plasma. 
}}\label{PositronVariationMADCriticalBeta}
\end{figure}

\textcolor{black}{In Fig. \ref{PositronVariationMADCriticalBeta}, the Critical $\beta$ image and $V/I$ map mirror the global structure in the $R-\beta$ case in Fig. \ref{PositronVariationMADRBeta}. They also share similar dependence of the circular polarization dependence on the free electron-positron fraction, and partial reversal of circular polariaztion sense in the flux eruption loop. } 
 
\subsubsection{MAD R-$\beta$ with Constant $\beta_E$ Jet}

\begin{figure}
\hspace{-0.0cm}
\includegraphics[height=150pt,width=250pt,trim = 6mm 1mm 0mm 1mm]{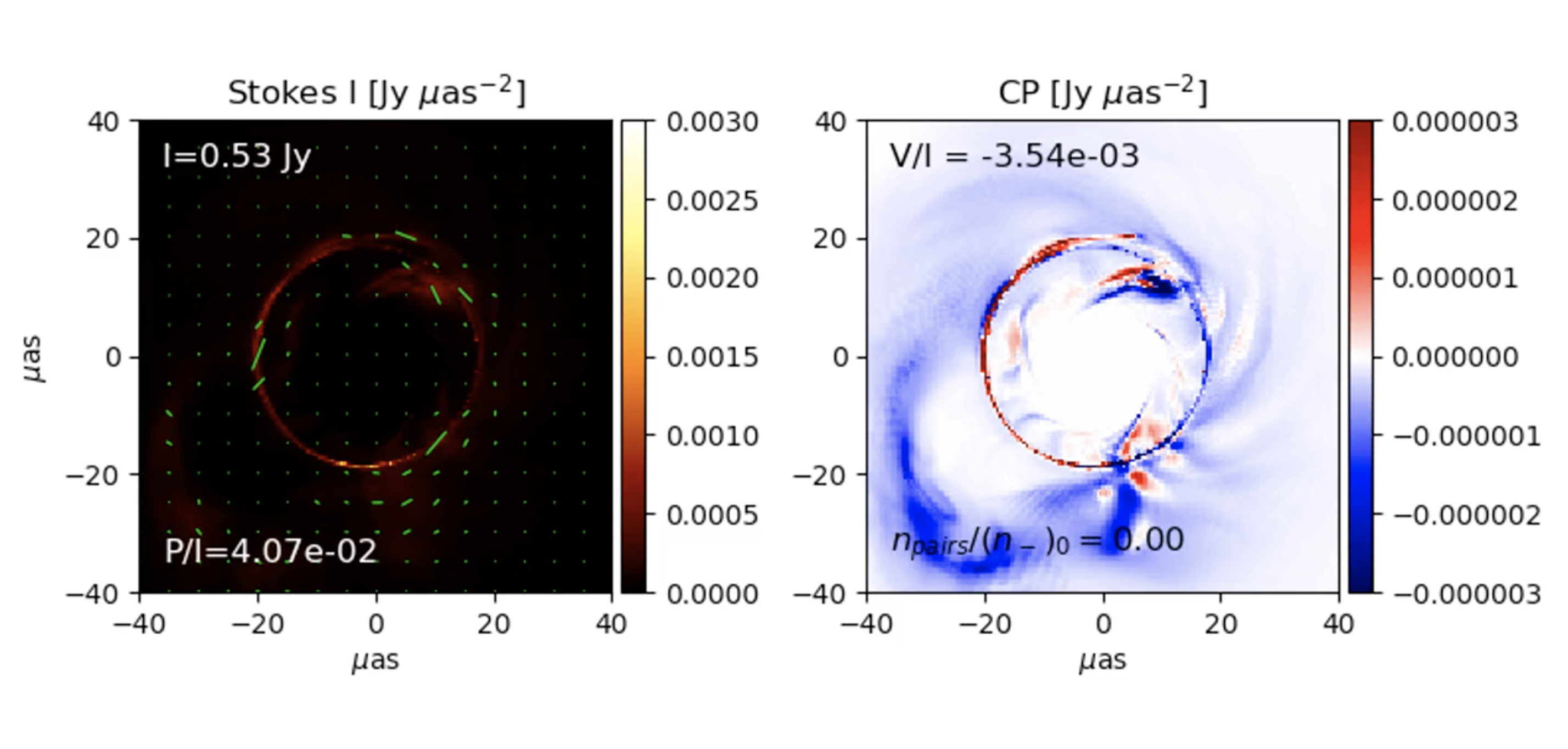}
\begin{align}\nonumber
& \hspace{-0.5cm} \includegraphics[trim = 6mm 1mm 0mm 0mm,  height=150pt,width=250pt
]{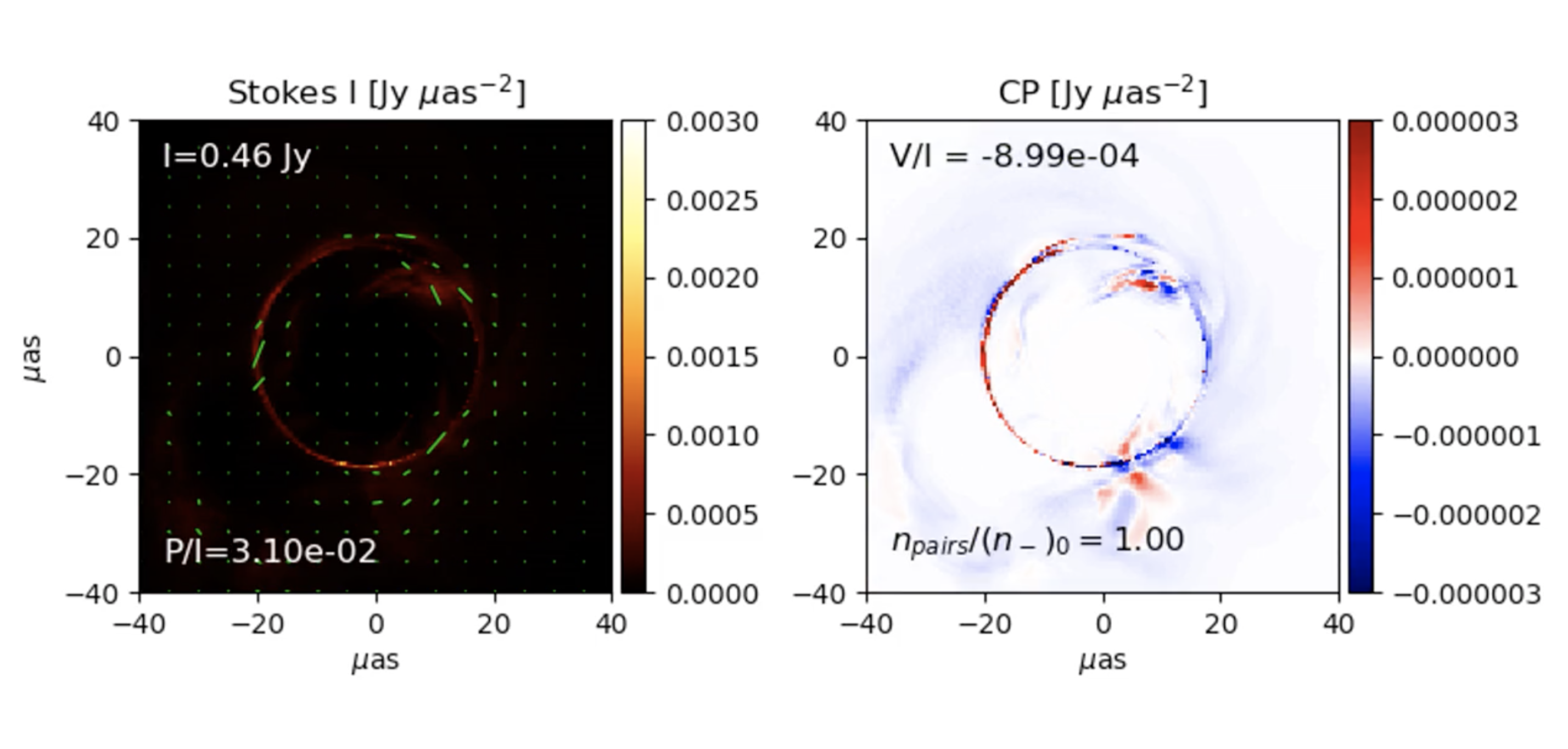}
\end{align}
\caption[R Beta Positron Effects]{\textcolor{black}{For the $a=-0.5$ MAD at $T=25,000M$: Top panel: R-Beta with $\beta_{e0}=0.01$ jet at 230 GHz without positrons
Bottom panel: R-Beta at 230 GHz for an even mix of pair and ionic plasma. 
}}\label{PositronVariationMADRBetaWithJet}
\end{figure}

\textcolor{black}{In Fig. \ref{PositronVariationMADRBetaWithJet}, the R-Beta model with jet maintains the prominent flux eruption loop as the above models. The presence of the Constant $\beta_e$ jets slightly reduced the circular polarization degree both with and without positrons.}
 
\subsubsection{MAD Critical $\beta$ with Constant $\beta_E$ Jet}

\begin{figure}
\hspace{-0.25cm}
\includegraphics[height=140pt,width=230pt,trim = 6mm 1mm 0mm 1mm]{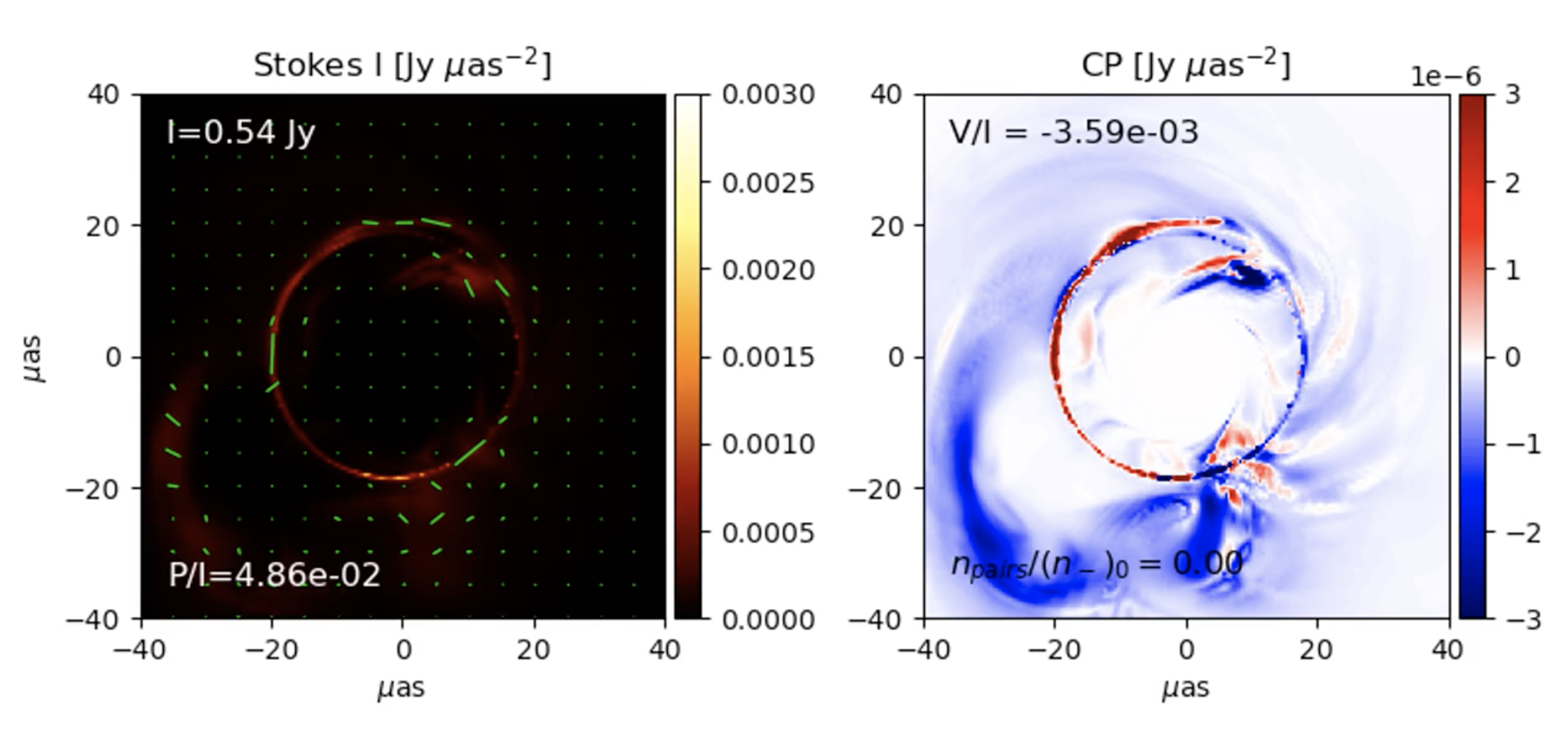}
\begin{align}\nonumber
& \hspace{-0.4cm} \includegraphics[trim = 6mm 1mm 0mm 0mm,  height=140pt,width=230pt
]{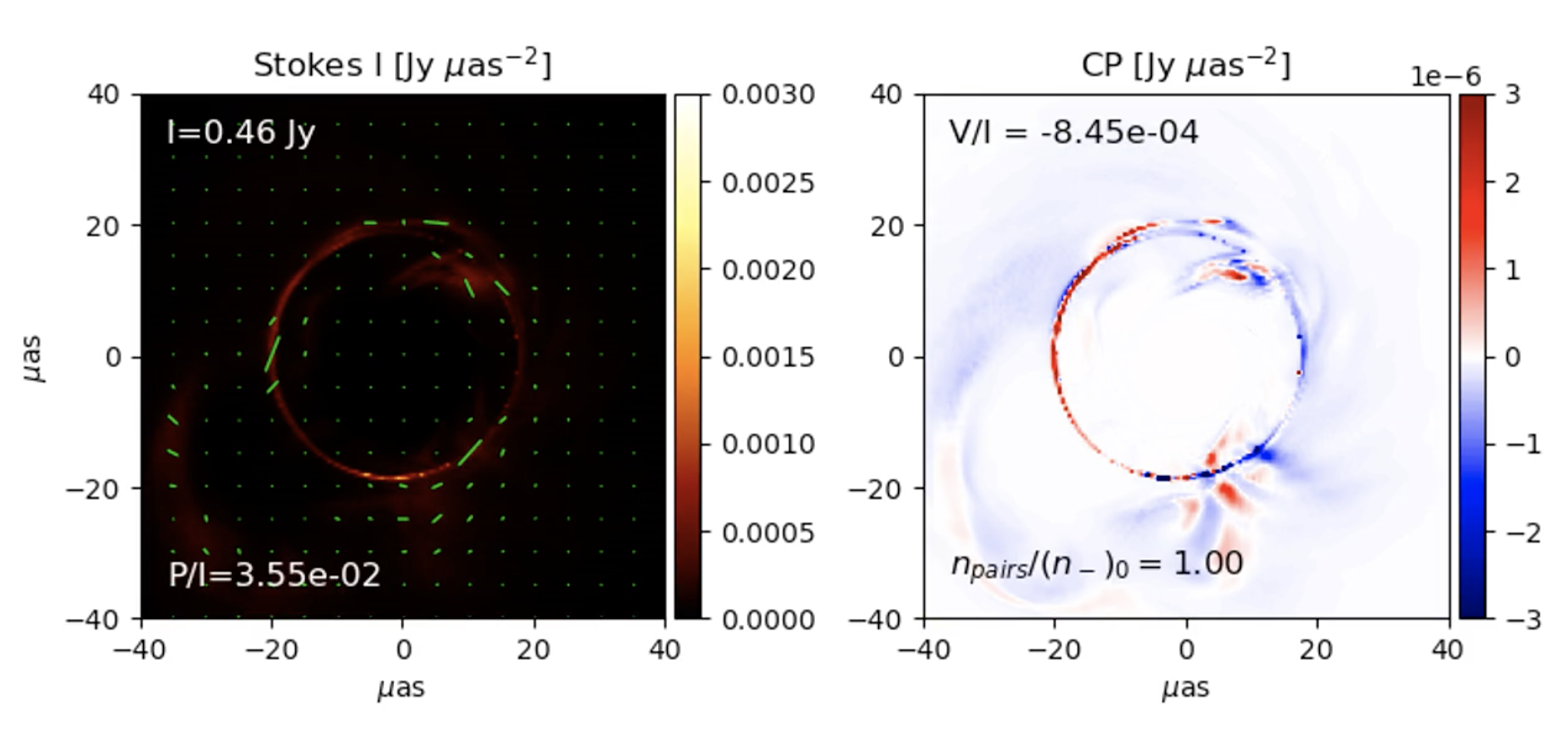}
\end{align}
\caption[Critical Beta Positron Effects]{\textcolor{black}{For the $a=-0.5$ MAD at $T=25,000M$: Top panel: Critical Beta with $\beta_{e0}=0.01$ jet at 230 GHz without positrons
Bottom panel: Critical Beta with $\beta_{e0}=0.01$ jet at 230 GHz for an even mix of pair and ionic plasma. 
}}\label{PositronVariationCriticalBetaWithJet}
\end{figure}

\textcolor{black}{In Fig. \ref{PositronVariationCriticalBetaWithJet} the Critical Beta model with jet 
exhibits the same trends as  its $R-\beta$ counterpart above.
}
 
\subsection{\textcolor{black}{Comparison of R-Beta and Critical Beta Models}}

\begin{figure}
\hspace{-0.25cm}
\includegraphics[height=150pt,width=250pt,trim = 6mm 1mm 0mm 1mm]{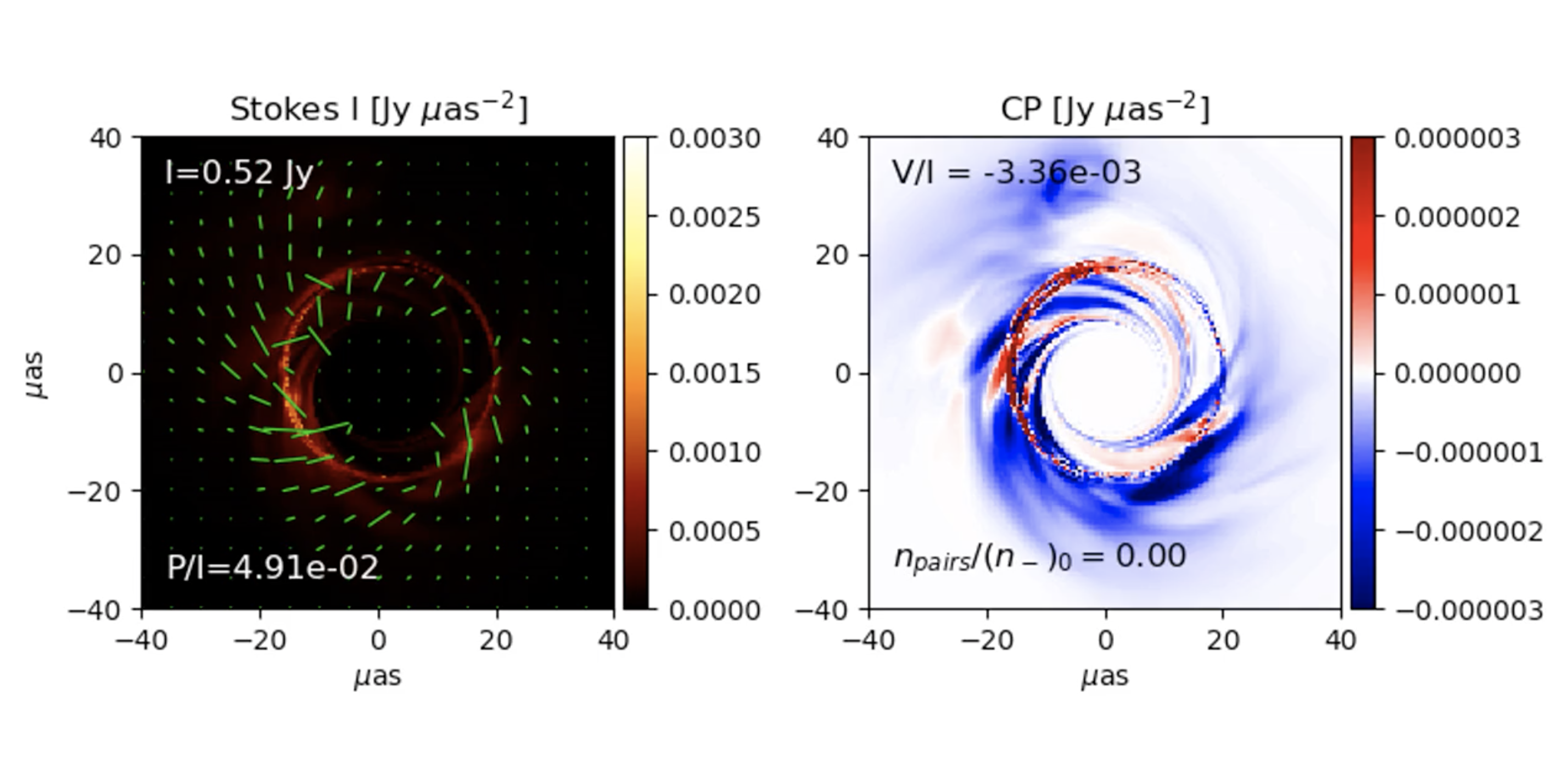
}
\begin{align}\nonumber
& \hspace{-0.25cm} \includegraphics[trim = 0mm 0mm 0mm 0mm, clip, height=150pt,width=250pt
]{fPt5BetaC1MADa+Pt94fPos0v2
.png}
\end{align}
\caption[R-Beta vs. Critical Beta]{\textcolor{black}{R-Beta vs. Critical Beta comparison for MAD  $a=+0.94$. Top panel: R-Beta at 230 GHz.
Bottom panel: Critical Beta }}\label{RBetavsCriticalBeta}
\end{figure} 

\textcolor{black}{In Fig. \ref{RBetavsCriticalBeta} we compare Critical Beta Model 
with $R-\beta$ Model for MAD $a/M=0.94$. Like comparisons between Figs. \ref{PositronVariationCriticalBeta} and \ref{PositronVariationRBetaWithJet} and between \ref{PositronVariationMADRBeta} and \ref{PositronVariationMADCriticalBeta} show, these models are quite degenerate at the level of total intensity morphology. 
However, the circular polarization of the Critical Beta $V/I$ map does exhibit more scrambling near the photon ring-- where a broad range of $\beta$'s may contribute given our shallow exponential parameter $\beta_c=0.5$.
}

\subsection{Extreme Positron Fractions}

\textcolor{black}{As mentioned in Sec. \ref{Sec:SimVsObs} 
increasing the pair fraction causes intrinsically emitted circular polarization to decrease, Faraday rotation to decrease, and Faraday conversion to increase.  Faraday rotation is essential for depolarizing these accretion flows.  Thus, dramatic effects occur when the pair fraction is raised high enough to turn an model from Faraday thick to Faraday thin.}

\begin{figure}
\hspace{-0.7cm}
\includegraphics[height=150pt,width=250pt,trim = 6mm 1mm 0mm 1mm]{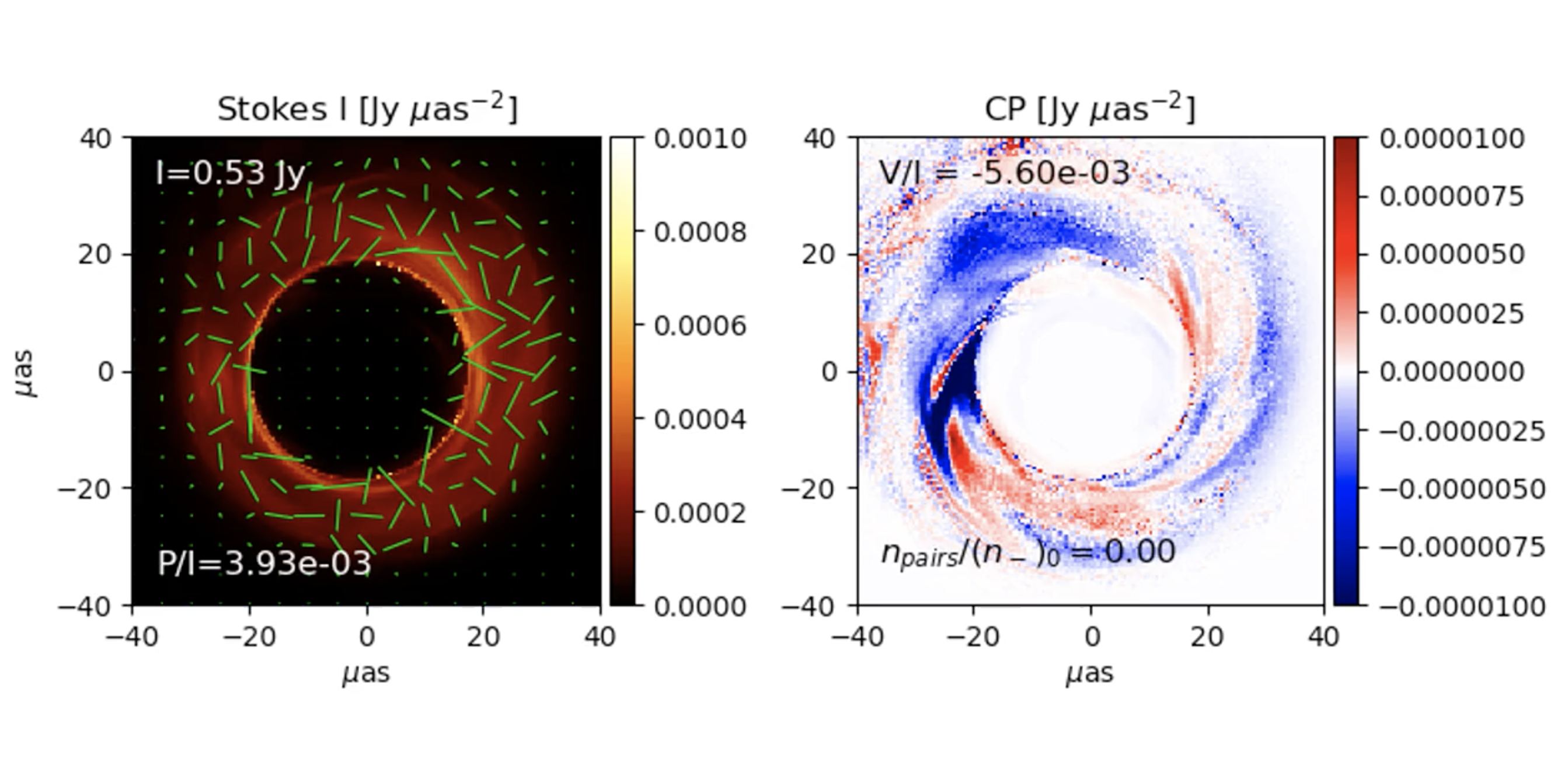
} 
\begin{align}\nonumber
& \hspace{-1.0cm} \includegraphics[trim = 0mm 0mm 0mm 0mm, clip, height=150pt,width=250pt
]{ SANEa-Pt5nPairsOnn100
.png}
\end{align}
\caption[R-Beta fPos SANE]{\textcolor{black}{Extreme positron variation comparison for SANE $a=-0.5$ R-Beta at 230 GHz:  fPos = 0 (Top Panel) vs. fPos = 100
(Bottom Panel).  }}\label{ExtremePositronSANEa-Pt5}
\end{figure}

\begin{figure}
\hspace{-0.25cm}
\includegraphics[height=150pt,width=250pt,trim = 6mm 1mm 0mm 1mm]{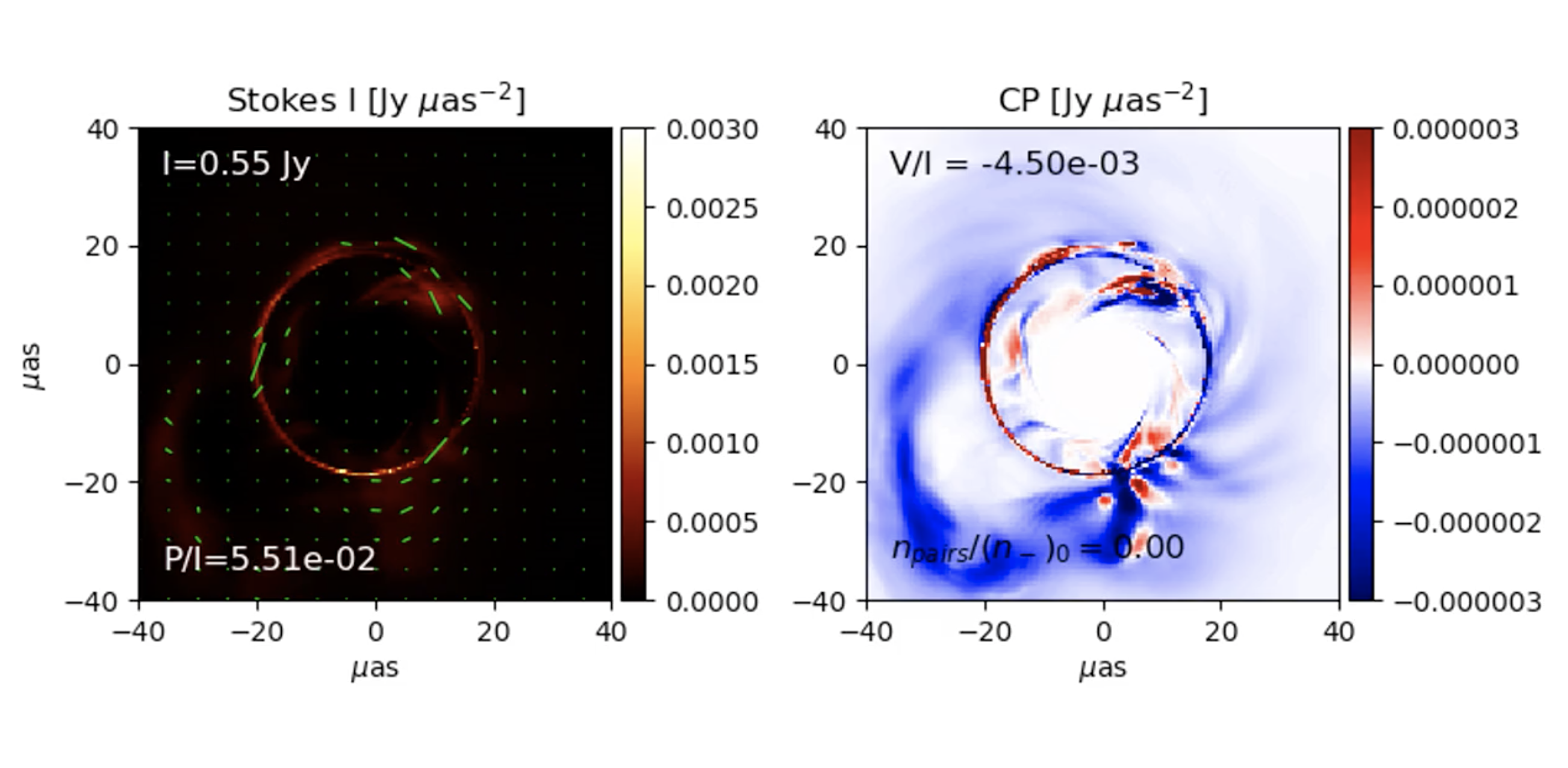
}
\begin{align}\nonumber
& \hspace{-0.25cm} \includegraphics[trim = 0mm 0mm 0mm 0mm, clip, height=150pt,width=250pt
]{MADa-Pt5nPairsOnn100
.png}
\end{align}
\caption[R-Beta fPos]{\textcolor{black}{
Extreme positron variation comparison for MAD $a=-0.5$ R-Beta at 230 GHz:  fPos = 0 (Top Panel) vs. fPos = 100
(Bottom Panel). }}\label{ExtremePositronMADa-Pt5}
\end{figure}

\begin{figure}
\hspace{-0.5cm}
\includegraphics[height=150pt,width=250pt,trim = 6mm 1mm 0mm 1mm]{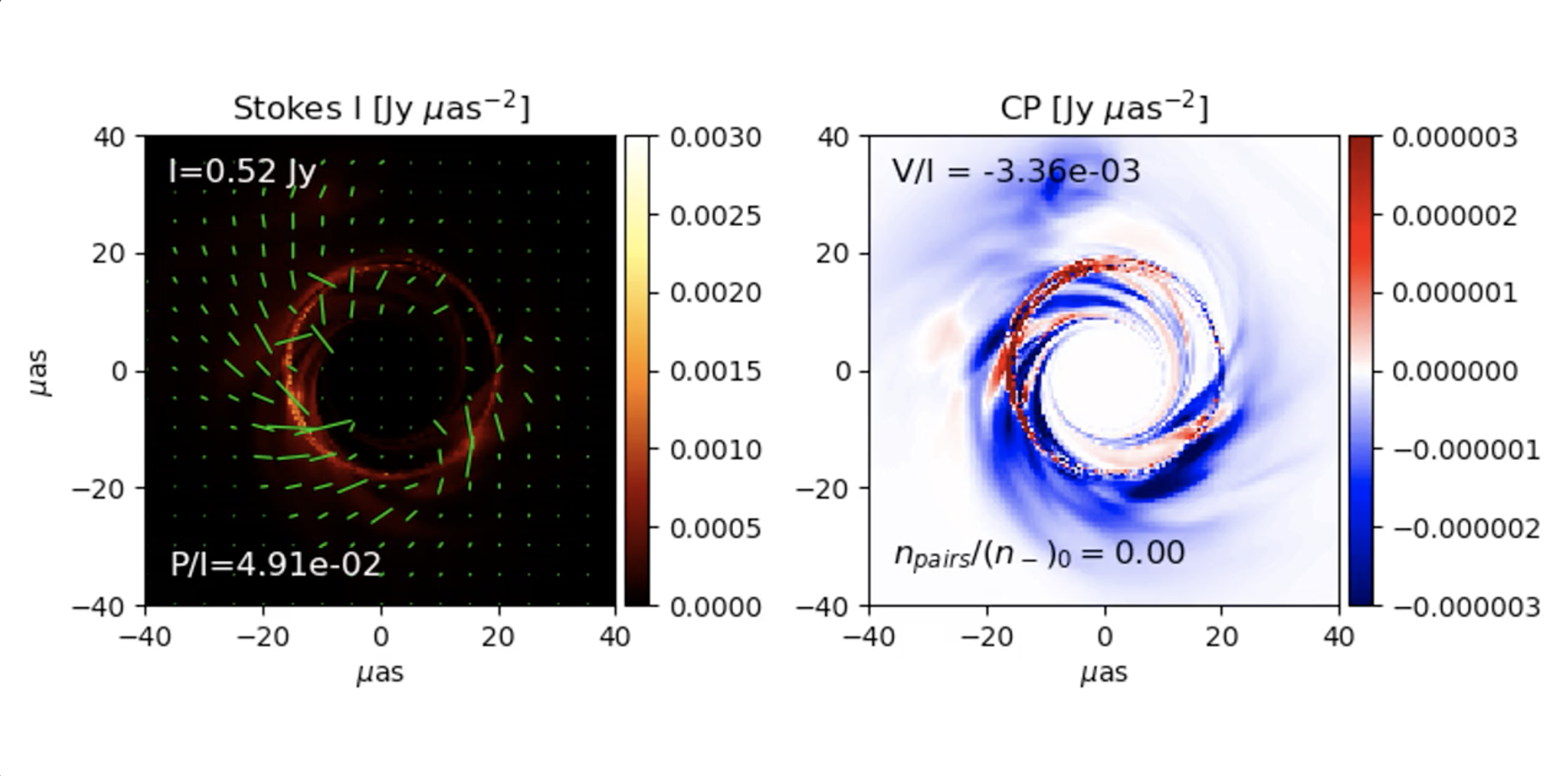
}
\begin{align}\nonumber
& \hspace{-.5cm} \includegraphics[trim = 0mm 0mm 0mm 0mm, clip, height=150pt,width=250pt
]{MADa+Pt94nPairsOnn100
.png}
\end{align}
\caption[R-Beta fPos]{\textcolor{black}{MAD $a=+0.94$ R-Beta at 230 GHz.Top panel: fPos = 0 
Bottom panel: fPos = 100 }}\label{ExtremePositronMADa+Pt94}
\end{figure}

\textcolor{black}{In Fig. \ref{ExtremePositronSANEa-Pt5} we show the effects of raising $n_\mathrm{pairs}/n_0=100$ for a SANE $a=-0.5$ simulation and contrast with MAD $a=-0.5$ and  $a=+0.94$ in  Figs. \ref{ExtremePositronMADa-Pt5} and
\ref{ExtremePositronMADa+Pt94},
respectively \textcolor{black}{(using the R-$\beta$ Model)}. 
The effect on the MAD simulation is subtle, characterized by a decrease in the intrinsically emitted circular polarization that dominates on large scales.  
\textcolor{black}{ Note that this is model-specific, and in fact Faraday conversion is the only source of circular polarization in 
pair plasma jet models in \cite{Anantua2020a}a }
Meanwhile, the effect is much stronger for the SANE simulation, which is intrinsically more Faraday thick.  After removing Faraday rotation, the simulation acquires a much more ordered linear polarization pattern.  In addition, very large circular polarization fractions are produced in the absence of depolarization by Faraday rotation. }


\subsection{\textcolor{black}{Comparison of Models with Polarization Constraints}}

\textcolor{black}{In Table \ref{Model_Linear_Polarization_Table}, we compare  observations from \cite{EHT2021VII} against fiducial model linear polarization: both summed from net (unresolved) $Q$, $U$ and $I$ across the image plane,
\begin{equation}
   |m|_\mathrm{net} = \frac{\sqrt{\left(\sum_i{Q_i}\right)^2+\left(\sum_i{U_i}\right)^2}}{\sum_i{I_i}}
\end{equation}
and its average local (resolved) magnitude
\begin{equation}
   <|m|> = \frac{\sum_i I_iP_i}{\sum_i{I_i}} = \frac{\sum_i\sqrt{Q_i^2+U_i^2}}{\sum_i{I_i}}.
\end{equation} SANE models tend to be less linearly polarized on net and MAD models more linearly polarized on net than the constraint-- though all models exceed the beam/resolution-dependent averaged polarization magnitude constraint. The  fiducial models satisfying the net linear polarization constraint for MAD $a=-0.5$ are the R-Beta, R-Beta with Jet and  Critical Beta with Jet, all with maximal positron fractions. For MAD $a=+0.94$, only the positron-free Critical Beta moddels (with and without jet) satisfy the $|m|_\mathrm{net} $ constraint.  }

\textcolor{black}{In Table }\textcolor{black}{
\ref{Model_Circular_Polarization_Table}, we project forward anticipating circular polarization measurements will be performed in the near future to compare with our models.}
\textcolor{black}{All of our models satisfy the preliminary $V_\mathrm{net}$ constraint in \cite{Goddi2021}. For the structure parameter $\beta_2$, however, Table \ref{Beta2_Polarization_Table} shows only MAD $a=0.94M$ pure turbulent heating models pass.}

\textcolor{black}{We may check the robustness of these tendencies by re-analyzing over the temporal evolution of the simulations (cf. Sect. \ref{TimeDependence}). Though little changes for the comparison against the more reliable unresolved linear polarization ($|m_\mathrm{net}|$) observations, 
the beam-dependent $(<|m|>)$ comparison
changes significantly with temporal evolution.}

\begin{table*}
\caption{\textcolor{black}{Linear polarization  $|m|_\mathrm{net}$ and $<|m|>$  for fiducial models at $T=25,000M$. The observational constraints from EHT M87 Paper VII take the form of the polarization ranges $0.01\le|m|_\mathrm{net} \le 0.037$ and $0.057< <|m|> < 0.107$.
} \textcolor{black}{Note that the bold values refer to fiducial models which satisfy the net linear polarization constraints.}
}
\makebox[ 0.83 \textwidth][c]{
\begin{tabular}{c|ccccc|ccccc|ccccc}
\hline
&$.$ & $\mathrm{SANE}$  & ($a=-0.5$) &  &  & $\mathrm{MAD}$ & ($a=-0.5$) 
\\ 
 &$R\mathrm{-Beta}$ &  $R\mathrm{-Beta}$  & $\mathrm{Crit.\ Beta}$ & $\mathrm{Crit.\ Beta}$&  $R\mathrm{-Beta}$ &  $R\mathrm{-Beta}$  & $\mathrm{Crit.\ Beta}$ & $\mathrm{Crit.\ Beta}$  \\ 
&&$\mathrm{w./\ Jet}$&&$\mathrm{w./\ Jet}$&&$\mathrm{w./\ Jet}$&&$\mathrm{w./\ Jet}$\\ 
 \hline
 $|m|_\mathrm{net}(f_\mathrm{pos,min})$&$3.93\cdot10^{-3}$ & $3.97\cdot10^{-3}$  & $8.51\cdot10^{-3}$  & $2.26\cdot10^{-3}$  & $5.51\cdot10^{-2}$ & $4.07\cdot10^{-2}$ & $6.93\cdot10^{-2}$  & $4.86\cdot10^{-2}$  \\ \hline
$|m|_\mathrm{net}(f_\mathrm{pos,max})$ &$1.62\cdot10^{-3}$ & $2.88\cdot10^{-3}$ & $2.73\cdot10^{-3}$  & $2.50\cdot10^{-3}$  & $\mathbf{3.67\cdot10^{-2}}$ & $\mathbf{3.10\cdot10^{-2}}$ & $5.21\cdot10^{-2}$ & $\mathbf{3.55\cdot10^{-2}}$  \\ \hline 
$<|m|>(f_\mathrm{pos,min})$ &$1.29\cdot10^{-1}$ & $1.38\cdot10^{-1}$ & $1.35\cdot10^{-1}$  & $1.44\cdot10^{-1}$  & $3.49\cdot10^{-1}$ & $3.30\cdot10^{-1}$ & $2.71\cdot10^{-1}$ & $2.53\cdot10^{-1}$  \\ \hline 
$<|m|>(f_\mathrm{pos,max})$ &$1.31\cdot10^{-1}$ & $1.45\cdot10^{-1}$ & $1.40\cdot10^{-1}$  & $1.47\cdot10^{-1}$  & $4.20\cdot10^{-1}$ & $3.60\cdot10^{-1}$ & $3.83\cdot10^{-1}$ & $3.46\cdot10^{-1}$  \\ \hline 
 &&  & &  &    &  $\mathrm{MAD}$ & ($a=+0.94$) 
\\ 
 & &  &  && $R\mathrm{-Beta}$ & $R\mathrm{-Beta}$& $\mathrm{Crit.\ Beta}$ & $\mathrm{Crit.\ Beta}$  \\ 
&&&&&&$\mathrm{w./\ Jet}$&&$\mathrm{w./\ Jet}$ \\ 
 \hline
 $|m|_\mathrm{net}(f_\mathrm{pos,min})$& & &  & & $4.91\cdot10^{-2}$  & $4.56\cdot10^{-2}$& $\mathbf{3.35\cdot10^{-2}}$ & $\mathbf{3.67\cdot10^{-2} }$ \\ \hline
$|m|_\mathrm{net}(f_\mathrm{pos,max})$ & & &  &  & $5.17\cdot10^{-2}$& $5.06\cdot10^{-2}$ & $4.59\cdot10^{-2}$ & $4.82\cdot10^{-2}$ \\ \hline
$<|m|>(f_\mathrm{pos,min})$ & & &  &  & $5.76\cdot10^{-1}$& $5.18\cdot10^{-1}$ & $5.18\cdot10^{-1}$ & $4.81\cdot10^{-1}$ \\ \hline
$<|m|>(f_\mathrm{pos,max})$ & & &  &  & $5.86\cdot10^{-1}$& $5.25\cdot10^{-1}$ & $5.81\cdot10^{-1}$ & $5.26\cdot10^{-1}$ \\ \hline
\end{tabular}
}
\label{Model_Linear_Polarization_Table}
\end{table*}

\begin{table*}
\caption{\textcolor{black}{Circular polarization  
 $|\textcolor{black}{v}|_\mathrm{net}$ and $<|\textcolor{black}{v}|>$  for fiducial models at $T=25,000M$. Note that all models satisfy the $\textcolor{black}{|v|}_\mathrm{net}$=0.008 EHT bound 
\citep{EHT2021VII,Goddi2021}
\textcolor{black}{and the newly resolved EHT circular polarization fraction constraint  $ <|v|> < 0.037$ \citep{EHT2023IX}.}
}
}
\makebox[ 0.83 \textwidth][c]{
\begin{tabular}{c|ccccc|ccccc|ccccc}
\hline
&$.$ & $\mathrm{SANE}$  & ($a=-0.5$) &  &  & $\mathrm{MAD}$ & ($a=-0.5$) 
\\ 
 &$R\mathrm{-Beta}$ &  $R\mathrm{-Beta}$  & $\mathrm{Crit.\ Beta}$ & $\mathrm{Crit.\ Beta}$&  $R\mathrm{-Beta}$ &  $R\mathrm{-Beta}$  & $\mathrm{Crit.\ Beta}$ & $\mathrm{Crit.\ Beta}$  \\ 
&&$\mathrm{w./\ Jet}$&&$\mathrm{w./\ Jet}$&&$\mathrm{w./\ Jet}$&&$\mathrm{w./\ Jet}$\\ 
 \hline
 $|V|_\mathrm{net}(f_\mathrm{pos,min})$&$-5.60\cdot10^{-3}$ & $-1.65\cdot10^{-3}$  & $-7.77\cdot10^{-3}$  & $-1.72\cdot10^{-3}$  & $-4.50\cdot10^{-3}$ & $-3.54\cdot10^{-3}$ & $-4.24\cdot10^{-3}$  & $-3.59\cdot10^{-3}$  \\ \hline
$|V|_\mathrm{net}(f_\mathrm{pos,max})$ &$-2.81\cdot10^{-3}$ & $2.67\cdot10^{-3}$ & $3.15\cdot10^{-3}$  & $7.44\cdot10^{-4}$  & $-1.39\cdot10^{-3}$ & $-8.99\cdot10^{-4}$ & $-1.19\cdot10^{-3}$ & $-8.45\cdot10^{-4}$  \\ \hline 
$<|V|>(f_\mathrm{pos,min})$ &$6.75\cdot10^{-3}$ & $5.14\cdot10^{-3}$ & $8.94\cdot10^{-3}$  & $6.62\cdot10^{-3}$  & $3.98\cdot10^{-3}$ & $2.96\cdot10^{-3}$ & $3.82\cdot10^{-3}$ & $3.03\cdot10^{-3}$  \\ \hline 
$<|V|>(f_\mathrm{pos,max})$ &$1.28\cdot10^{-2}$ & $7.65\cdot10^{-3}$ & $2.32\cdot10^{-2}$  & $9.62\cdot10^{-3}$  & $9.04\cdot10^{-4}$ & $5.75\cdot10^{-4}$ & $8.57\cdot10^{-4}$ & $5.66\cdot10^{-4}$  \\ \hline 
 &&  & &  &    &  $\mathrm{MAD}$ & ($a=+0.94$) 
\\ 
 & &  &  && $R\mathrm{-Beta}$ & $R\mathrm{-Beta}$& $\mathrm{Crit.\ Beta}$ & $\mathrm{Crit.\ Beta}$  \\ 
&&&&&&$\mathrm{w./\ Jet}$&&$\mathrm{w./\ Jet}$ \\ 
 \hline
 $|V|_\mathrm{net}(f_\mathrm{pos,min})$& & &  & & $-3.36\cdot10^{-3}$  & $2.33\cdot10^{-3}$& $-3.02\cdot10^{-3}$ & $-2.41\cdot10^{-3}$ \\ \hline
$|V|_\mathrm{net}(f_\mathrm{pos,max})$ & & &  &  & $-1.13\cdot10^{-2}$& $7.37\cdot10^{-4}$ & $-1.19\cdot10^{-3}$ & $-7.29\cdot10^{-4}$ \\ \hline
$<|V|>(f_\mathrm{pos,min})$ & & &  &  & $3.18\cdot10^{-3}$& $2.17\cdot10^{-3}$ & $2.85\cdot10^{-3}$ & $2.19\cdot10^{-3}$ \\ \hline
$<|V|>(f_\mathrm{pos,max})$ & & &  &  & $7.62\cdot10^{-4}$& $5.02\cdot10^{-4}$ & $7.74\cdot10^{-4}$ & $4.86\cdot10^{-4}$ \\ \hline
\end{tabular}
}
\label{Model_Circular_Polarization_Table}
\end{table*}

\begin{table*}
\caption{\textcolor{black}{Azimuthal structure mode $\beta_2$ for fiducial models at $T=25,000M$. The observational constraints from EHT M87 Paper VII are in the range $0.04\le |\beta_2| \le 0.07$. \textcolor{black}{Note that the bold values refer to fiducial models which satisfy the observational constraints.}}
}
\makebox[ 0.83 \textwidth][c]{
\begin{tabular}{c|ccccc|ccccc|ccccc}
\hline
&$.$ & $\mathrm{SANE}$  & ($a=-0.5$) &  &  & $\mathrm{MAD}$ & ($a=-0.5$) 
\\ 
 &$R\mathrm{-Beta}$ &  $R\mathrm{-Beta}$  & $\mathrm{Crit.\ Beta}$ & $\mathrm{Crit.\ Beta}$&  $R\mathrm{-Beta}$ &  $R\mathrm{-Beta}$  & $\mathrm{Crit.\ Beta}$ & $\mathrm{Crit.\ Beta}$  \\ 
&&$\mathrm{w./\ Jet}$&&$\mathrm{w./\ Jet}$&&$\mathrm{w./\ Jet}$&&$\mathrm{w./\ Jet}$\\ 
 \hline
 $\beta_2(f_\mathrm{pos,min})$&$3.51\cdot10^{-3}$ & $2.82\cdot10^{-3}$  & $3.84\cdot10^{-3}$  & $3.80\cdot10^{-3}$  & $6.22\cdot10^{-3}$ & $6.31\cdot10^{-3}$ & $1.17\cdot10^{-2}$  & $6.18\cdot10^{-3}$  \\ \hline
$\beta_2(f_\mathrm{pos,max})$ &$2.96\cdot10^{-3}$ & $3.03\cdot10^{-3}$ & $4.24\cdot10^{-3}$  & $9.86\cdot10^{-4}$  & $1.47\cdot10^{-2}$
& $9.83\cdot10^{-3}$ & $2.15\cdot10^{-2}$ & $1.32\cdot10^{-2}$  \\ \hline 
 &&  & &  &    &  $\mathrm{MAD}$ & ($a=+0.94$) 
\\ 
 & &  &  && $R\mathrm{-Beta}$ & $R\mathrm{-Beta}$& $\mathrm{Crit.\ Beta}$ & $\mathrm{Crit.\ Beta}$  \\ 
&&&&&&$\mathrm{w./\ Jet}$&&$\mathrm{w./\ Jet}$ \\ 
 \hline
 $\beta_2(f_\mathrm{pos,min})$& & &  & & $3.23\cdot10^{-2}$  & $2.42\cdot10^{-2}$& $\mathbf{3.58\cdot10^{-2}}$ & $2.77\cdot10^{-2} $ \\ \hline
$\beta_2(f_\mathrm{pos,max})$ & & &  &  & $\mathbf{3.93\cdot10^{-2}}$& $2.72\cdot10^{-2}$ & $\mathbf{3.66\cdot10^{-2}}$ & $2.72\cdot10^{-2}$ \\ \hline
\end{tabular}
}
\label{Beta2_Polarization_Table}
\end{table*}


\subsection{Faraday Effects} 

\textcolor{black}{
As linear polarization travels through a magnetized plasma, its EVPA is rotated by Faraday rotation, interchanging Stokes $Q$ and $U$.  Similarly, Faraday conversion exchanges linear and circular polarization, interchanging Stokes $U$ and $V$.  Both of these effects can be significant in accreting black hole systems.  In particular, Faraday rotation is believed to be extremely important for reducing the linear polarization fraction in models of M87* to the observed values.  Typically, SANEs have larger Faraday rotation and conversion depths than MADs.  This is largely because SANE models require larger mass densities to match the observed flux of M87*.  They also have lower temperatures, which increases the efficiency of Faraday effects.}

\subsubsection{Faraday Rotation}

\textcolor{black}{Table \ref{Farady-Rotation-Depth_Table
} of fiducial model Faraday rotation depths shows a pronounced gap between a marginal effect in MAD simulations relative to the corresponding effect which is 3 orders of magnitude larger in SANE simulations. Varying positron content even at the percent level leads to large EVPA rotational swings for SANE plasmas due to the large absolute response of the Faraday rotation measure to the increased fraction of positrons. This naturally leads to a profoundly discriminating probe of plasma magnetic inflow properties in regions of changing positron fraction. Even when the plasma composition is in steady state, we may identify the rapid spatial variation of circular polarization as a signature of high Faraday rotation depth characteristic of SANE accretion flows. }

\newpage
\begin{table*}
\caption{\textcolor{black}{Faraday rotation depth at min and max fpos for fiducial models at $T=25,000M$:}
}
\makebox[ 0.83 \textwidth][c]{
\begin{tabular}{c|ccccc|ccccc|ccccc}
\hline
&$.$ & $\mathrm{SANE}$  & ($a=-0.5$) &  &  & $\mathrm{MAD}$ & ($a=-0.5$) 
\\ 
 &$R\mathrm{-Beta}$ &  $R\mathrm{-Beta}$  & $\mathrm{Crit.\ Beta}$ & $\mathrm{Crit.\ Beta}$&  $R\mathrm{-Beta}$ &  $R\mathrm{-Beta}$  & $\mathrm{Crit.\ Beta}$ & $\mathrm{Crit.\ Beta}$  \\ 
&&$\mathrm{w./\ Jet}$&&$\mathrm{w./\ Jet}$&&$\mathrm{w./\ Jet}$&&$\mathrm{w./\ Jet}$\\ 
 \hline
$\tau_V(\mathrm{f_{pos,min}})$&$2.73\cdot10^{3}$ & $1.30\cdot10^{3}$  & $7.74\cdot10^{3}$  & $3.40\cdot10^{3}$  & 1.88 & 1.55 & 4.64 & 4.38  \\ \hline
$\tau_V(\mathrm{f_{pos,max}})$&$5.92\cdot10^{2}$ & $2.01\cdot10^{2}$ & $1.52\cdot10^{3}$  & $6.19\cdot10^{2}$  & $3.31\cdot10^{-2}$ & $2.44\cdot10^{-1}$ & $9.93\cdot10^{-1}$ & $7.00\cdot10^{-1}$  \\ \hline 
 &&  & &  &    &  $\mathrm{MAD}$ & ($a=+0.94$) 
\\ 
 & &  &  && $R\mathrm{-Beta}$ & $R\mathrm{-Beta}$& $\mathrm{Crit.\ Beta}$ & $\mathrm{Crit.\ Beta}$  \\ 
&&&&&&$\mathrm{w./\ Jet}$&&$\mathrm{w./\ Jet}$ \\ 
 \hline
$\tau_V(\mathrm{f_{pos,min}})$& & &  & & $6.48\cdot10^{-1}$  & $4.78\cdot10^{-1}$& 2.66 & 2.22  \\ \hline
 $\tau_V(\mathrm{f_{pos,max}})$ & & &  &  & $1.16\cdot10^{-1}$& $9.6\cdot10^{-2}$ & $5.25\cdot10^{-1}$ & $3.79\cdot10^{-1}$ \\ \hline
\end{tabular}
}
\label{Farady-Rotation-Depth_Table}
\end{table*}

\begin{table*}
\caption{\textcolor{black}{Faraday conversion depth at min and max fpos for fiducial models at $T=25,000M$:}
}
\makebox[ 0.83 \textwidth][c]{
\begin{tabular}{c|ccccc|ccccc|ccccc}
\hline
&$.$ & $\mathrm{SANE}$  & ($a=-0.5$) &  &  & $\mathrm{MAD}$ & ($a=-0.5$) 
\\ 
 &$R\mathrm{-Beta}$ &  $R\mathrm{-Beta}$  & $\mathrm{Crit.\ Beta}$ & $\mathrm{Crit.\ Beta}$&  $R\mathrm{-Beta}$ &  $R\mathrm{-Beta}$  & $\mathrm{Crit.\ Beta}$ & $\mathrm{Crit.\ Beta}$  \\ 
&&$\mathrm{w./\ Jet}$&&$\mathrm{w./\ Jet}$&&$\mathrm{w./\ Jet}$&&$\mathrm{w./\ Jet}$\\ 
 \hline
$\tau_Q(\mathrm{f_{pos,min}})$&$4.30\cdot10^{0}$ & $4.66\cdot10^{0}$  & $2.11\cdot10^{0}$  & $9.22\cdot10^{-1}$  & $3.88\cdot10^{-2}$ & $2.12\cdot10^{-2}$ & $2.84\cdot10^{-2}$  & $2.08\cdot10^{-2}$  \\ \hline
$\tau_Q(\mathrm{f_{pos,max}})$&$4.19\cdot10^{0}$ & $4.54\cdot10^{0}$ & $1.75\cdot10^{0}$  & $5.41\cdot10^{-1}$  & $3.54\cdot10^{-2}$ & $1.49\cdot10^{-2}$ & $3.22\cdot10^{-2}$ & $1.60\cdot10^{-2}$  \\ \hline 
 &&  & &  &    &  $\mathrm{MAD}$ & ($a=+0.94$) 
\\ 
 & &  &  && $R\mathrm{-Beta}$ & $R\mathrm{-Beta}$& $\mathrm{Crit.\ Beta}$ & $\mathrm{Crit.\ Beta}$  \\ 
&&&&&&$\mathrm{w./\ Jet}$&&$\mathrm{w./\ Jet}$ \\ 
 \hline
$\tau_Q(\mathrm{f_{pos,min}})$& & &  & & $4.03\cdot10^{-2}$  & $2.06\cdot10^{-2}$& $2.73\cdot10^{-2}$ & $1.84\cdot10^{-2}$ \\ \hline
 $\tau_Q(\mathrm{f_{pos,max}})$ & & &  &  & $3.77\cdot10^{-2}$& $2.04\cdot10^{-2}$ & $2.90\cdot10^{-2}$ & $1.56\cdot10^{-2}$ \\ \hline
\end{tabular}
}
\label{Farady-Conversion-Depth_Table}
\end{table*}

\textcolor{black}{\subsubsection{Faraday Conversion}}

\textcolor{black}{Table \ref{Farady-Conversion-Depth_Table
} shows fiducial model Faraday conversion depths for SANEs are 1-2 orders greater than those for MADs.  Faraday conversion depths tend to be lower than Faraday rotation depths: by 3 orders of magnitude for SANEs and 1-2 orders for MADs. However, because Faraday conversion results in the direct production of circular polarization (from linear), it may result in a significant contribution of $V$. 
\textcolor{black}{Faraday conversion can produce CP even in a pure pair plasma as long as the magnetic field twists along the line of sight \citep{Wardle2003,Ricarte+2021}.}
}

$\qquad$

\subsection{Frequency \textcolor{black}{and Inclination} Dependence}

\begin{figure}
\hspace{-0.4cm}
\includegraphics[height=150pt,width=250pt,trim = 6mm 1mm 0mm 1mm]{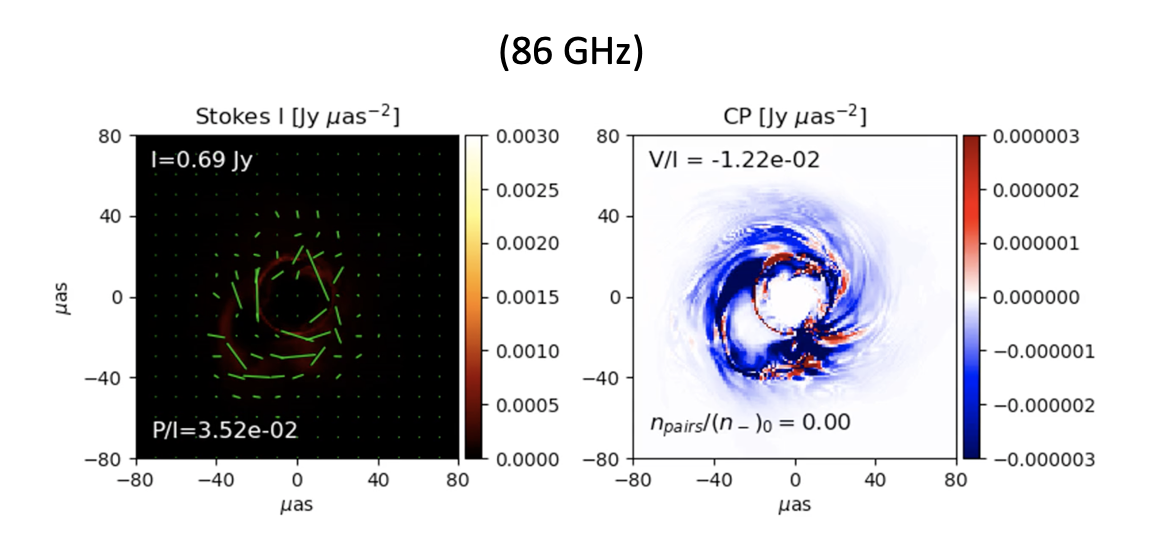}
\begin{align}\nonumber
& \hspace{-0.5cm} \includegraphics[trim = 6mm 1mm 0mm 0mm,  height=150pt,width=250pt
]{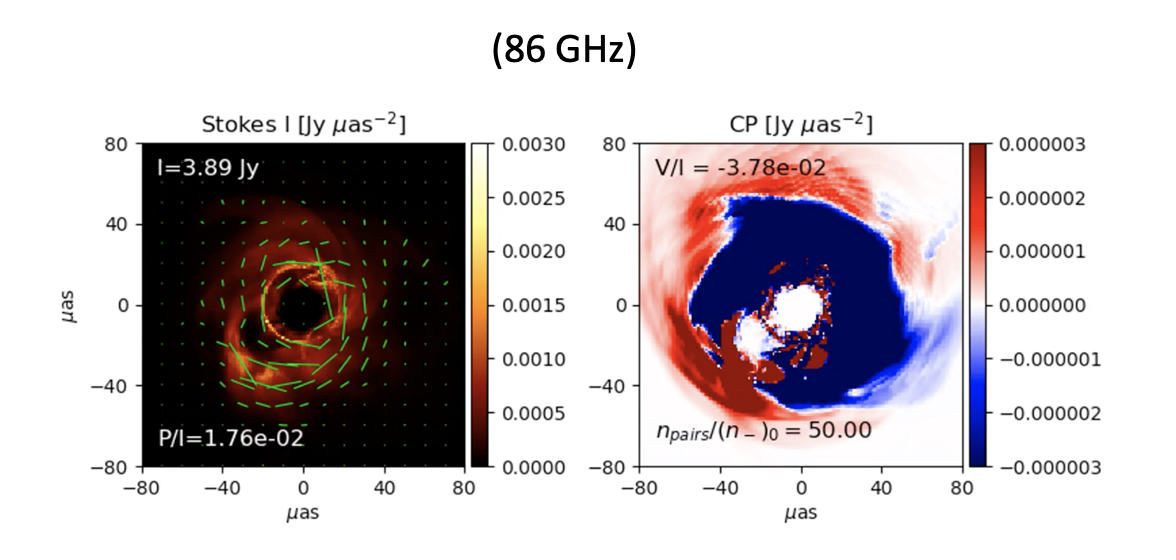}
\end{align}
\caption[Critical Beta Positron Effects]{\textcolor{black}{For the MAD $a=-0.5$ at T=25,000M at 86 GHz for the  R-Beta model with $\beta_{e0}=0.01$ jet: Top panel: Ionic plasma (fPos = 0); 
Bottom panel: fPos = 50.  }
}\label{MADRBetaWjet86GHz}
\end{figure}

\begin{figure}
\hspace{-0.25cm}
\includegraphics[height=150pt,width=250pt,trim = 6mm 1mm 0mm 1mm]{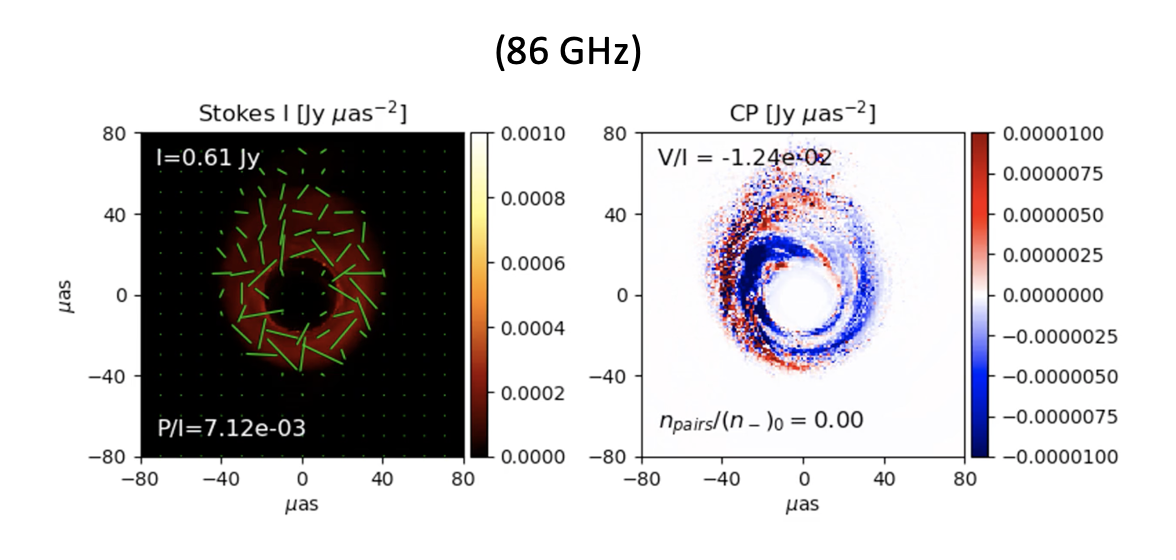}
\begin{align}\nonumber
& \hspace{-0.25cm} \includegraphics[trim = 6mm 1mm 0mm 0mm,  height=150pt,width=250pt
]{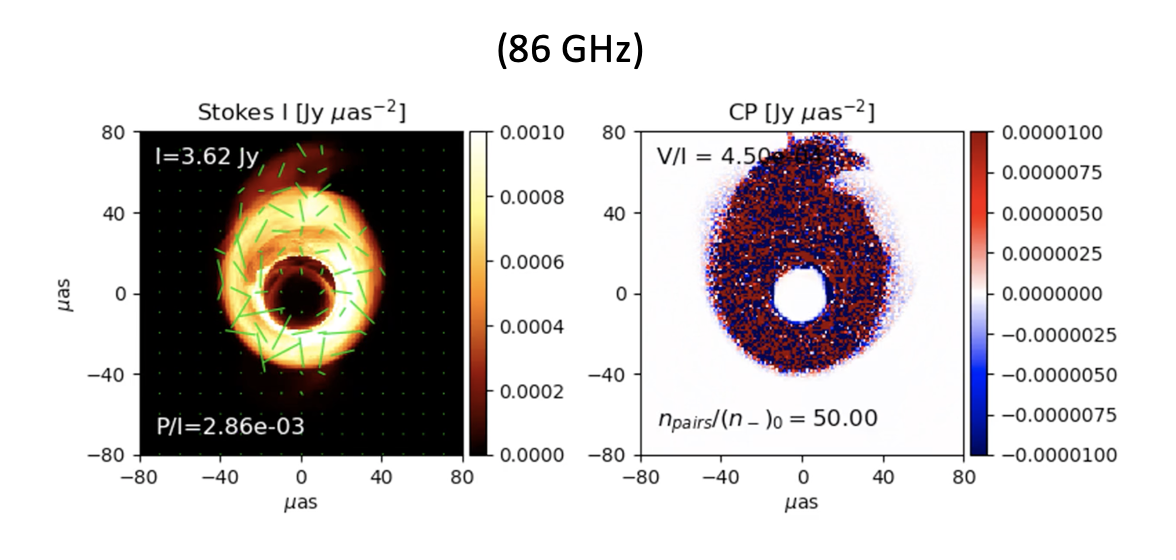}
\end{align}
\caption[Critical Beta Positron Effects]{\textcolor{black}{For the SANE $a=-0.5$ at T=25,000M at 86 GHz for the  R-Beta model with $\beta_{e0}=0.01$ jet: Top panel: Ionic plasma (fPos = 0); 
Bottom panel: fPos = 50.  }
}\label{SANERBetaWJet86GHz}
\end{figure}

\textcolor{black}{In Figs. \ref{MADRBetaWjet86GHz} and \ref{SANERBetaWJet86GHz}, we search for extended structure at 86 GHz in the R-$\beta$ with $\beta_{e0}=0.01$ Model in the MAD and SANE cases, respectively. }\textcolor{black}{In our 86 GHz images, we use the same Munit that normalized the 230 GHz images to .5 Jy, though now we have more flux with a larger field of view and shifting emitting regions at low frequency.} \textcolor{black}{The SANE Fig. \ref{SANERBetaWJet86GHz} in particular shows an upwardly extended feature reminiscent of the M87 jet base in \cite{2023arXiv230413252L}. }

\begin{figure}
\hspace{-0.25cm}
\includegraphics[height=150pt,width=250pt,trim = 6mm 1mm 0mm 0mm]{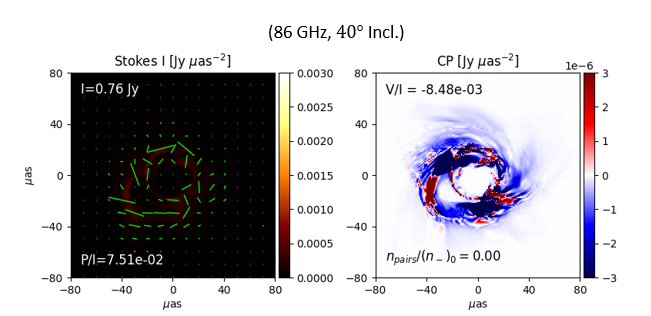}
\begin{align}\nonumber
& \hspace{-0.15cm} \includegraphics[trim = 6mm 1mm 0mm 1mm,  height=150pt,width=250pt
]{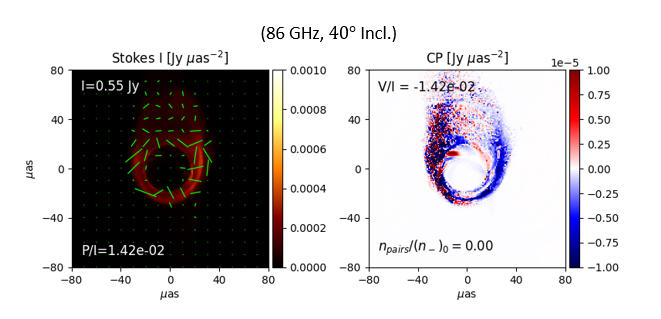}
\end{align}
\caption{\textcolor{black}{For $a=-0.5$ at $T=25,000M$ and 86 GHz for the  R-Beta model with $\beta_{e0}=0.01$ jet: a MAD (Top) vs. SANE (Bottom) comparison now with $40^\circ$ inclination. 
}}
\label{MAD-SANEpositron86GHz}
\end{figure}

\textcolor{black}{Changing observer inclination induces further distinctive image morphological features, as shown in  the 86 GHz maps in Fig. \ref{MAD-SANEpositron86GHz} with inclination angle $40^\circ$ (instead of the default M87 inclination  $17^\circ$ used throughout this work). In Fig. \ref{MAD-SANEpositron86GHz}, the  $R-\beta$ model with $\beta_{e0}=0.01$ jet has image plane projection horizontally elongated in the MAD case and vertically elongated in the SANE case relative to the default orientation. The jet collimation profile generally broadens as the jet inclination tilts away from the line of sight; the broader jet is more like observed in  \cite{2023arXiv230413252L}. More edge-on morphologies are expected to break degeneracies in face-on images due to general relativistic strong lensing effects.}

\subsection{Temporal Evolution}
\addtocounter{subsection}{1}\label{TimeDependence}

\begin{figure}
\begin{align}\nonumber\hspace{-0.7cm}
\includegraphics[height=135pt,width=245pt,trim = 6mm 1mm 0mm 1mm]{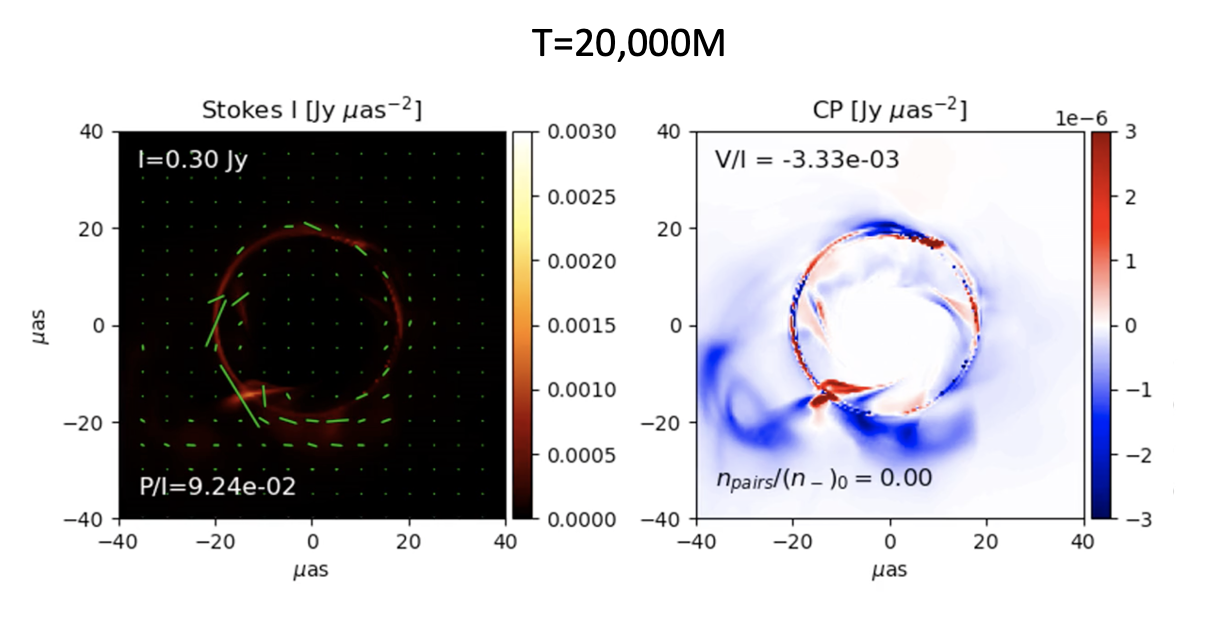}& \\
%
\hspace{0.3cm} \includegraphics[trim = 6mm 1mm 0mm 0mm,  height=140pt,width=250pt
]{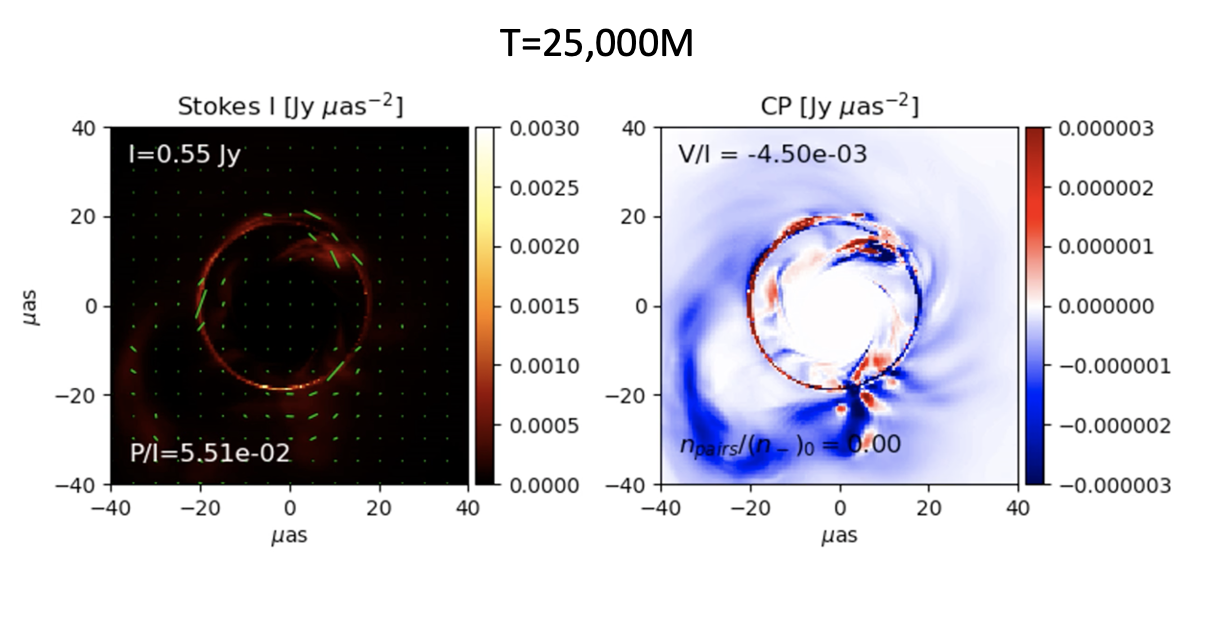}& \nonumber \\
\hspace{0.3cm} \includegraphics[trim = 6mm 1mm 0mm 0mm,  height=140pt,width=250pt
]{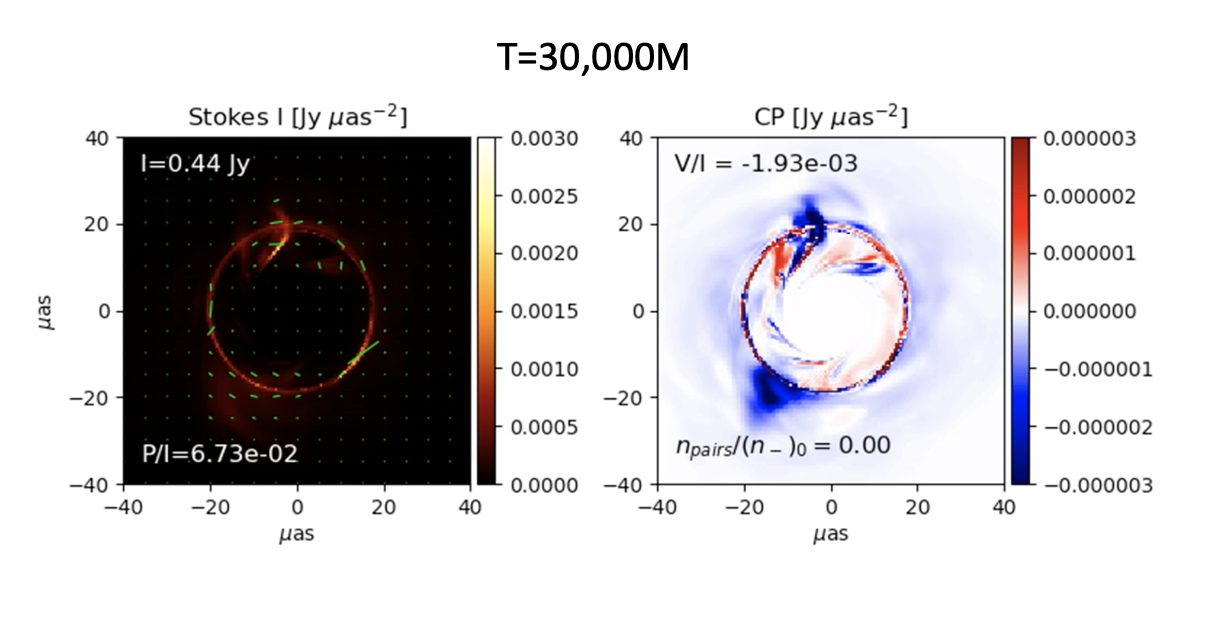} \nonumber &
\end{align}
\caption[R Beta Temporal Effects]{\textcolor{black}{For the $a=-0.5$ MAD at $T=25,000M$: R Beta at 230 GHz without positrons at $T=20,000M$ (Top), $T=25,000M$ (Middle) and $T=30,000M$ (Bottom).}}\label{TemporalVariationMADRBeta}
\end{figure}

\begin{figure}
\begin{align}\nonumber\hspace{-0.55cm}
\includegraphics[height=140pt,width=240pt,trim = 6mm 1mm 0mm 1mm]{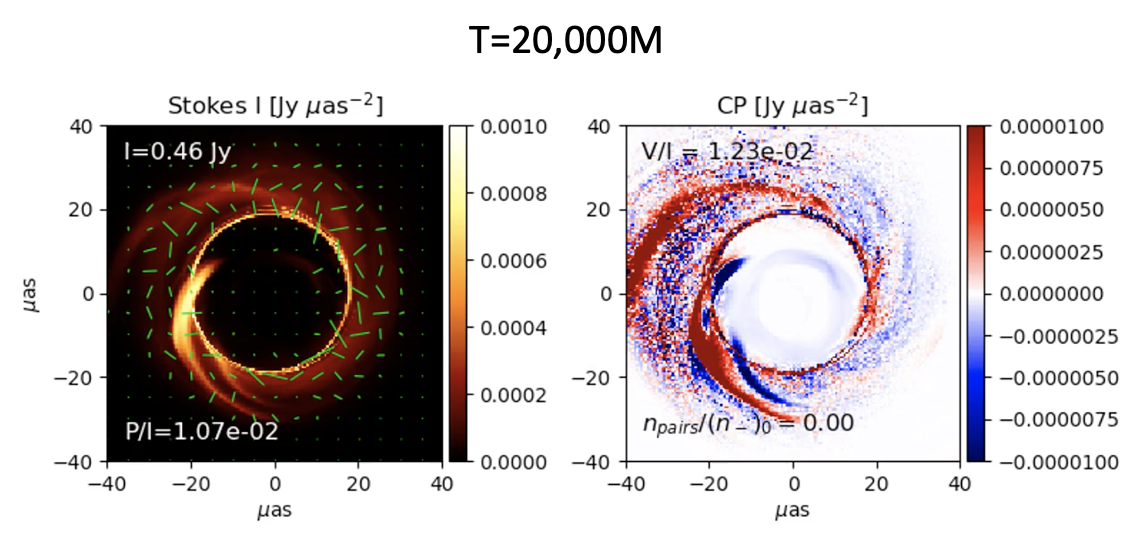}& \\
%
\hspace{-0.55cm} \includegraphics[trim = 6mm 1mm 0mm 0mm,  height=140pt,width=240pt
]{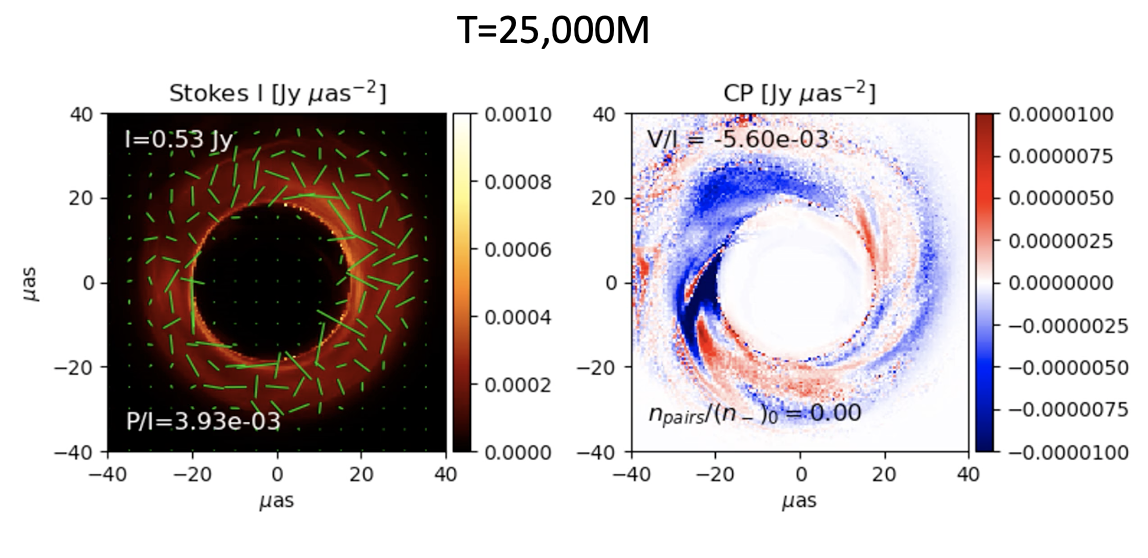}& \nonumber \\
\hspace{-0.55cm} \includegraphics[trim = 6mm 1mm 0mm 0mm,  height=140pt,width=240pt
]{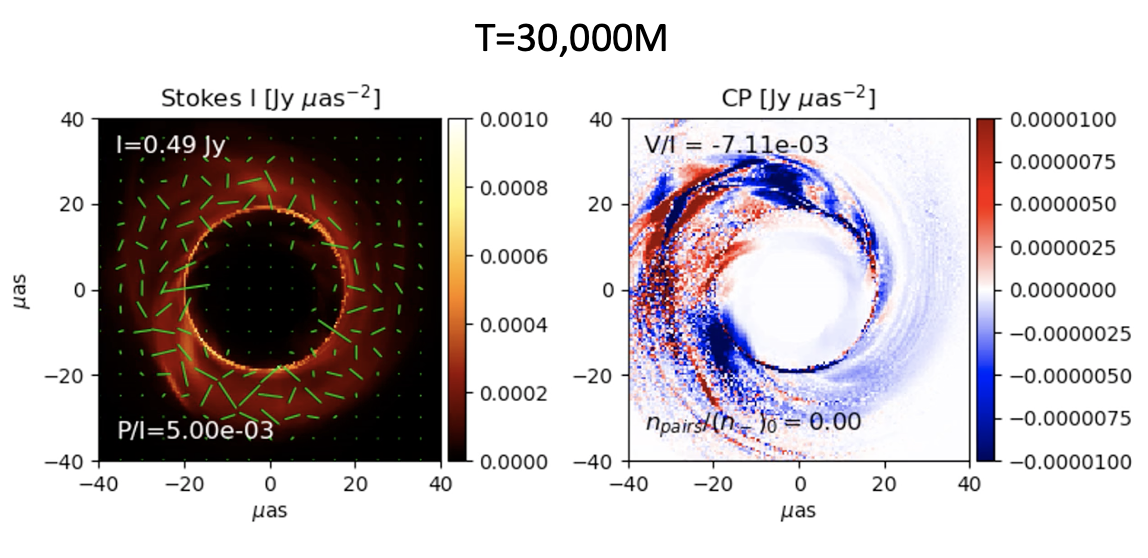}& \nonumber
\end{align}
\caption[R Beta Temporal Effects]{\textcolor{black}{For the $a=-0.5$ SANE at $T=25,000M$: R Beta at 230 GHz without positrons at $T=20,000M$ (Top), $T=25,000M$ (Middle) and $T=30,000M$ (Bottom).}}\label{TemporalVariationSANERBeta}
\end{figure}

\textcolor{black}{Figs. \ref{TemporalVariationMADRBeta} and  \ref{TemporalVariationSANERBeta} show temporal variation of the GRMHD simulations for the MAD and SANE cases, respectively. Both intensity and polarization  morphology vary noticeably over timescales of thousands of gravitational radii (years), as expected from observations \cite{Wielgus2020} in which bright spots appear at various azimuths throughout the M87 emitting ring. The flux eruption feature outside the photon ring at $T=25,000M$ is partially replaced by a spiral arm at $T=20,000M$ and a smaller extrusion at $T=30,000M$ in the circular polarization maps. }
 
\begin{figure}
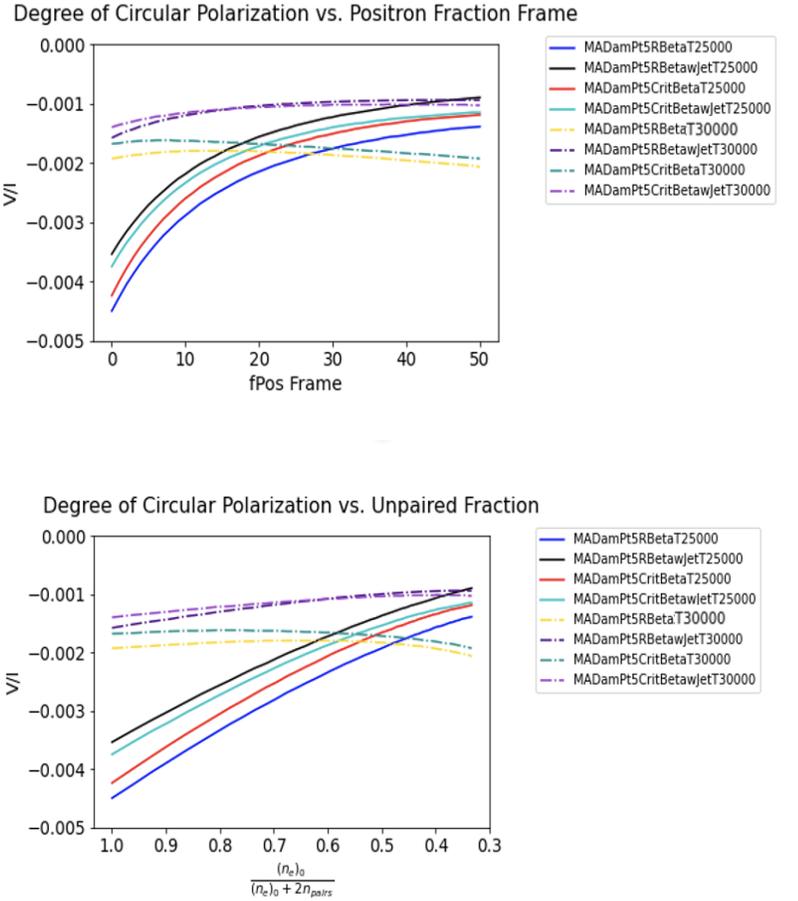

\hspace{-0.5cm}
\includegraphics[height=180pt,width=300pt
,trim = 6mm 1mm 0mm 1mm]{LaniM87VIv2
.png}
\begin{align}\nonumber
& \hspace{-0.5cm} \includegraphics[trim = 6mm 1mm 0mm 0mm,  height=190pt,width=300pt
]{LaniVIvUnpairedv3
.png}
\end{align}
\caption[Time Evolution  Effects]{\textcolor{black}{ Degree of circular polarization as a function of $\frac{n_\mathrm{pairs}}{(n_e)_0}$ (Top Panel) and unpaired synchrotron emitter fraction $\frac{(n_e)_0}{(n_e)_0+2n_\mathrm{pairs}}$  (Bottom Panel). The fiducial time is $T=25,000M$ and has characteristically different monotonicity behavior $5,000M$ after.}}\label{VOnITemporalEvolution}
\end{figure}

\textcolor{black}{Figure \ref{VOnITemporalEvolution} compares the evolution of circular polarization fraction with positron fraction at fiducial time $T=25,000M$ with that of a later snapshot of the simulation at $30,000M$. \textcolor{black}{Fifty-one different positron fractions are used in the series of frames  producing these curves representing the variation of $V/I$ with positron fraction.}  The fiducial time with a prominent flux ejection loop has $V/I$  monotonically going from the most negatively polarized to approaching 1/3 this value linearly in the fraction of unpaired emitters (slightly slower than linearly around $\frac{(n_e)_0}{(n_e)_0+2n_\mathrm{pairs}}=1/3$ where the plasma is an even mix of electrons, positrons and protons).  }

\begin{figure}
\hspace{-0.5cm}
\includegraphics[height=130pt,width=270pt,trim = 6mm 1mm 0mm 1mm]{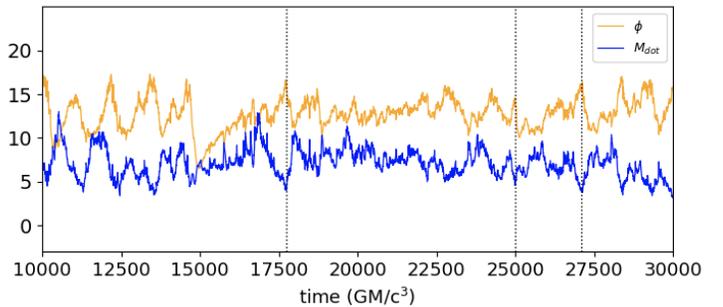}
\caption[Time Evolution  Effects]{\textcolor{black}{Time series of fiducial MAD $a=-0.5$ simulation horizon threading flux $\phi$ (Yellow) and mass accretion rate $\dot{m}$ (Blue) for $10000M<T<30000M$. The vertical lines at $T=17,730M,25,000M$ and $27,110M$ correspond to sharp peaks in $\phi$, troughs in $\dot{m}$ and flux eruption loops.}}\label{PhiMDotTimeSeries}
\end{figure}

\textcolor{black}{The flux loop occurrence at fiducial time $T=25,000M$ of the MAD $a=-0.5$ simulation may be representative of a broader episodic phenomenon occurring throughout the evolution of the flow. The Fig \ref{PhiMDotTimeSeries} time series of mass accretion rate $\dot{m}$ and horizon-threading flux $\phi$ from  $10,000M<T<30,000M$ reveals at $T=25,000M$ a sharp rise in $\phi$ accompanied by a sharp decrease in $\dot{m}$, which accords with the flux eruption scenario where a highly polarized magnetic flux loop is added to a magnetically arrested disk. Similar loop morphologies were observed for $T=
17,730M$ and $
27,110M$ where the simulation time series have similar peaks in $\phi$ and troughs in  $\dot{m}$  as at $T=25,000M$. }

\textcolor{black}{
The MAD advantage found at the fiducial time $T=25,000M$ for our emission models with respect to our polarization constraints is 
fairly consistent 
in time, 
with the cavaet that we continue to rely on the net linear polarization $|m|_\mathrm{net}$ measurement less subject to uncertainties related to the EHT beam resolution than $<|m|>)$. 
For evenly spaced samples in the range $20,000M\le T\le 30,000M$, MADs pass a higher percentage of $|m|_\mathrm{net}$ 
constraints for $82\%$ of times while the corresponding spin SANEs outperform MADs for $18\%$ of times (note for the resolved case, 
however, 
SANEs pass more frequently than MADs 73\% of times). The SANE/MAD dichotomy in $|m|_\mathrm{net}$ favoring MAD models while $<|m|>$ favors SANEs manifests in Table \ref{Model_Linear_Polarization_Table_Median} comparing the median values in a sample of evenly spaced times from $20,000M\le T\le 30,000M$.
}

\textcolor{black}{New circular polarization observations by the EHT \citep{EHT2023IX} more are more one-sided (favoring MADs) in our model parameter space. Performing a similar analysis for circular polarization, we find  $55\%$ of times MAD has a larger fraction of models passing the $|v|_\mathrm{net}$ constraint,  55$\%$ of times MAD has a larger fraction of models passing the $<|v|>$ constraint and $0\%$ of times SANE outperforms MAD with respect to these constraints (all other cases having ties due to all models passing the circular polarization constraints in such cases). In median CP Table \ref{Model_Circular_Polarization_Table_Median} over  $20,000M\le T\le 30,000M$, the only median that fails is for thee SANE $a/M=-0.5$ Critical Betw Model positrons with respect to the  $<|v|>$ constraint.}

\textcolor{black}{One of the most discriminating constraints for MAD versus SANE in EHT’s relatively large simulation libraries represented in the Pass-Fail tables of EHT M87 Paper V \citep{EHT2019V}  and M87 Polarization Paper VIII \citep{EHT2021VIII} is the jet power, which heavily 
excludes SANEs, 
is observationally uncertain (ranging between $10^{42}$ to and $10^{45}$ erg/s) and which we do not consider here with our limited number of simulations. The 
overwhelming 
MAD advantage found by EHT depends on both the choices for simulations/emission models to be included in the library and the particular constraints chosen to make the comparison, underscoring the need for alternative investigations such as this exploring new regions of parameter space. 
}

\begin{table*}
\caption{\textcolor{black}{\textcolor{black}{Median linear polarization tables for simulation times 20,000$M\le T\le 30,000M$. Note that the bold values refer to SANE and MAD fiducial models with $a/M=-0.5$ which satisfy the 
linear polarization  $|m|_\mathrm{net}$ and $<|m|>$  
constraints from EHT M87 Paper VII, 
which 
take the form of the polarization ranges $0.01\le|m|_\mathrm{net} \le 0.037$ and $0.057< <|m|> < 0.107$.
} \textcolor{black}{
We 
indicate models
satisfying 
polarization constraints 
only at the number of constraint significant figures in italics.}
}
}
\makebox[ 0.83 \textwidth][c]{
\begin{tabular}{c|ccccc|ccccc|ccccc}
\hline
&$.$ & $\mathrm{SANE}$  & ($a=-0.5$) &  &  & $\mathrm{MAD}$ & ($a=-0.5$) 
\\ 
 &$R\mathrm{-Beta}$ &  $R\mathrm{-Beta}$  & $\mathrm{Crit.\ Beta}$ & $\mathrm{Crit.\ Beta}$&  $R\mathrm{-Beta}$ &  $R\mathrm{-Beta}$  & $\mathrm{Crit.\ Beta}$ & $\mathrm{Crit.\ Beta}$  \\ 
&&$\mathrm{w./\ Jet}$&&$\mathrm{w./\ Jet}$&&$\mathrm{w./\ Jet}$&&$\mathrm{w./\ Jet}$\\ 
 \hline
 $|m|_\mathrm{net}(f_\mathrm{pos,min})$&$5.00\cdot10^{-3}$ & $7.26\cdot10^{-3}$  & $3.46\cdot10^{-3}$  & $5.84\cdot10^{-3}$  & $4.92\cdot10^{-2}$ & $\mathbf{2.90\cdot10^{-2}}$ & $4.77\cdot10^{-2}$  & $3.44\cdot10^{-2}$  \\ \hline
$|m|_\mathrm{net}(f_\mathrm{pos,max})$ &$3.81\cdot10^{-3}$ & $6.56\cdot10^{-3}$ & $2.74\cdot10^{-3}$  & $6.14\cdot10^{-3}$  & $\textcolor{black
}{\textbf{\textit{3.71}}\cdot\textbf{\textit{10}}^{\textbf{\textit{-2}}}}$ & $\mathbf{3.10\cdot10^{-2}}$ & $4.07\cdot10^{-2}$ & $\mathbf{3.31\cdot10^{-2}}$  \\ \hline 
$<|m|>(f_\mathrm{pos,min})$ &$\mathbf{9.43\cdot10^{-2}}$ & $\mathbf{1.02\cdot10^{-1}}$ & $\mathbf{6.17\cdot10^{-2}}$  & $\mathbf{8.85\cdot10^{-2}}$  & $3.49\cdot10^{-1}$ & $3.30\cdot10^{-1}$ & $2.71\cdot10^{-1}$ & $2.61\cdot10^{-1}$  \\ \hline 
$<|m|>(f_\mathrm{pos,max})$ &$\mathbf{9.49\cdot10^{-2}}$ & ${1.08\cdot10^{-1}}$ & $\mathbf{6.55\cdot10^{-2}}$  & $\mathbf{9.72\cdot10^{-2}}$  & $4.49\cdot10^{-1}$ & $4.31\cdot10^{-1}$ & $3.88\cdot10^{-1}$ & $3.91\cdot10^{-1}$  \\ \hline 
\end{tabular}
}
\label{Model_Linear_Polarization_Table_Median}
\end{table*}

\begin{table*}
\caption{\textcolor{black}{Median circular polarization tables for simulation times 20,000$M\le T\le 30,000M$. Note that the italic values refer to SANE and MAD fiducial models with $a/M=-0.5$ that fail to satisfy the constraints $|v|_\mathrm{net}\le 0.008$ and $<|v|>\le 0.037$  
%
%
}
}
\makebox[ 0.83 \textwidth][c]{
\begin{tabular}{c|ccccc|ccccc|ccccc}
\hline
&$.$ & $\mathrm{SANE}$  & ($a=-0.5$) &  &  & $\mathrm{MAD}$ & ($a=-0.5$) 
\\ 
 &$R\mathrm{-Beta}$ &  $R\mathrm{-Beta}$  & $\mathrm{Crit.\ Beta}$ & $\mathrm{Crit.\ Beta}$&  $R\mathrm{-Beta}$ &  $R\mathrm{-Beta}$  & $\mathrm{Crit.\ Beta}$ & $\mathrm{Crit.\ Beta}$  \\ 
&&$\mathrm{w./\ Jet}$&&$\mathrm{w./\ Jet}$&&$\mathrm{w./\ Jet}$&&$\mathrm{w./\ Jet}$\\ 
 \hline
 $|V|_\mathrm{net}(f_\mathrm{pos,min})$&$-3.53\cdot10^{-3}$ & $-2.26\cdot10^{-3}$  & $-4.53\cdot10^{-3}$  & $-1.42\cdot10^{-3}$  & $-3.85\cdot10^{-3}$ & $-2.61\cdot10^{-3}$ & $-3.27\cdot10^{-3}$  & $2.34\cdot10^{-3}$  \\ \hline
$|V|_\mathrm{net}(f_\mathrm{pos,max})$ &$-2.81\cdot10^{-3}$ & $1.76\cdot10^{-3}$ & $2.02\cdot10^{-3}$  & $1.85\cdot10^{-3}$  & $-3.14\cdot10^{-3}$ & $-1.36\cdot10^{-3}$ & $-3.19\cdot10^{-3}$ & $-1.54\cdot10^{-3}$  \\ \hline 
$<|V|>(f_\mathrm{pos,min})$ &$8.39\cdot10^{-3}$ & $7.07\cdot10^{-3}$ &$1.67\cdot10^{-2}$  & $8.12\cdot10^{-3}$  & $3.24\cdot10^{-3}$ & $2.26\cdot10^{-3}$ & $3.05\cdot10^{-3}$ & $2.32\cdot10^{-3}$  \\ \hline 
$<|V|>(f_\mathrm{pos,max})$ &$1.42\cdot10^{-2}$ & $1.04\cdot10^{-2}$ & $\mathit{3.99\cdot10^{-2}}$  & $1.86\cdot10^{-2}$  & $1.23\cdot10^{-3}$ & $6.11\cdot10^{-4}$ & $1.17\cdot10^{-3}$ & $6.75\cdot10^{-4}$  \\ \hline 
\\ 
\end{tabular}
}
\label{Model_Circular_Polarization_Table_Median}
\end{table*}

$\qquad$

\section{Discussion and Conclusions}


There have been remarkable advances in imaging and simulating AGN jets over the past \textcolor{black}{couple of decades.} 
Despite this progress there are potentially vital components  --- 
the jet composition and relativistic particle acceleration remains
\textcolor{black}{which remain controversial.} Our methodology \textcolor{black}{to address these} is to focus on one well-studied source, M87, and one region of the electromagnetic spectrum, radio, millimeter and submillimeter, and to incorporate different phenomenological prescriptions to bridge this divide into the simulations and then ``observe'' them. The actual observations, especially \textcolor{black}{those} 
from the Event Horizon Telescope, can then be used to discern empirically some of the rules that govern jet formation, collimation\textcolor{black}{, polarization} and dissipation. This approach can be extended using more sources, frequencies and simulations and statistical comparisons can also be conducted. These extensions will be discussed in future publications \textcolor{black}{including the completion of this series on Sgr A*, M87 and 3C 279}.  

The \textcolor{black}{G}RMHD model that we have used to develop a more generally applicable set of techniques is quite specific in terms of spin (\textcolor{black}{$a/M_H=-0.5,0.94$})
 and disposition of the surrounding gas (dense orbiting torus with non-relativistic wind at high latitude outside the jet). The magnetic flux density and polarity were consequences of the conditions of the simulations. Given these boundary conditions, the concentration of horizon-crossing magnetic flux and the formation of an electromagnetic \textcolor{black}{outflow or} jet are inevitable.

Within the Bondi radius ($\sim10^5M$), the jet profile is roughly parabolic, consistent with other simulations, e.g., \cite{2013MNRAS.436.3741P}. The form of the flux and  velocity variation across the jet should also be reasonably generic, though the stability properties and entrainment at the jet surface is probably sensitive to the numerical details. In conclusion, we should have a pretty representative \textcolor{black}{suite of simulations} of the flow and the field to link to the highest resolution \textcolor{black}{mm-observations.}

%


The 
\textcolor{black}{\say{Observing} JAB simulations} methodology reproduces a surprising number of observed signatures of M87 morphology and dynamics. Starting with \textcolor{black}{turbulent heating models including that used by the EHT, R-$\beta$, and the Critical $\beta$ model of \cite{Anantua2020a}a, we have the expected  ring-like global structure for intensity and EVPAs strongest in local maxima of intensity. Adding} equipartition-inspired models, the jet magnetic substructure for \textcolor{black}{Constant Electron Beta} 
models characterized by constant \textcolor{black}{electron-}gas\textcolor{black}{-to-magnetic} pressure along the jet  
\textcolor{black}{gives a more broadly distributed profile. }

In the case of M87, 
the radio emissivity is not simply a function of the gas and/or the magnetic pressure. So the rule for particle acceleration must depend upon other factors (e.g. $\beta$ and $\Gamma{\bf v}$). \textcolor{black}{We have implemented models where the emissivity is governed by total plasma $\beta$ in the turbulent plasma and by conversion of magnetic-to-particle energy (parametrized by the contribution $\beta_e$ of radiating electrons and positrons) in the relativistic jet. }


\textcolor{black}{Our models also go beyond what is currently directly observable in simulating the effects of incrementally changing positron fraction; however, SANE and MAD produce a sharp enough dichotomy to currently be distinguished. The key finding is that polarization is a sharp cleaver distinguishing SANE and MAD accretion flows. Particularly, we find distinct polarized emission signatures that depend on the positrons content in radically different ways for  SANE and MAD simulations.}

In summary, the primary findings of the \say{observing} simulations methodology \textcolor{black}{applied to M87} include:

\begin{itemize}




\item{\textcolor{black}{Both $R-\beta$ and Critical $\beta$ turbulent heating models produce ring-like intensity profiles with some 
\textcolor{black}{MAD} 
cases satisfying linear polarization constraints and all satisfying preliminary circular polarization upper bounds. } 
}

\item{\textcolor{black}{The piecewise addition of a Constant $\beta_e$ jet tends to broader annular emission profiles.} 
}

\item{\textcolor{black}{MAD and SANE images with polarization at constant overall flux have markedly different morphological properties. The MAD can exhibit a prominent flux eruption in intensity and linear polazrization. } 
}

\item{
\textcolor{black}{The Faraday depths of the SANE are 2-3 orders greater than for MAD. The SANE linear polarization is more disordered and circular polarization structure is completely scrambled.}}

\item{\textcolor{black}{The circular polarization degree for MAD maps dominated by intrinsic $V/I$ exhibit a linear vanishing of $V/I$ in the fraction of paired emitters. }
}

\end{itemize}



\textcolor{black}{
The AGN environment is  certainly a messy and chaotic one-- replete with winds, gas, dust and molecular clouds to name a few. The task of emission modeling jet/accretion flow/black hole systems in such an uncertain setting, on the other hand, is a fertile wonderland for the creation of theoretical models and the discovery of new phenomenology. With few constraints on black hole spin or jet composition, vast libraries of GRMHD libraries remain viable for even the most well studied sources like M87.
The ``Observing" JAB Simulations approach embraces this uncertainty by using piecewise models and generic plasma compositions to allow for complex interactions leading to unexpected results such as the positron-mediated Faraday effects leading to the sharp SANE-MAD dichotomy in polarization signatures illustrated in this work.
 The present application leaves us not only closer characterizing M87 as polarized MAD flow near horizon scales, but also to narrow the possible plasma descriptions for other JAB systems such as the jetted AGN 3C 279 that will be the third work of this series-- and  the vast universe of future horizons to be discovered.
}

\section{Future Directions}

\textcolor{black}{With our suite of turbulent and sub-equipartition heating models with positrons, we have taken a key step in bridging rapidly advancing GRMHD simulations and observations.  The stark SANE-MAD dichotomy found in polarized intensity spatial distribution and time evolution presents a key opportunity to  rule out SANE models of M87 by comparing variability, e.g., EVPA rotation rate, with the results of M87 2017 combining later observing campaigns. } 

\textcolor{black}{It has been demonstrated that  prescriptions involving dissipation as a function of effective magnetic field $B_e=\mathcal{D}|\vec{n}\times\vec{B}|$ exhibit violation of bilateral symmetry across the jet axis both in the stationary, axisymmetric, self-similar semi-analytic model 
\citep{Blandford2017,Anantua2020a}a \textcolor{black}{(with general relativistic ray tracing in \cite{emami2021positron})}, and in the time-dependent 3D \textcolor{black}{GRMHD} simulation in \cite{Anantua2018}. Though barely visible in 
M87 observations, e.g., at 86 GHz \cite{2018arXiv180502478K}, this is predicted to be a robust -- albeit, generic -- observation for EHT, with details depending on whether EHT sees a jet or disk-jet in the inner few gravitational radii from the hole.  Signs of this bilateral asymmetry from ``Observing" JAB Simulations have appeared in 230 GHz EHT observations of 3C 279 \citep{Kim2020}. We may implement prescriptions in $B_e$ in future emission modeling work to reproduce bilateral asymmetry-- particularly for 3C 279. }

\textcolor{black}{FRI disk wind momentum carries jet, while FRII jet momentum carries the wind. 
\textcolor{black}{Another way jets exchange momentum with their surroundings is through currents. } We can apply the current density model \citep{Anantua2018} 
to investigate whether
 current sheets  account for limb-brightening past 100$M$. The $B$-field alone struggles to remain toroidal past $100M$ unless it’s replenished by the disk.}
\textcolor{black}{In addition to currents,}
\textcolor{black}{we may systematically associate the dissipation in JAB systems with a number of plausible physical mechanisms, such as 
Shakura-Sunyaev momentum transport and Newtonian shear as outlined in the Appendix. These phenomenological models give firm theoretical intuition behind the physical mechanisms powering jets. }

\textcolor{black}{In future work, we will also incorporate positrons in a broader range of emission models. We may use positron production rates from \cite{Wong2021} to evolve the local positron fraction-- a key advance over the single-positron-ratio maps used here. The computational expense of a three-fluid ($e^-,e^+,p$) simulation may be mitigated by spatial symmetry and temporal stationarity of some simulated flows. 
}

A key feature of the \say{Observing} JAB simulations exercise presented here is its generality: a simulation of a general relativistic magnetohydrodynamic flow onto a compact object is flexible enough to model jets from neutron stars, black hole/x-ray binaries and AGN alike. 
In this work, we started with a 
\textcolor{black}{suite of simulations} fairly representative of an AGN in that it exhibited the commonly occurring combination of a thick ion torus confining electromagnetic flux from a polar outflow from a rotating black hole, then fine-tuned it to M87 to emulate its observed 
\textcolor{black}{JAB system polarized} substructure. Disk emission 
\textcolor{black}{has been} emphasized in 
\textcolor{black}{other } work, starting with Sgr A* at our Galactic Center, replete with new near horizon observations of photon rings courtesy of EHT. Our models will also be applied to near-horizon emission in future EHT observational targets such as the highly variable quasar 3C 279 \textcolor{black}{in the last work of this trilogy}.

\section*{Acknowledgments}
Richard Jude Anantua \textcolor{black}{was} supported by the California Alliance \textcolor{black}{at the outset of this investigation and the Oak Ridge Associated Universities Powe Award for Junior Faculty Enhancement 
towards the end}.\textcolor{black}{ This work was supported by a grant from the Simons Foundation (00001470, RA, LO, JD and BC).}  Roman Shcherbakov \textcolor{black}{and Alexander Tchekhovskoy have} provided excellent guidance and mentorship at the beginning of this investigation. 
\textcolor{black}{UTSA undergraduate Noah \textcolor{black}{Heridia} 
has been helpful through graphic-related activities. \textcolor{black}{BASIS Shavano San Antonio student Luke Fehlis provided valuable input on data analysis. }
\textcolor{black}{Daniel Palumbo provided observational guidance.} 
Angelo Ricarte was supported by the Black Hole Initiative at Harvard University, made possible through the support of grants from the Gordon and Betty Moore Foundation and the John Templeton Foundation. The opinions expressed in this publication are those of the author(s) and do not necessarily reflect the views of the Moore or Templeton Foundations.}
{
Razieh Emami acknowledges the support by the Institute for Theory and Computation at the Center for Astrophysics as well as grant numbers 21-atp21-0077, NSF AST-1816420 and HST-GO-16173.001-A.}

\section{Data Availability}
The data underlying this article will be shared on reasonable request to the corresponding author.

\newpage
\nocite{*}
\bibliographystyle{mn2e}
\bibliography{M87PaperReferences}

\newpage
\section*{Appendix A. Alternative Phenomenological Emission Models}
\textcolor{black}{To extend the range of phenomenology covered by emission models in this work, we add some of the models introduced in \cite{AnantuaThesis2016}.These models are in terms of the partial pressure due to electrons that contribute to emission near the observed frequency.}

\subsection*{A.1 Magnetic Bias Model}

The calculated intensity in simple constant electron $\beta$ models typically declines with radius far faster than the observations. This can be addressed by stipulating that $\beta\propto P^{-N}$, where the exponent $N$ is typically in the range $\sim0.25-0.5$ and the constant of proportionality can be set at a fiducial radius.
%

We can incorporate these considerations in 
a Magnetic Bias Model by scaling $P_e\propto P_{\rm mag}^n \propto b^{2n}$, which near the black hole reduces to $P_e\propto P^n\propto P^{1-N}$. In this regime, Bias Models for $n=0$, i.e., $N=1$, have constant pressure along the jet. 
$\qquad$
\newline
\noindent
\subsection*{A.2 $\alpha$ Model}
A more 
detailed 
approach to setting the partial pressure is to suppose that the free energy in the jet is dissipated at a specific rate $W'$ in the comoving frame and that the associated 
energy density is proportional to $u_e$ $\sim0.1W' t_{\rm loss}'$ where $t_{\rm loss}$ is the smaller of the radiative cooling time and the expansion time. There are many possible models for the shear stress but, as with accretion disks (\citep{1973A&A....24..337S}), it suffices to assume that the shear stress $\tau$ is a fixed fraction ($\alpha$) of the total pressure, $P=P_{\rm mag}+P_g+...$. Close to the hole $P\sim P_{\rm mag}$. The dominant component of the rate of velocity shear evaluated in the comoving frame is $S'\sim(1-v_z^2)^{-1}|dv_z/ds|$ so that $W'=\frac12\tau'S'$.

\subsection*{A.3 Shear or Velocity Gradient ($S^2$) Model}

The shear, or velocity gradient model, which we shall also term $S^2$-model 
\textcolor{black}{due to its dependence on $S$ both via shear stress and velocity shear, is based on the Newtonian model for viscosity using a} 
characteristic length scale $L_S$ and 
\textcolor{black}{setting } 
kinematic viscosity $\nu'=cL/3$. The dynamic viscosity is then obtained by multiplying $\nu'$ by density (or energy density divided by $c^2$)
\begin{equation}
\mu'=\frac{L_S}{3c}\sqrt{\left(\rho c^2+\frac{b_\mu b^\mu}{2\mu_0}\right)\left(\frac{u_g}{3}+\frac{b_\mu b^\mu}{2\mu_0}\right)}
\end{equation}
Since the shear stress $\tau'=\mu' S'$, the dissipation rate in this model is then quadratic in the shear stress: $W'=\frac{1}{2}S'\tau'=\frac{1}{2}\mu'S'^2$, and $\tilde{P}_e=W'\min[t_\mathrm{exp}',t_\mathrm{rad}']$ as above.
 
$\qquad$
\newline
\noindent
$\mathbf{A.4\ Summary\ of\ 
Pressure\mathrm{-}to\mathrm{-}Dissipation\ Prescriptions}$
$\qquad$
\newline 
$\qquad$
\newline
\noindent
We now write complete formulas for the partial pressures in our models with explicit dissipation. Putting the parts of the $\alpha$, $S^2$ and $j^2$ model all together
\begin{equation}
\begin{split}
P_\mathrm{e,\ \alpha\ model}=\frac{1}{2}\alpha\left(\frac{b_\mu b^\mu}{2\mu_0}\right)\left(\gamma^2|\frac{dv_z}{ds}|\right)t' \hspace{4cm}\\
P_\mathrm{e,\ S^2\ model}=\frac{L}{6c}\sqrt{\left(\rho c^2+\frac{b_\mu b^\mu}{2\mu_0}\right)\left(\frac{u_g}{3}+\frac{b_\mu b^\mu}{2\mu_0}\right)}\left(\gamma^2\frac{dv_z}{ds}\right)^2t' \hspace{2cm}\\
P_\mathrm{e,\ j^2\ model}=\mu_0cL|j_\mu j^\mu|t' \mathrm{,\ \ where\ }\hspace{5cm}
\\t'= \mathrm{min}\left\lbrace\frac{1}{\gamma|\gamma\vec{\triangledown}\cdot\vec{v}+\vec{v}\cdot\vec{\triangledown}\gamma|},\frac{\mu_0c}{\sigma_T}\sqrt{\frac{3em_e}{2\pi}}\mathcal{D}^{-1}B_e^{-2/3}\nu^{-1/2}\right\rbrace\hspace{2cm}
\end{split}
\end{equation}
(note the $\alpha$ model is assumed to be dominated by magnetic pressure).

\textcolor{black}{Numerical implemetation of these modes is facilitated by expressing the $s-$derivative in velocity shear as}
\begin{equation}
\left(\frac{dv_z}{ds}\right)^2=\left(\frac{dv_z}{dx}\right)^2+\left(\frac{dv_z}{dy}\right)^2
\end{equation} 
\textcolor{black}{avoiding division by 0 at the 
axis
}
and 
\textcolor{black}{expressing }
\begin{equation}
\vec{\triangledown}\gamma=
\begin{pmatrix} \frac{\partial }{\partial x}  \\\frac{\partial }{\partial y}   \\ \frac{\partial }{\partial z}   
  \end{pmatrix}
\left(1-\frac{v^2}{c^2}\right)^{-\frac{1}{2}}=\frac{1}{c^2}\gamma^3v
\begin{pmatrix} \frac{\partial v}{\partial x}  \\\frac{\partial v}{\partial y}   \\ \frac{\partial v}{\partial z}   
  \end{pmatrix}
\end{equation}
\textcolor{black}{ where the derivatives along the axial directions is now}
\begin{equation}
\frac{\partial v}{\partial x_i}\sim  \frac{v\begin{pmatrix}\vec{x}+x_\mathrm{i\ Step}\hat{e}_i
\end{pmatrix}-v\begin{pmatrix}
\vec{x} 
\end{pmatrix}}{x_\mathrm{i\ Step}}
\end{equation}

$\qquad$
\newline
\noindent
\section*{Appendix B. Alternative Positron Distribution Models}
\subsection*{B.1 Positron Proxies by Plasma Variables}


\textcolor{black}{One strategy for estimating the local distribution of positrons is to relate positron density to plasma variables that positrons may trace. The radiation energy density, magnetic pressure, electron temperature and functions thereof may play this role. Inspired by successful parametrizations of electron temperature, we may take as a proxy for positron fraction $\frac{n_{e^+}}{n_{e^-}}$}

\begin{equation}
f_{e^+}(\beta)=f_\mathrm{max}e^{-\frac{\beta}{\beta_c}}
\end{equation}
or

\begin{equation}
R_{e^+}(\beta)=\frac{\beta^2}{1+\beta^2}f_\mathrm{min}+\frac{1}{1+\beta^2}f_\mathrm{max}
\end{equation}
so that

\begin{equation}
    P_e=P_{e^-}+P_{e^+}=\beta_{e0}(1+(f_{e^+}(\beta) \mathrm{\ OR\ }  R_{e^+}(\beta)))P_B.
\end{equation}

\subsection*{B.2 Positron Distribution from Geometry}


\textcolor{black}{We also propose a geometric model separating jet inner and outer (or funnel) regions by a critical parabolic streamline such that 
\begin{itemize}
    \item Inner  jet is supplied by pairs from the stagnation surface
    \item Outer funnel wall jet is supplied by the  analytic prescription for pair production given 
    above
\end{itemize}
}

$\mathit{Inner-Outer\ Jet\ Model\ Precriptions}$

We specify inner- and outer-jet emission prescriptions here:

\cite{Broderick2015} argue models in which pairs are produced near the horizon and $u_{e\pm}\propto B^2$ overproduce core emission relative to jet emission. We may use the spark gap estimate for lepton number density from \cite{Broderick2015} and $u_{e\pm}=\gamma_{e\pm}n_{e\pm}m_ec^2$ to construct inner jet emissivity
$$ 
j_{\nu,\mathrm{in}}=\begin{cases}
u_{e\pm} P_B^\frac{3+\alpha}{2}\nu^{-\alpha},\ z<z_{\mathrm{max}},\ 0< \xi<\frac{1}{2}\xi_{\mathrm{max}}
\\
\beta_{\mathrm{e0}}P_B^\frac{3+\alpha}{2}\nu^{-\alpha},\ z\ge z_{\mathrm{max}},\ 0< \xi<\frac{1}{2}\xi_{\mathrm{max}}
\end{cases}
$$
where $z_\mathrm{max}\approx 10r_g$, the altitude of the stagnation surface.


For the outer jet, we use the model of \cite{Moscibrodzka2011}
,  
where 
synchrotron photons from a radiatively inefficient accretion flow collide with the jet funnel wall, where  they undergo photon annihilation via Breit-Wheeler process. We may adopt this as a model for the outer jet source of emitting jet electrons and positrons.
$$ 
j_\nu=\begin{cases}
j_{\nu,\mathrm{in}}
,\ 0< \xi<\frac{1}{2}\xi_{\mathrm{max}}
\\
\frac{\sqrt{2}\pi e^2n_e\nu_s}{3cK_2(\Theta_e^{-1})}(X^{1/2}+2^{11/12}X^{1/6})^2e^{-X^{1/3}}
,\ \frac{1}{2}\xi_{\mathrm{max}}< \xi<\xi_{\mathrm{max}} 
\end{cases}
$$
, where $X\equiv\nu/\nu_s,  \nu_s=\frac{2}{9} \frac{eB}{2\pi m_ec}\Theta_e^2\sin\theta$, $\xi=s^2/z$ and $\alpha=\frac{p-1}{2}$ .

\label{lastpage}

\end{document}